Free University of Bozen-Bolzano

Doctoral Thesis

# Towards a Theory of Affect and Software Developers' Performance

*Author:*

**Daniel Graziotin**

*Supervisors:*

Xiaofeng Wang

Pekka Abrahamsson

*Reviewers:*

Torgeir Dingsøyr

Robert Feldt

*Defense commission members:*

Tommi Mikkonen

Barbara Russo

Darja Šmite

*A thesis submitted in partial fulfillment of the requirements*
*for the degree of Doctor of Philosophy*

*in the*

Faculty of Computer Science

January 12, 2016

*"Junior programmer's bookshelf: 90% APIs and programming languages; Senior programmer's bookshelf: 80% applied psychology.."*

J. B. (Joe) Rainsberger, 1st July 2015, .

# *Abstract*

**Towards a Theory of Affect and Software Developers' Performance**

by Daniel GRAZIOTIN

For more than thirty years, it has been claimed that a way to improve software developers' productivity and software quality is to focus on people. This claim has been echoed by agile software development in the value "individuals and interactions over processes and tools". We should "build projects around motivated individuals. Give them the environment and support they need, and trust them to get the job done". High-tech companies have followed this advice by providing incentives and benefits to their workers. The underlying assumption seems to be that "happy and satisfied software developers perform better". More specifically, affects—emotions and moods—have an impact on cognitive activities and the working performance of individuals. Development tasks are undertaken heavily through cognitive processes, yet software engineering research lacks theory on affects and their impact on software development activities. This PhD dissertation supports the advocates of studying the human and social aspects of software engineering research and the psychology of programming. This dissertation aims to theorize on the link between affects and software development performance. A mixed method research approach was employed, which comprises studies of the literature in psychology and software engineering, quantitative experiments, and a qualitative study, for constructing a multifaceted theory of the link between affects and programming performance. The theory explicates the linkage between affects and analytical problem-solving performance of developers, their software development task productivity, and the process behind the linkage. The results are novel in the domains of software engineering and psychology, and they fill an important lack that had been raised by both previous research and by practitioners. The implications of this PhD lie in setting out the basic building blocks for researching and understanding the affect of software developers, and how it is related to software development performance. Overall, the evidence hints that happy software developers perform better in analytic problem solving, are more productive while developing software, are prone to share their feelings in order to let researchers and managers understand them, and are susceptible to interventions for enhancing their affects on the job. Finally, this work has sparked the guidelines for "psychoempirical software engineering"—that is empirical software engineering research with psychology theory, methodologies, and measurements—which, when applied to the existing proposals to the psychology of software engineering, will reveal a new understanding of the complex software construction process.

# *Acknowledgements*

Roughly 12 years ago, there was this boy in a scientific lyceum. The boy had felt out of place for a long time. His aspiration was to become a mechanic and to fix engines and cars. He did not even want to be in that school. The student found textbooks and classic frontal lectures to be extremely boring. His opinions of any knowledge-related job such as teaching or researching were pretty negative and undermined by misconceptions and ignorance. As a consequence of these happenings, he was not really considered a clever student. His marks were fairly below average. His behavior was not even among the bests one could have. Needless to say, his family has suffered from this. They were worried about his future. However, there was this lecturer for a course that was taught on that school year only. The lecturer had a PhD (or he was about to have one, I am not sure), so he knew *something* about researching, investigating, finding sources and working with them, criticizing, building on top of them, and so on. The lecturer somehow understood that the boy was pissed off at the way people taught at him. The lecturer told the boy he knew that he was not stupid at all, and he gave him a task to do for the next week. The Iraq war had recently started. The extra homework demanded the student to read the major national newspapers—of different political sides—for some days. The student needed to look for news articles describing the same *facts* on the Iraq war, and to compare them. How boring, the student thought, as he had to read the *same thing* four, five, even six times a day for some days. Eventually, the student realized that reading newspapers was less boring than reading the school's textbooks and, surprise, these *facts* were depicted differently by the different newspapers! The results of the comparison of the news articles were reported in a little paper that the student wrote using LaTeX(as the homework was a good opportunity for learning it). The report was actually nothing too special. Yet, the process that brought the student to write it, which consisted in acquiring sources, comparing them, and realizing how what he considered as *fact* was fuzzy and varied according to who was describing it, was the most exciting thing the boy had ever done at school. Experiencing the process sparked him a never-ending thirst of knowledge that perhaps brought him later on to engage in several research activities towards a MSc degree and a PhD. The boy knew, several years later, that the lecturer kept the report and had showed it to several of his classes. The lecturer wanted to illustrate that there were processes of reasoning that were beyond the frontal lectures and what was generally taught in schools. The boy could realize what happened during and after that homework only a decade later and, as it is often the case, he could never thank the lecturer for it. Now it is the time for it. Thank you so much, Pietro Frigato, for showing me the way. This work is dedicated to you.

First and foremost, I would like to express my gratitude to my supervisors Xiaofeng Wang and Pekka Abrahamsson. This journey started way before the PhD cycle, and you have been academically parenting me together since. Thank you for supporting my choices no matter how weird they were and for understanding my needs to enhance my happiness at work. You are something I wish to happen to every PhD student.

I wish to thank the reviewers of my thesis, Robert Feldt and Torgeir Dingsøyr, for the time spent in reading and reviewing my thesis. Their valuable comments have helped me to revise the text of this dissertation for the best. I also want to thank Darja Šmite, Barbara Russo, and Tommi Mikkonen in advance of their commission duties for the dissertation defense. I look forward to the discussion. Finally, I want to thank all the peers who constructively reviewed my papers, no matter what the editorial decision was.

My work could never happen without the precious support of dozens of people, such as the participants to my studies and the researchers who provided their feedback when I designed the studies. I took care of thanking them in each of my published papers, as I predicted that I could never remember all of them when the time for writing the dissertation would come. Still, I want to thank and apologize whoever was possibly left out from those lists because of my weak memory. I want to repeat myself with the administration of the Faculty of Computer Science at the Free University of Bozen-Bolzano. You are the real engine of the entire faculty.

I wish to thank my family, who supported my choices no matter how much they liked them or even understood them. Thank you for being the best parents and sisters I could wish for. You never stopped worrying about me, and I truly hope that the day will come when you will never have to.

There are people who did not directly support my work in terms of research activities, but supported me in other ways that still allowed the accomplishment of my research activities. I consider them my friends. Among them, I wish to thank Amantia Pano for her constant presence and for being my dearest friend for nearly a decade. Then there is that weirdo from München. Daniel Méndez, you are possibly the most obliging and big-hearted person I will ever meet during my career. Thank you for listening to my rants and for offering your help in every unimaginable way.

Finally, thank you Elena Borgogno. I consider myself the luckiest man in the world to have you acknowledged both in this dissertation and in my MSc thesis. You tirelessly contributed to every single work I did during this PhD cycle. But most importantly, you changed my life for the better, and I hope you will keep changing it for the rest of it.

# Contents

# List of figures

# List of tables

# Abbreviations

| | |
|---|---|
| **ACM** | **A**ssociation for **C**omputing **M**achinery |
| **ACR** | **A**verage **CR**eativity |
| **AET** | **A**ffective **E**vents **T**heory |
| **AIS** | **A**ssociation for **I**nformation **S**ystems |
| **APS** | **A**nalytical **P**roblem **S**olving score |
| **BCR** | **B**est **CR**eativity |
| **BSc** | **B**achelor of **S**cience |
| **CHASE** | **C**ooperative and **H**uman **A**spects of **S**oftware **E**ngineering, International Workshop on |
| **CI** | **C**onfidence **I**nterval |
| **IDE** | **I**ntegrated **D**evelopment **E**nvironment |
| **IEEE** | **I**nstitute of **E**lectrical and **E**lectronics **E**ngineers |
| **IEC** | **I**nternational **E**lectrotechnical **C**ommission |
| **ISERN** | **I**nternational **S**oftware **E**ngineering **R**esearch **N**etwork, Annual meeting of |
| **ISO** | **I**nternational **O**rganization for **S**tandardization |
| **IT** | **I**nformation **T**echnology |
| **MSc** | **M**aster of **SC**ience |
| **NCR** | **N**umber of **CR**eative solutions |
| **PAD** | **P**leasure-**A**rousal-**D**ominance |
| **PhD** | **Ph**ilosophiae **D**octor, read Doctor of Philosophy |
| **PANAS** | **P**ositive **A**nd **N**egative **A**ffect **S**chedule |
| **PPIG** | **P**sychology of **P**rogramming **I**nterest **G**roup |
| **RQ** | **R**esearch **Q**uestion |
| **SAM** | **S**elf-**A**ssessment **M**aniking |
| **SD** | **S**tandard **D**eviation |
| **SPANE** | **S**cale of **P**ositive And **N**egative **E**xperience |
| **SPANE-B** | **S**cale of **P**ositive And **N**egative **E**xperience, Affect **B**alance |
| **SPANE-N** | **S**cale of **P**ositive And **N**egative **E**xperience, Negative **F**eelings |
| **SPANE-P** | **S**cale of **P**ositive And **N**egative **E**xperience, Positive **F**eelings |
| **SSE** | **S**ocial **S**oftware **E**ngineering, International Workshop on |
| **TOL** | **T**ower **o**f **L**ondon Game |
| **PTS** | **P**lanning **T**ime **S**core |
| **TOLSS** | **T**ower **o**f **L**ondon **S**um of **S**cores |
| **SWEBOK** | **S**oftware **E**ngineering **B**ody of **K**nowledge |

# Areas of research

The Areas of Research of this PhD study have been classified according to the ACM Computing Classification System (Association for Computing Machinery, 1998).

- D. Software
  - D.2. SOFTWARE ENGINEERING
    * D.2.9. Management
      · Subject Descriptor: Productivity, Programming Teams

- H. Information Systems
  - H.1. MODELS AND PRINCIPLES
    * H.1.2. User/Machine Systems
      · Subject Descriptor: Human Factors, Software psychology

- J. Computer Applications
  - J.4. SOCIAL AND BEHAVIORAL SCIENCES
    * Subject Descriptor: psychology

# List of original publications

This dissertation is based on the following research papers.

Graziotin, D., Wang, X., and Abrahamsson, P. Software Developers, Moods, Emotions, and Performance. *IEEE Software*, 31(4):24–27 (2014c). doi: 10.1109/MS.2014.94. Cited in this thesis as Paper I.

Graziotin, D., Wang, X., and Abrahamsson, P. Understanding the affect of developers: theoretical background and guidelines for psychoempirical software engineering. In *Proceedings of the 7th International Workshop on Social Software Engineering - SSE 2015*, pp. 25–32. ACM Press, New York, New York, USA (2015d). ISBN 9781450338189. doi: 10.1145/2804381.2804386. Cited in this thesis as Paper II.

Graziotin, D., Wang, X., and Abrahamsson, P. The Affect of Software Developers: Common Misconceptions and Measurements. In *2015 IEEE/ACM 8th International Workshop on Cooperative and Human Aspects of Software Engineering*, pp. 123–124. IEEE, Firenze, Italy (2015c). ISBN 9781467370318. doi: 10.1109/CHASE.2015.23. Cited in this thesis as Paper III.

Graziotin, D., Wang, X., and Abrahamsson, P. Happy software developers solve problems better: psychological measurements in empirical software engineering. *PeerJ*, 2(1):e289 (2014b). doi: 10.7717/peerj.289[1]. Cited in this thesis as Paper IV.

Graziotin, D., Wang, X., and Abrahamsson, P. Are Happy Developers More Productive? In *14th International Conference on Product-Focused Software Process Improvement (PROFES 2013)*, volume 7983 LNCS, pp. 50–64. Paphos, Cyprus (2013). ISBN 9783642392580. doi: 10.1007/978-3-642-39259-7_7[2]. Cited in this thesis as Paper V.

Graziotin, D., Wang, X., and Abrahamsson, P. Do feelings matter? On the correlation of affects and the self-assessed productivity in software engineering. *Journal of Software: Evolution and Process*, 27(7):467–487 (2015a). doi: 10.1002/smr.1673[2]. Cited in this thesis as Paper VI.

Graziotin, D., Wang, X., and Abrahamsson, P. How do you feel, developer? An explanatory theory of the impact of affects on programming performance. *PeerJ Computer Science*, 1:e18 (2015b). doi: 10.7717/peerj-cs.18. Cited in this thesis as Paper VII.

During this PhD's cycle, the present author has worked on several research articles that are not related to this dissertation. These articles are listed in Appendix D.

---

[1]This article has been selected, among 471 articles, in the journal's top 20 most memorable and interesting peer-reviewed articles from 2014. The article is in the top 5% of all 4M+ articles ever tracked by Altmetric.com, which assessed the impact of articles in terms of social media shares and mentions.

[2]Paper VI is a revised, extended, and re-reviewed journal article based on the conference article in Paper V. The article was included in a special issue of the Journal of Software: Evolution and Process as one of the best two papers of PROFES 2013.

# Chapter 1

# Introduction

For more than thirty years, it has been claimed that a way to improve software developers' productivity and software quality is to focus on people (Boehm and Papaccio, 1988). Some suggested strategies to achieve low-cost but high-quality software have involved assigning developers private offices, creating a working environment to support creativity, and providing incentives (Boehm and Papaccio, 1988). That is, to make software developers happy (Whitaker, 1997). High-tech companies like Google, Facebook, and Supercell have followed this advice by providing incentives and the so called *perks*– e.g., having fun things to do and good food to eat during working hours–to make their developers happy (Drell, 2011; Google Inc., 2014; Stangel, 2013) and, allegedly, more productive (Marino and Zabojnik, 2008). The underlying assumption seems to be that "happy and satisfied software developers perform better".

Developers have indeed asked managers, employers, and customers to value them more as *human individuals* in software development, up to engrave the value "Individuals and interactions over processes and tools" in the Agile manifesto (Beck *et al.*, 2001). Cockburn and Highsmith (2001) have further stressed this issue by stating that "if the people on the project are good enough, they can use almost any process and accomplish their assignment. If they are not good enough, no process will repair their inadequacy– 'people trump process' is one way to say this." (p. 131).

But how can we believe this? There has not been evidence supporting this claim in software development context. Furthermore, there has not been any theoretical foundation to set the basis for understanding such a claim. Evidence and theory to support this desirable, yet still anecdotal claim are especially significant for those who don't work in the "fun" companies. The paradisaical environment usually described by tech-giant workers is contradicted by horror stories told by their own frustrated developers on a daily basis [1]. It seems that the managers of non-hero companies, which are indeed the majority of the companies out there, need to be reminded of the "obvious" claim that happy developers work better.

For understanding the complex underlying constructs, such as the happiness of software developers or their motivation, the software engineering field faces an additional challenge compared with more traditional engineering fields. Software development is substantially more complex than industrial processes. Software development activities are creative and autonomous (Knobelsdorf and Romeike, 2008). Many of the tasks that software developers engage in require problem solving. For example, software developers need to plan strategies to find a possible solution to a given problem or to generate multiple creative and innovative ideas. Therefore, among the many skills required for software development, developers need to possess high analytical problem-solving skills and creativity (Dybå, 2000; De La Barra and Crawford, 2007; Gu and Tong, 2004; Johnson and Ekstedt, 2015). As the environment of software development is all but simple and predictable, environmental turbulence requires creativity to make sense of the changing environment (Dybå, 2000). Much change occurs while software is being developed, and agility is required to adapt and respond (Williams and Cockburn, 2003). Furthermore, software is usually developed alone but the context of software development is often of a team working together, if not a network of teams within an organization. In a software company, the source of events is vast and complex. There are team dynamics with complicated, and even historical, relationships that are hard to grasp. After all, software developers are human beings.

As research has shown, human-related factors (sometimes called human aspects) play an important role in the execution of software processes and the resulting products (Colomo-Palacios *et al.*, 2010; Feldt *et al.*, 2010; Sommerville and Rodden, 1996; Fagerholm and Pagels, 2014). Several human factors, such as the management quality, motivation, and team skills, are linked to software development productivity and performance (Sampaio

---

[1]see **http://thedailywtf.com**, **http://clientsfromhell.net**, **https://www.reddit.com/r/talesfromtechsupport/**, **http://www.reddit.com/r/IIIIIIIIIITTTTTTTTTTTTT**, **http://www.reddit.com/r/shittyprogramming/**, **http://www.reddit.com/r/ProgrammerHumor/** for some fresh examples

*et al.*, 2010) [2]. Yet, when we focus on those human factors, we realize that the majority of them are difficult to be made concrete because they are part of the human mind. This is because software development happens in our brains mostly. However, even though it has been established that software development is carried out mainly through intellectual undertakings, thus cognitive activities (Feldt *et al.*, 2010; Fischer, 1987; Khan *et al.*, 2010; Sommerville and Rodden, 1996), research in software engineering has forgotten that affects—for now emotions and moods—have an impact on the cognitive activities of individuals (Khan *et al.*, 2010). Software engineering as a research field has explored related constructs such as job satisfaction (Melnik and Maurer, 2006; Tessem and Maurer, 2007), motivation (França *et al.*, 2014; Verner *et al.*, 2014; Sharp *et al.*, 2009), and commitment (Abrahamsson, 2001, 2002). However, as it is shown in the next chapter, little or no research has explored the affects of software developers—which are related to motivation, job satisfaction, and commitment but are not coinciding with them—and how they impact on the performance of software developers while developing software.

Kajko-Mattsson (2012) has argued that software engineering research has been suffering from researchers continuously jumping from trend to trend in terms of research topics. The tendency is to isolate small research problems to be solved and to jump to a completely different one right after fixing the first one. As a result of this, the software engineering community is lacking an understanding of the software development life-cycle. Johnson *et al.* (2012) have reasoned that the majority of disciplines, from circuit theory to organizational theory and physics, are very concerned with their theories; yet, they continue, software engineering has not appeared to be concerned with theory. The present author agrees with Johnson *et al.* (2012) that software engineering research could be *more* concerned about theory than it has been. Furthermore, Johnson *et al.* (2012) have cautioned about a lack of a general theory of software engineering, but the present author would like to extend it to include a lack of general theories in software engineering field. On the other hand, Sjøberg *et al.* (2008); Jorgensen and Sjoberg (2004) have greatly been concerned about theory building in software engineering as they wrote guidelines about it. Proposals of general theories of software engineering have been made recently, such as the SEMAT initiative (Jacobson *et al.*, 2013) with its workshops on general theories of software engineering (Ralph *et al.*, 2013; Johnson *et al.*,

---

[2]My stance for this thesis is that performance and productivity are two interchangeable terms, in line with several authors and as explained in Section 2.2

2013). Other proposals have been made by (Johnson and Ekstedt, 2015) and by Wohlin *et al.* (2015). It appears that a "theoretical" line of research of software engineering is slowly emerging.

Another issue on software engineering as a research field is that it has been dominated by the assumption that agents are rational, as much as it had happened to theory in social sciences and economics (Kahneman, 1997). The Nobel prize winner Daniel Kahneman has pointed out that it is unrealistic to limit our understanding of human behaviors solely through rational models (Kahneman, 2003). Yet, software engineering research has been known to be too confined in the fallacy of the rationality-above-everything paradigm (Murgia *et al.*, 2014), to miss out the possibility to be a social discipline (Sjøberg *et al.*, 2008), and to focus too much on the domains of technical nature while neglecting human-related topics (Lenberg *et al.*, 2015).

This PhD dissertation advocates the human and social aspects of software engineering research. The research activities undertaken during this PhD cycle have had the aim to set the building blocks for understanding software developers and their performance under the lens of how they feel. That is, this PhD's research activities aimed to theorize the impact of affects on software development performance.

The following two sections will provide more details about why the affects of software developers were taken into consideration, and motivate our research activities through evidence gathered from practitioners. After stating the motivation of our research, the chapter will proceed by offering the research questions of this dissertation, the scope of the investigation, and the thesis structure.

## 1.1 Motivation

We are human beings, and, as such, we behave based on affects as we encounter the world through them (Ciborra, 2002). Ciborra (2002) argued that affects enable what matters in our experiences by "indelibly coloring our being in the situation". Affects define how we perceive the world around us, as they enable how we feel about it. Therefore, they enable the "mattering of things" (pp. 159–165). Things happen around us, but how they matter to us—negatively or positively—is defined by our affect.

The affects pervade organizations because they influence worker's thoughts and actions, and they are influenced by the worker's thoughts and actions (Brief and Weiss, 2002). Affects have a role in the relationships between workers, deadlines, work motivation, sense-making, and human-resource processes (Barsade and Gibson, 2007).

Although emotions have been historically neglected in the studies of industrial and organizational psychology (Muchinsky, 2000), an interest in the role of affect on job outcomes has increased over the past decade in management and psychology research (Fisher, 2000; Ashkanasy and Daus, 2002; Oswald *et al.*, 2015; Zelenski *et al.*, 2008).

Diener *et al.* (1999) and Lyubomirsky *et al.* (2005) reported that numerous studies have shown that the happiness of an individual is related to achievement in various life domains, including work achievements. Indeed, emotions play a role in daily jobs. The relationship between affect on the job and work-related achievements, including performance (Barsade and Gibson, 2007; Miner and Glomb, 2010; Shockley *et al.*, 2012) and problem-solving processes, such as creativity, (Amabile *et al.*, 2005; Amabile, 1996) has been of interest for recent research.

Some evidence has been found that happier employees are more productive (Fisher and Noble, 2004; Oswald *et al.*, 2015; Zelenski *et al.*, 2008). Other studies instead suggested that affects relate to job performance differently with respect to the various aspects or types of jobs (Kaplan *et al.*, 2009b). However, still little is known about the productivity of individual programmers (Scacchi, 1995) and the link between affects and the performance of developers (Khan *et al.*, 2010; Shaw, 2004). It is necessary to understand how affects play a role in software development as real-time correlations of performance and affects are often overlooked (Beal *et al.*, 2005; Fisher and Noble, 2004).

Denning (2012) recently argued that the recognition of moods of developers is essential to professional success in software development. There are also several calls for research on the role of emotions in software engineering (Colomo-Palacios *et al.*, 2010; Denning, 2012; Khan *et al.*, 2010; Shaw, 2004). However, the actual research in the field is very limited.

Despite the fact that the ability to sense the moods and emotions of software developers may be essential for the success of an Information Technology firm (Denning, 2012), software engineering research lacks an understanding of the role of emotions in the software development process (Khan *et al.*, 2010; Shaw, 2004). In software engineering research, the affective states of software developers have been investigated rarely in spite of the fact that affective states have been a subject of other computer science disciplines, such as human-computer interaction and computational intelligence (Lewis *et al.*, 2011; Tsonos *et al.*, 2008). Furthermore, Johnson *et al.* (2012) have recently identified a crisis in software engineering in terms of theory building studies. So far, there are no theoretical frameworks for guiding research on the affects of developers.

Based on the existing literature, it is believed that studying the affects of software developers will provide new insights about their performance. This PhD thesis reports our research activities towards a new understanding of the complex relationship between the affects of developers and their performance while they develop software. This PhD contribution is a three-faceted theory of the relationship of affects and the performance of software developers. The theory is expressed into two variance facets and one process based facet. Overall, the facets express what is the relationship of the pre-existing affects of software developers and their problem solving performance, the relationship of the immediate, real-time affects and the productivity of a software development task, and the process that explains how affects impact the performance while developing software.

So far, we have motivated our research activities by reviewing the academic literature, which has strongly encouraged a deeper understanding of the affects of developers. But would practitioners care about an added understanding about their affects, if not even being studied from this point of view? We will show in Chapter 8, Section 8.5.1 that the answer is strongly positive, to the point of having these studies called for by practitioners.

## 1.2 Research questions

Based on the literature review, which is available in the next chapter but was briefly touched upon in this chapter in its motivational part, it was established that the linkage between the affects of software developers and their performance while developing software was a novel topic that deserved much attention from researchers. Therefore, the main research question of this dissertation follows.

RQ: How are affects of software developers related to their performance?

The main research question is divided into the following questions:

**RQ1** What are the theoretical foundations regarding the affects of software developers?

**RQ2** How do affects indicate problem-solving performance among developers?

**RQ3** How are affects correlated to in-foci task productivity of software developers?

**RQ4** How do software developers experience affects and their impact on their development performance?

The nature of this dissertation requires a cross-disciplinary understanding of the underlying constructs and concepts (the theoretical foundations), as they are not familiar to the readers from the fields of software engineering research. One example is the apparent misuse of the terms performance and productivity, which is made clearer in Chapter 2 as a strategy to have high level definitions that can be operationalized in different studies. I invite the reader to come back to the research questions of this dissertation after reading the following chapter.

RQ2 has been mainly derived from the literature review in Section 2.2.2, where it is shown that a clear consensus on the relationship between problem-solving performance and affect has yet to be reached. The research question has arisen from the classical studies from psychology, and it investigates a static representation of the relationship, where time is not concerned.

RQ3 has been derived mainly from the literature review in Section 2.2.1, where industrial and organizational psychology and organizational behavior studies have investigated the dynamic nature of the relationship between affect and job performance over time. Hence, the in-foci task productivity.

RQ4 is concerned about the underlying process that regulates the complex relationship between affect and performance while developing software. This requires an understanding of how developers experience their affect and the effects on their own software development performance.

## 1.3   Scope of the research

This doctoral thesis combines the research fields of software engineering and psychology, together with organizational behavior[3] and human-computer interaction.

Recently, the discipline of software engineering has begun to adopt a multi-disciplinary view and has embraced theories from more established disciplines, such as psychology, organizational research, and human-computer interaction. For example, Feldt *et al.* (2008) proposed that the human factors of software engineering could be studied empirically by "collecting psychometrics". Although this proposal has begun to gain traction, limited research has been conducted on the role of emotion and mood on software developers' skills and productivity.

We share Lenberg *et al.* (2014) view that software engineering should also be studied from a behavioral perspective. Therefore, we have employed theories and measurement instruments from psychology to understand how affects have an impact on software developers' performance under a quantitative strategy using experiments. However, in order to understand the human behavior behind affects and software development, there is a need to observe software developers in-action and perform interviews. So far, research has not produced qualitative insights on the mechanism behind the impact of affects on the performance of developers. We have called for such studies in the past (Paper VI).

It is now necessary to state what this thesis is and is not about. In brief, the research activities undertaken for this PhD thesis are (1) for understanding the impact of affects on the performance of a programming task, and not for understanding the performance itself, (2) interested in the analysis unit of individuals, and (3) focused on the software development activity of programming.

The aim of this study is to offer a theory of the impact of affects on performance while programming rather than proposing a performance or productivity theory. A plethora of factors influence the performance of developers—see Wagner and Ruhe (2008); Sampaio *et al.* (2010) for a comprehensive review of the factors—and affects are one of them, as it is shown in the next chapter, although they are not yet part of any review paper. At the same time, software development performance is composed of several complex interrelated constructs—see Petersen (2011) for a review of productivity measurements—to which we add those driven by cognitive processes and *also* influenced by affects, e.g., creativity and analytic problem solving (Paper IV). Moreover, we limit the scope of the research to affect itself, and exclude the disposition of individuals to experience affect, namely personality characteristics, attitudes, and traits.

---

[3]For the scopes of this work, we consider the two disciplines of organizational behavior and I/O (Industrial/Organizational) psychology as equivalent, in line with Jex and Britt (2014), who claim that their main difference relies in where these disciplines are thought–business schools the first, psychology departments the second–and that organizational behavior is less individual-centric than I/O. Still, most authors in the disciplines cross-publish in journals of both disciplines.

When designing research activities, it is important to define an unit of analysis (Trochim and Donnelly, 2007). The unit of analysis is the entity that is being analyzed in a scientific research (Dolma, 2010). In general research, we consider as units of analysis the individual, group, and organization. Specific fields add further units, e.g., Trochim and Donnelly (2007) have argued that the most common units of analysis in social science are individual, group, artifact (e.g., books, photos, newspapers), geographical location, and social interaction (e.g., dyadic relations, divorces, arrests). In the field of software engineering, Easterbrook *et al.* (2008) have argued that the unit of analysis "might be a company, a project, a team, an individual developer, a particular episode or event, a specific work product, etc." (pp. 297). We chose to put our focus on individual developers, regardless of them being part of a team or not. One reason is that we identified a lack of theory in software engineering on the affects of developers; therefore, our strategy for constructing such missing building blocks has been to confine our understanding to developers as single individuals. A broader, perhaps more severe reason is that the studies of performance in software engineering have so far mostly ignored individual programmers, by focusing more on teams and organizations (Meyer *et al.*, 2014). This lack in software engineering research was highlighted in the review study by Scacchi (1995), which was conducted twenty years ago, and it was renewed in the recent article by Meyer *et al.* (2014). That is, little is still known about the productivity of individual programmers (Scacchi, 1995).

Besides focusing on developers as single individuals, our strategy has been to specialize our research with respect to the software development process. Various software development life-cycles have offered different activities and ways of approaching them. The ISO/IEC TR 19759:2005, namely the Software Engineering Body of Knowledge (SWEBOK) (Bourque and Fairley, 2014), has defined 15 knowledge areas or the software engineering body of knowledge, e.g., software requirements, software design, software testing, and software quality. Throughout the text, we use the terms software programming, coding, development, and construction interchangeably. However, this dissertation's research activities have focused on the SWEBOK (Bourque and Fairley, 2014) knowledge area of software construction, that is the "detailed creation of working software through a combination of coding, verification, unit testing, integration testing, and debugging" (p. 66). The strategy here has been similar to the one for selecting individual programmers as the unit of analysis. There is the need to set the building blocks in software engineering research on the affects of software developers; therefore, we focused on software construction as we consider it the central activity of any software development endeavor, in line with Sommerville (2011).

To summarize, this work aimed to understand the linkage between the affects of software developers, as individuals, and their performance while they develop software. This work (1) does not study any other construct such as motivation, job-satisfaction, or commitment, (2) does not analyze at the team and organization units, (3) does not analyze any software development activity that produces artifacts other than source code, such as requirements, design, or architecture artifacts.

## 1.4 Thesis structure

This PhD thesis is based on seven original research papers, which have been outlined at page xiii. Each article has contributed to the understanding of the phenomena under the study.

The structure of the thesis has been organized by following three criteria. The first criterion reflected the widely recognized structure of a scientific paper, namely introducing the topic and the problem, presenting the state of the art of research about the topic and the problem, providing a research design to further understand the problem, reporting the results of executing the design, and discussing the results.

The second structuring criterion was dictated by the organization of the research activities. We anticipate here [4] that the research approach in the study of this dissertation was organized into three main parts, namely knowledge acquisition and theory translation, variance theory testing, and process based theory development. These are the phases of research that the present writer has come with retrospectively, and they make sense of the research activities. The chapters are ordered by following the three phases.

---

[4]The research phases are described in detail in section 3.5.

The third and final criterion of structuring the dissertation was clarity. Given the unconventional topics and output of this dissertation, the strategy for structuring the dissertation was to report the concept and the activities in ascending order of details, from general to detailed. This strategy can be observed within chapters and between chapters. The literature review of this dissertation starts with the broad concepts of the human mind and it ends with the related work of studies of affect and performance in software engineering. The final part of Chapter 3 outlines the research design of the three main empirical studies of this PhD's research activities. It is followed by the three chapters about the studies, which go into detail about the study designs and results. Chapter 7 highlights the results of the three main studies of this PhD's research activities.

Chapter 1 introduced the key concepts, which have been studied during the research activities. It provided the motivation for our studies, laid down the research questions of this PhD's research activities, and provided the scope of this dissertation. Chapter 1 was inspired by mapped to Paper I, which creates parts of the motivation for our research activities.

Chapter 2 reports an extensive literature review of the constructs under study—affects in psychology and their measurement, the related work in software engineering, the common misconceptions of affects, performance in terms of cognitive activities and on the job, and human aspects in software engineering productivity studies. Chapter 2 was mapped to Paper II and Paper III, which form part of the literature review that was performed for understanding the phenomenon under study and for constructing the research activities. Another contribution of Paper II was the proposal of guidelines for psychoempirical software engineering, which is our umbrella term for the studies of empirical software engineering with psychology theory, methodologies, and measurements. These guidelines for psychoempirical software engineering are a high-level structure to be followed when conducting studies in software engineering research using psychology theory and measurements. While the proposals are an important output of this PhD's research activities, they were included in Appendix C as the main focus of this dissertation is not about guidelines.

Chapter 3 lays down the empirical research design of this dissertation, i.e., the research approach, the employed research methods, and the context in which the research activities took place. It also maps the research questions to the thesis chapters and the studies. Finally, it describes the three phases of research of this PhD research activities, and provides a summary of the subsequent three studies' designs.

The following three chapters, namely 4, 5, and 6, report the three major empirical studies undergone through this PhD's research activities. Each chapter provides the detailed description of the research design, the execution of the design, and the obtained results.

Chapter 4 tests a variance based theory of affect by studying the relationship between two sub-constructs of the produced theory, namely the affects and the problem -setting and -solving performance of developers. The chapter was mapped to Paper IV, which assesses the linkage between the affects of software developers and their problem-setting-and-solving skills, in terms of analytic thinking and creativity.

Chapter 5 tests a variance based theory of affect by providing a correlation study of the affects of software developers and their productivity. The reported study quantifies the correlation. The chapter is built upon Paper V and Paper VI.

Chapter 6 presents a study, which produced an explanatory process based theory of the impact of affects on the programming performance. The study is presented in Paper VII, so the chapter draws upon that article.

Chapter 7 summarizes the results of this dissertation by offering an outline of the various outputs of the research activities.

Chapter 8 discusses the research findings, and provides the theoretical and practical implications of them. Section 8.4 was mapped to Paper I for reporting how much practitioners care about research on their affects.

Finally, Chapter 9 concludes with a summary of the results, answers the research questions, addresses limitations of the study, and provides suggestions for future research.

## A note on the writing style

This dissertation was written by following established guidelines for scientific writing style (Hofmann, 2010; American Psychological Association, 2010). Some further clarification on the writing style is provided here.

This document uses the first person plural *we* when referring to research activities conducted in collaboration. The reason is that the present author has agreed with Arunachalam (2008) that the construction of knowledge is a community-oriented activity, no matter if in cooperative or competitive ways. As a consequence of this reasoning, the majority of the thesis employs the first person plural.

On the other hand, there are several points where the present author's opinions and choices have to be explicated, for example when stating the ontology and epistemology of this work. In such cases, the text will deviate from its use of the first person plural and use forms such as *the present author* (or candidate).

The passive voice is also employed throughout the text, in both cases mentioned above, in order to avoid too much personal language and for offering a more variegated writing style to this dissertation of more than 200 pages.

# Chapter 2

# Theoretical foundations and related work

This chapter provides the theoretical foundations of the constructs of affect, performance, their relationship, and the review of the related work in software engineering research. This chapter is a contribution for answering RQ1—*What are the theoretical foundations regarding the affects of software developers?*.

The chapter begins with a brief overview of the conceptualization of the mind in psychology research which has the affects as the base of a tripartite system. Section 2.1 starts by reporting how providing a definition for affect, emotions, and moods has been a challenge even for psychology research. The section continues by extensively exploring the two major frameworks from psychology and organizational behavior for categorizing affects, and a recent unifying framework and theory. The section provides the common ways to measure the affects and how to cope with the several issues that arise when employing psychological measurements in software engineering. Section 2.2 reports a review of the theory about performance in psychology and organizational behavior research, and in software engineering research. The section deals with performance on the job, problem solving performance, and productivity and performance studies in software engineering. Section 2.3 summarizes the theoretical foundations of the relationship between affects and performance in psychology and organizational behavior. Section 2.4 summarizes the works advocating the use of psychology research in software engineering research. Section 2.5 reviews the works related with this PhD dissertation. It first reports the works on the affects of software developers. Then, the section provides the related work about affects and performance in the context of software development. Finally, the section ends with a discussion of the common misconceptions when dealing with the affects of software developers.

The mind has been historically conceptualized as a tripartite system of primary parts, namely *cognitive*, *conative* (or motivation, sometimes also called behavior), and *affective* (Hilgard, 1980). Conation originates from the Latin term *conatus*, meaning any natural tendency, impulse, striving, or directed effort and behavior (OED Online, 2015c). Cognition originates from the Latin verb *cognosco*, and it is an umbrella term encompassing mental processing abilities such as knowledge, attention, working memory, all of which are related to our action or faculty of knowing (OED Online, 2015b). Finally, the affective part is related to the manner in which one is inclined or disposed and a mental state, mood, emotion, or feeling (OED Online, 2015a)[1].

Nowadays, several additions and alternatives have been proposed to the tripartite system, including the public and private personalities (Singer, 1984), the structural sets of id, ego, and superego (Freud, 1961), and the dual-process accounts of reasoning (Stanovich and West, 2000; Kahneman, 2011). Yet, the tripartite system has been rooted since 1750 (Mayer, 2001), has been much validated, e.g., (Insko and Schopler, 1967), and still considered as the major foundation for teaching psychology (McLeod, 2009; Tallon, 1997; Huitt, 1999). As reported in this chapter, affects have strong influences on the other two systems of the mind—as much as the other systems influence each other in a complex interrelated dance—thus making them an interesting focus of study for each intellect-pervaded discipline, including software engineering.

---

[1]The present author is aware that these definitions are of dictionary nature and do not represent the myriad of research activities behind them. They would deserve their own space. For the purposes of this dissertation, the dictionary definitions of the cognitive and conative parts are sufficient. Affect, on the other hand, is extensively defined in Section 2.1

## 2.1 Affects [2]

### Emotions, moods, and the struggle for a definition

Any good researcher and supervisor will advice a novice researcher to always provide precise definitions for the constructs, which are the subjects of a research activity. Our research activities base their grounds on the concept of affective states (or affects), which at this point appear to be related to emotions, moods, and feelings. The concepts of emotions and alike terms have puzzled us since a remarkably long time, beginning with the philosopher James (1884) interminably asked question, *What is an emotion?* We anticipate that this question is still unanswered (Russell and Barrett, 1999). As it is shown below, the fields of psychology have failed to agree on definitions of affects and the related terms emotions, moods, and feelings (Ortony *et al.*, 1990; Russell, 2003). Before the disappointment kills the mood while reading this chapter, the reader might be comforted by two facts; (1) it might not matter that these constructs cannot be firmly defined as expected by technical research; and (2) there is a recent unifying theory by Russell (2003) that, as reported by Ekkekakis (2012), is satisfying the majority of the researchers in this field due to a considerable convergence among the different views. This section will provide some proposed definitions for these terms, the related issues and the non-agreement among researchers, and how these issues are not a problem at all for psychology and for the present research.

Let us start by stating that the term *affective states* (or affects) has been defined as "any type of emotional state [...] often used in situations where emotions dominate the person's awareness" (VandenBos, 2013). This definition is problematic as it contains the term *emotion*, which has not yet been defined, and it does not help in defining the (now apparently) more basic concept of affects. Indeed, the term *affects* is often associated in the literature with *emotions* and *moods.* The majority of the authors in specialized fields have used these terms interchangeably, e.g., (Schwarz and Clore, 1983; Schwarz, 1990; Wegge *et al.*, 2006; De Dreu *et al.*, 2011). To further complicate the issue of defining what appear to be the building blocks of affects, some researchers consider emotions and moods *as* affects, e.g., (Weiss and Cropanzano, 1996; Fisher, 2000; Fisher and Ashkanasy, 2000; Oswald *et al.*, 2015; Khan *et al.*, 2010). We now are left with three terms, which look remarkably similar but cause the irritant need to differentiate them.

Plutchik and Kellerman (1980) have defined *emotions* as the states of mind that are raised by external stimuli and are directed toward the stimulus in the environment by which they are raised. However, Kleinginna and Kleinginna (1981) reported one year later that more than 90 definitions have been produced for this term, and no consensus in the literature has been reached. The term has been taken for granted and often defined with references to a list, e.g. anger, fear, joy, surprise (Cabanac, 2002). To worsen this, *emotion* itself is not universally employed, as it is a word that does not exist in all languages and cultures (Russell, 1991). The terms anger, fear, joy, surprise, do not seem to be extraneous to a mood classification, as well.

*Moods* have been defined as emotional states in which the individual feels good or bad, and either likes or dislikes what is happening around him/her (Parkinson *et al.*, 1996). Yet again, a definition of one construct contains another construct of our interest.

How do emotions and moods differ, then? While for some researchers certain moods are emotions and vice-versa (DeLancey, 2006), it has been suggested that a distinction is not necessary for studying cognitive responses that are not strictly connected to the origin of the mood or emotion (Weiss and Cropanzano, 1996). As put by Ortony *et al.* (1990) emotion study is a "very confused and confusing field of study" (p. 2) because an agreement on what affect, emotions, and moods actually are has yet to be reached, although I will show in this section that the issue does not compromise our understanding of affect.

Distinctions between emotions and moods are clouded, because both may feel very much the same from the perspective of an individual experiencing either (Beedie *et al.*, 2005). The concepts of affects, emotions, and moods have provided theories, understanding, and predictions in psychology, and now they are part of common sense (Russell, 2003). They are embedded in psychologists' questions and, as a consequence, answers. Reisenzein (2007) argued that we do not actually need a consensual definition of moods and emotions for conducting research; rather, a consensual definition is a revisable research hypothesis that will be likely ended by scientific research. Additionally, it has been proposed that an emotion is an emergent construction rather than a latent entity; therefore there is a call for a shift in the literature (Clore and Ortony, 2008; Minsky, 2008). We agree with Beedie *et al.* (2005), who have argued that a difference between emotion and mood might be of semantic nature only. These concepts are linked to our language and culture (Russell, 1991), and the fact that we have these two specific words does not entail that emotion and mood refer to two different constructs or to the same one (Beedie *et al.*, 2005). Finally, Beedie *et al.* (2005) conclude that it a job for psychology researchers to attempt to clarify the nature of moods and emotions and their relationship, if any.

Therefore, for the purposes of several of our published research, we have adopted the same stance of several researchers in the various fields and employed the noun *affective states (affects)* as an umbrella term for emotions and moods. However, a recent theory in psychology has proposed to unify these concepts while allowing a distinction among them, so that it is possible to study affects at different levels of understanding and dimensionality (Russell, 2009). The theory is compatible with our stance of studying affects regardless of them being moods or emotions.

### 2.1.1   The two major frameworks for affect theories

Huang (2001), identified four major theories for emotions (moods, affects) in psychology, namely the Differential Emotions theory, the Circular Model of Emotions, the Pleasure-Arousal-Dominance (PAD) model of affect, and the Positive And Negative Affect Schedule (PANAS). Our literature review has suggested that many more theories have been proposed, and at least six theories can be considered established in the literature. We also noted that the theories could fall into two frameworks, namely the discrete framework and the dimensional framework.

One framework, namely the discrete approach, collects a set of basic affective states that can be distinguished uniquely (Plutchik and Kellerman, 1980), and that possess high cross-cultural agreement when evaluated by people in literate and preliterate cultures (Ekman, 1971) (Matsumoto and Hwang, 2011).

The Differential Emotions Theory (Izard, 1977) states that the human motivation system is based on ten fundamental emotions (interest, joy, surprise, distress, anger, disgust, contempt, fear, shame , and guilt). These fundamental emotions function for the survival of human beings, possess an own neural network in the brain, and an own behavioral response. Finally, these emotions interact with each other simultaneously.

Ekman (1971) proposed a set of basic affects, which include anger, happiness, surprise, disgust, sadness, and fear. However, the list has received critique, leading to an extended version with other eleven elements (Ekman, 1992). They include amusement, embarrassment, relief, and shame.

In the Circular Model of Emotion (Plutchik and Kellerman, 1980; Plutchik, 1980), a structure describing the interrelations among emotions is proposed. Eight primary, bipolar affective states were presented as coupled pairs: joy versus sadness, anger versus fear, trust versus disgust, and surprise versus anticipation. These eight basic emotions vary in intensity and can be combined with each other, to form secondary emotions. For example, joy has been set as the midpoint between serenity and ecstasy, whereas sadness has been shown to be the midpoint between pensiveness and grief. Emotions can vary in intensity and persistence (to form moods, for example) Under this theory, they serve an adaptive role in dealing with survival issues. Developing a minimal list of basic affective states appears to be difficult with the discrete approach. Subsequent studies have come to the point where more than 100 basic emotions have been proposed (Shaver *et al.*, 1987).

The other framework groups affective states in major dimensions that allow a clear distinction among them (Russell, 1980; Lane *et al.*, 1999).

In the Positive and Negative Affect Schedule (PANAS) (Watson and Tellegen, 1985; Watson *et al.*, 1988a,b), the positive and negative affects are considered as the two primary emotional dimensions. However, these two dimensions are the result of the self-evaluation of a number of words and phrases that describe different feelings and emotions. That is, discrete emotions are rated but two dimensions are evaluated. This theory is designed to present a mood scale. Finally, positive and negative affects are mutually independent.

In the PAD models, three dimensions of Pleasure-displeasure, Arousal-nonarousal, and Dominance-submissiveness (Russell and Mehrabian, 1977; Russell, 1980; Mehrabian, 1996) characterize the emotional states of humans. *Valence* (or pleasure) is the attractiveness (or adverseness) of an event, object, or situation (Lewin, 1935) (Lang *et al.*, 1993). The term refers to the "direction of a behavioral activation associated toward (appetitive motivation) or away (aversive motivation) from a stimulus" (Lane *et al.*, 1999). *Arousal* represents the intensity of emotional activation (Lane *et al.*, 1999). It is the sensation of being mentally awake and reactive to stimuli, i.e., vigor and energy or fatigue and tiredness (Zajenkowski *et al.*, 2012). *Dominance* (or control, over-learning) represents a change in the sensation of the control of a situation (Bradley and Lang, 1994). It is the sensation by which an individual's skills are perceived to be higher than the challenge level for a task (Csikszentmihalyi, 1997). Emotional states under the PAD models include moods, feelings, and any other feeling-related concepts. The three dimensions are bipolar, indicating that the presence of pleasure excludes the possibility of displeasure. Some variations of these model have been proposed using different notations but without changing the core meaning (Russell, 2003), some of which omit the dominance dimension (Lane *et al.*, 1999). The dimensional approach has been distinguished from the discrete approach in its lesser number of elements to be evaluated. Thus, it has been deemed useful in tasks where affective states must be evaluated quickly and preferably often. Indeed, it has been commonly adopted to assess affective states triggered by an immediate stimulus (Bradley and Lang, 1994; Ong *et al.*, 2011), especially when repeated assessments of affective states are needed in very short time-spans.

## 2.1.2 Core affect: a unifying framework and theory

Russell and Barrett (1999); Russell (2003, 2009) have proposed the concept of *core affect* to unify the theories of emotions and moods in psychology. What follows has been sourced from the previously stated references, which will be repeated only in the case of a direct quotation.

Core affect is "a pre-conceptual primitive process, a neurophysiological state, accessible to consciousness as a simple non-reflective feeling that is an integral blend of hedonic (valence) and arousal values" ((Russell, 2003), p. 147). In other words, core affect is the atomic level of feeling, almost at an idealized level. Examples are feeling good or bad, feeling lethargic or energized (Russell, 2003). The state is accessible at a consciousness level as the simplest raw feelings, which is distinct in moods and emotions. A feeling is an assessment of one's current condition. An affect is raw, as explained next. *Pride* can be thought of as feeling good about oneself. The "feeling good" is core affect and the "about oneself" is an additional (cognitive) component. Affect *per se* is in the mind but it is not cognitive and it is not reflective (Zajonc, 2000). Core affect is simply not about anything. That is, affect can be felt in relation to no obvious stimulus—in a free-floating form—as moods are perceived. Indeed, mood is defined as a prolonged core affect without an object, i.e., an unattributed affect.

Changes in affects result from a combination of happenings, such as stressful events on the job. Sometimes the cause of the change is obvious. However, sometimes one can undergo a change in core affect without understanding the reasons. The individuals possess a limited ability to track this complex causality connection. Instead, a person makes attributions and interpretations of core affect.

When an affect is attributed to an object (also known as *attributed affect*), a change in affect becomes linked to the perceived cause of the change; the cause could be anything possible, e.g., an event, a person, a location, a physical object, a situation and any virtual object. It is important to understand that the attribution is an individual's perception of causal links between events. Therefore, it allows room for individual and cultural differences. On the other hand, research has demonstrated that individuals commit misattributions, as well. An attributed affect is constituted by three components: (a) a change in the individual's core affect, (b) an object (or stimulus), and the (c) individual's attribution of the core affect to the object. Attributed affects are common in every day's life, for example being afraid of bees, feeling sad at a personal loss, liking a new song, and so on. Attributed affect steers the attention to the attributed object, as well as the behavior directed at the attributed object, regardless if and when misattribution happens. Finally, attributed affect is the principal route to the affective quality of an object.

Affective quality is a property of the object (i.e., the stimulus). It is the object's capacity to change an individual's core affect. Affective quality can be perceived without it taking action. Indeed, core affect does not need to change to know that sunset is lovely or the wild bear scary. It is not the experience of a change in core affect. Affective quality is the anticipation of a change in core affect.

In Russel's theory, emotions are episodes instead of simple psychological objects. An emotion is a complex set of interrelated sub-events about a specific object. An emotional episode prototype is composed by (1) an obvious external antecedent event, (2) a perception of the affective quality of the antecedent event, (3) a change in the core affect, (4) an attribution of the core affect to the antecedent, which becomes the object, (5) a cognitive appraisal of the object itself (e.g., what is it, what does it mean to me, etc.), (6) an instrumental action directed to the object (e.g., the formation of a goal), (7) physiological and expressive changes (e.g., a smile, a frown), (8) subjective conscious experiences (e.g., urgency, indecision, confusion, etc.), (9) the emotional meta-experience and categorization (e.g., I realize that I am afraid), and (10) emotion regulation (e.g., self-placing with respect of social norms and roles). The reader should note that an emotional episode prototype is typical but not universal. In borderline cases, which are non-prototypical, the core affect can be extreme before the appearance of the antecedent. One can enjoy what one appraises as dangerous; there might be misattribution of the antecedent; there might be atypical appraisal (e.g., being afraid of harmless objects), etc.

> For the purposes of this investigation, we adopt Russell (2003) definition of **affect** as "a neurophysiological state that is consciously accessible as a simple, non-reflective feeling that is an integral blend of hedonic (valence) and arousal values." (p. 147).

Although we do not neglect moods and emotions *per se*, we chose to understand the states of minds of software developers at the *affective* level only, that is underlying moods and emotions.

Adhering to Russell (2003) theory, we consider *emotions* and *moods* as affects [3] in line with several other authors, e.g., (Weiss and Cropanzano, 1996; Fisher, 2000; Fisher and Ashkanasy, 2000; Oswald *et al.*, 2015; Khan *et al.*, 2010).

That is, in this dissertation we do not distinguish between emotions and moods, nor do we investigate them *directly* in our measurements. We will always discuss them as aggregated affects. Also, sometimes we use the terms emotions, moods, affects, and feelings interchangeably, in line with many authors (Schwarz and Clore, 1983; Schwarz, 1990; Wegge *et al.*, 2006; De Dreu *et al.*, 2011).

---

[3] In our conceptualization, affect can be considered as the atomic unit upon which moods and emotions can be constructed. By keeping Russell (2003) definitions, we consider moods as prolonged, unattributed affects, and we consider emotions as interrelated events, i.e., an episode, about a psychological object.

Russell (2003) theory is interesting for this study because it unifies the previous theories, and it maintains compatibility with the majority of the existing measurement instruments, regardless of them being about moods or emotions. In psychology research, there exist both theoretical works on affect that are tied up to their measurement instrument (e.g., Watson *et al.* (1988a,b)) and theoretical pieces that that attempt to be neutral to measurement instruments. The Core Affect theory attempts rather to end the "emotions vs. moods. vs. the definitions" war by suggesting (core) affect as the building block for emotions and moods. Core affect theory is compatible with most existing measurement instruments including several of those listed in the next section, and those that we selected. Furthermore, core affect is satisfying the majority of the researchers in psychology field due to a considerable convergence among the different views (Ekkekakis, 2012). Core affect is a framework for understanding the constructs of affect, emotions, and moods, and for interpreting the measurements for the constructs. Therefore, the values for the measurement instruments can be interpreted as *pre-existing affect* rather than mood and *affect raised by a stimulus* instead of emotions.

### 2.1.3   Measuring affects

Psychology studies have often grouped participants according to their affective states, in terms of negative, neutral (less often), or positive. In the case of controlled experiments, the grouping has been based on the treatments to induce affective states. In the case of quasi-experiments and natural experiments, the grouping has been based on the values of affective state metrics, usually employed in questionnaires. Several techniques have been employed to induce affective states on participants, such as showing films, playing certain types of music, showing pictures and photographs, or letting participants remember happy or sad events in their lives (Westermann and Spies, 1996; Lewis *et al.*, 2011). Recent studies have questioned the effects of mood-induction techniques, especially when studying pre-existing affective states of the participants (Forgeard, 2011). Alternately, some studies have used quasi-experimental designs that select participants with various affective states, which have usually been based on answers to questionnaires.

One of the most notable measurement instruments for affective states is the Positive and Negative Affect Schedule (PANAS) (Watson *et al.*, 1988a,b). It is a 20-item survey that represents positive affects (positive affects) and negative affects (negative affects).

However, several shortcomings have been criticized for this instrument: PANAS reportedly omits core emotions such as *bad* and *joy* while including items that are not considered emotions, like *strong*, *alert*, and *determined* (Diener *et al.*, 2009a; Li *et al.*, 2013). Another limitation has been reported for its non-consideration of the differences in desirability of emotions and feelings in various cultures (Tsai *et al.*, 2006; Li *et al.*, 2013). Furthermore, a considerable redundancy has been found in PANAS items (Crawford and Henry, 2004; Thompson, 2007; Li *et al.*, 2013). PANAS has also been reported to capture only high-arousal feelings in general (Diener *et al.*, 2009a).

**Scale of Positive and Negative Experience**

Recent, modern scales have been proposed to reduce the number of the PANAS scale items and to overcome some of its shortcomings. Diener *et al.* (2009a) developed the Scale of Positive and Negative Experience (SPANE).

SPANE assesses a broad range of pleasant and unpleasant emotions by asking the participants to report them in terms of their frequency during the last four weeks. It is a 12-itemized scale, divided into two sub-scales. Six items assess positive affective states and form the SPANE-P scale. The other six assess negative affective states and form the SPANE-N scale. The answers to the items are given on a five-point scale ranging from 1 (*very rarely or never*) to 5 (*very often or always*). For example, a score of five for the *joyful* item means that the respondent experienced this affective state *very often* or *always* during the last four weeks.

The SPANE-P and SPANE-N scores are the sum of the scores given to their respective six items. Therefore, they range from 6 to 30. The two scores can be further combined by subtracting SPANE-N from SPANE-P, resulting in the Affect Balance Score (SPANE-B).

SPANE-B is an indicator of the pleasant and unpleasant affective states caused by how often positive and negative affective states have been felt by the participant. SPANE-B ranges from -24 (*completely negative*) to +24 (*completely positive*). Even if the SPANE-B score is a fuzzy indication of the affective states felt by individuals, it could be employed to split participants into groups using a median split. It is common to adopt the split technique on affective states measures (Hughes and Stoney, 2000; Berna *et al.*, 2010; Forgeard, 2011). Regression analysis is also possible if the data are suitable for it.

The SPANE measurement instrument has been reported to be capable of measuring positive and negative affective states regardless of their sources, arousal level or cultural context, and it captures feelings from the emotion circumplex (Diener *et al.*, 2009a; Li *et al.*, 2013).

The time-span of four weeks was chosen in SPANE in order to provide a balance between the sampling adequacy of feelings and the accuracy of memory (Li *et al.*, 2013), and to decrease the ambiguity of people's understanding of the scale itself (Diener *et al.*, 2009a). Therefore, four weeks is considered as a good candidate for the assessment of pre-existing affect of the participants.

The SPANE has been validated to converge to other affective states measurement instruments, including PANAS (Diener *et al.*, 2009a), despite of its shorter length. The scale provided good psychometric properties in the introductory research (Diener *et al.*, 2009a) and in numerous follow-ups, with up to twenty-one thousand participants in a single study (Silva and Caetano, 2011; Dogan *et al.*, 2013; Li *et al.*, 2013; Jovanović, 2015). Additionally, the scale proved consistency across full-time workers and students (Silva and Caetano, 2011).

The SPANE questionnaire has been employed in our research activities for assessing the pre-existing affects of the participants in the long run. That is, the natural affects of the participants have been gathered using SPANE.

**Self-Assessment Manikin**

One of the most notable instruments is the Affect Grid Russell *et al.* (1989), which is a grid generated by intersecting the axes of valence and arousal accompanied by four discrete affects, i.e. depression-relaxation and stress-excitement, to guide the participant in pointing where the emotional reaction is located. The affect grid has been employed in SE research, e.g. in Colomo-Palacios *et al.* (2011). Yet, the grid was shown to have only moderate validity Killgore (1998), thus other measurement instruments would be more desirable.

Thus comes the Self-Assessment Manikin (SAM, Bradley and Lang (1994); Lang *et al.* (1999)). SAM is a pictorial, i.e. non-verbal, assessment method. SAM measures valence, arousal, and dominance associated with a person's affective reaction to an object (or a stimulus) Bradley and Lang (1994). SAM is reproduced in Figure 2.1. The figures of the first row range from a frown to a smile, representing the valence dimension. The second row depicts a figure showing a serene, peaceful, or passionless face to an explosive, anxious, or excited face. It represents the arousal dimension. The third row ranges from a very little, insignificant figure to a ubiquitous, pervasive figure. It represents the dominance affective dimension.

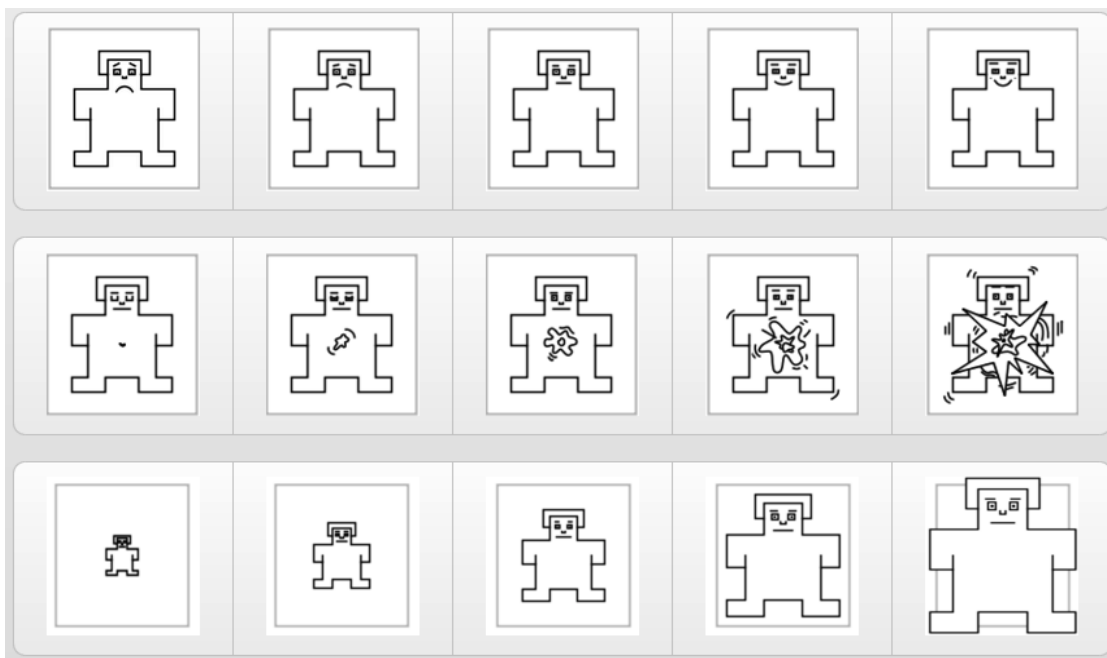

FIGURE 2.1: *The Self-Assessment Manikin.*

The values of the affective state constructs range from one to five. A value of three means "perfect balance" or "average" between the most negative (1) and the most positive value (5). For example, the value of one for the valence variable means "complete absence of attractiveness". A value of five for valence means "very high attractiveness and pleasure towards the stimulus".

The SAM has been under scholarly scrutiny, as well. As reported in Kim *et al.* (2002), SAM has the advantage of eliminating the cognitive processes associated with verbal measures but it is still very quick and simple to use. The original article describing SAM already reports good psychometric properties Bradley and Lang (1994). A very high correlation was found between the SAM items and those of other verbal-based measurement instruments Morris and Waine (1993); Morris (1995), including high reliability across age Backs *et al.* (2007). Therefore, SAM is one of the most reliable measurement instruments for affective reactions Kim *et al.* (2002).

SAM has been chosen in our research activities for assessing the affects raised by an object (stimulus) because of its peculiarities, for example, the pictorial items and the shortness of the questionnaire itself. Finally, the several psychometric studies have evaluated SAM as a reliable and valid tool for affect assessment.

**Issues when measuring affects**

These scales, and similar other psychological measures, present issues when employed within (and between) subjects analyses of repeated measurements, which is the case one in one of our studies. First, there is not a stable and shared metric for assessing the affects across persons. For example, a score of one in valence for a person may be equal to a score of three for another person. However, a participant scoring two for valence at time $t$ and five at time $t+x$ unquestionably indicates that the participant's valence increased. As stated by Hektner *et al.* (2007), "*it is sensible to assume that there is a reasonable stable metric within persons*" (p. 10). In order to have comparable measurements, the raw scores of each participant are typically transformed into z-scores (also known as standard scores). The z-score transformation is given in Equation (2.1):

$$z_{score}(x_{pc}) = \frac{x_{pc} - \bar{x}_{pc}}{s_{pc}} \qquad (2.1)$$

where $x_{pc}$ represents a participant's measured construct, $\bar{x}_{pc}$ is the average value of all the construct measurements of the participant, and $s_{pc}$ is the standard deviation for the participant's construct measurements

A z-score transformation is such that a participant's mean score for a variable is zero, and scores for the same variable that lie one standard deviation above or below the mean have the value equivalent to their deviation. One observation is translated to how many standard deviations the observation itself is above or below the mean of the individual's observations. Therefore, the participants' measurements become dimensionless and comparable with each other, because the z-scores indicate how much the values are spread. (Larson and Csikszentmihalyi, 1983; Hektner *et al.*, 2007).

As the variables are transformed to z-scores, their values will follow a normal distribution in their range. The three-sigma rule states that 99.73% of the values lie within three standard deviations of the mean in a normal distribution (Pukelsheim, 1994). Therefore, the range of the variables, while theoretically infinite, is practically the interval [-3, +3].

Second, the repeated measurements employed in contexts, like those of two of our studies, present dependencies of data at the participants' level and the time level grouped by the participant. The analysis of variance (ANOVA) family provides rANOVA as a variant for repeated measurements. However, rANOVA and general ANOVA procedures are discouraged (Gueorguieva and Krystal, 2004) in favor of mixed-effects models, which are robust and specifically designed for repeated, within-participant longitudinal data (Laird and Ware, 1982; Gueorguieva and Krystal, 2004; Baayen *et al.*, 2008).

Linear mixed-effects models are the most valuable tool to be employed in such cases. A linear mixed-effects model is a linear model that contains both fixed effects and random effects. The definition of a linear mixed-effects model by Robinson (1991) is in Equation (2.2)

$$y = X\beta + Zu + \varepsilon \tag{2.2}$$

where $y$ is a vector of observable random variables, $\beta$ is a vector of unknown parameters with fixed values (i.e., fixed effects), $u$ is a vector of random variables (i.e., random effects) with mean $E(u) = 0$ and variance-covariance matrix $\text{var}(u) = G$, $X$ and $Z$ are known matrices of regressors relating the observations $y$ to $\beta$ and $u$ , and $\varepsilon$ is a vector of independent and identically distributed random error terms with mean $E(\varepsilon) = 0$ and variance $\text{var}(\varepsilon) = 0$.

The estimation of the significance of the effects for mixed models is an open debate (Bates, 2006; R Community, 2006). One proposed way is to employ likelihood ratio tests (*ANOVA*) as a way to attain *p-values* (Winter, 2013). With this approach, a model is constructed by adding one factor (or interaction) at a time and performing a likelihood ratio test between the null model and the one-factor model. By adding one factor or interaction at a time, it would be possible to construct the best fitting possible model. However, this technique is time-consuming and prone to human errors. Another proposed approach is to construct the full model instead. A way to express the significance of the parameters is to provide upper and lower bound *p-values*. Upper-bound *p-values* are computed by using as denominator degrees of freedom the number of data points minus the number of fixed effects and the intercept. Lower bound (or conservative) *p-values* are computed by using as denominator degrees of freedom the number of data points minus the within-participant slopes and intercepts multiplied per the number of participants. The reader is advised to read the technical manual (Tremblay and Ransijn, 2013) for additional details. This is the approach that was followed in our studies.

Lastly, agreement measurements exist to provide validation of repeated affective states measurements. Cronbach (1951) developed the Cronbach's alpha as a coefficient of internal consistency and interrelatedness specially designed for psychological tests. It considers the variance specific to individual items. The value of Cronbach's alpha ranges from 0.00 to 1.00, where values near 1.00 mean excellent consistency (Cronbach, 1951; Cortina, 1993).

Our last note is on the misuse of the term *psychometrics in* previous software engineering research. So far psychometrics has been employed to mean *psychological measurements* (Feldt *et al.*, 2008). However, *psychometrics* is the field of study concerned with the implementation and validation of psychological measurements. A measurement instrument in psychology has to possess acceptable validity and reliability properties, which are provided in psychometric studies of the measurement instrument. A modification to an existing measurement instrument (e.g., adding, deleting, or rewording items) often requires a new psychometric study. For similar reasons and for ensuring higher reproducibility, participants' instructions should be made available with a paper, because the instructions might influence the participants' responses.

## 2.2 Performance

Performance has been defined in the Oxford Dictionary as "the accomplishment [...] of something commanded or undertaken" but also as "the quality of execution of such an action, operation, or process; the competence or effectiveness of a person or thing in performing an action; spec. the capabilities, productivity, or success of a machine, product, or person when measured against a standard" (OED, 2013). It follows that performance can mean several things even in common use of the word itself.

Typically software team performance is associated with *productivity* (Kettunen, 2013), and our stance for this investigation is that performance and productivity are two interchangeable terms, in line with several other authors, e.g., Kettunen (2013); Fagerholm *et al.* (2015, 2014); Petersen (2011); Meyer *et al.* (2014); Kaplan *et al.* (2009a).

In science and research, performance is a multi-faceted construct, which has been employed at different organizational levels to mean different things (Freeman and Beale, 1992). As stated by Kettunen and Moilanen (2012), "there is no one universal measure of software [...] performance. [...] the performance is relative to the organizational environment of the team." (p. 79). This issue is reflected by our review of the most prominent productivity and performance studies in software engineering, which we show in the following section.

### 2.2.1   Job performance

Job performance has been investigated in several disciplines, especially in organizational behavior and industrial-organizational (I/O) psychology. Over the last two decades, research within I/O psychology has reached a consensus that job performance is multidimensional (Rotundo, 2002). As summarized by Kaplan *et al.* (2009a), the most prominent sub-dimensions of job performance are task performance, organizational citizenship behaviors, counterproductive work behaviors, and work withdrawal.

*Task performance* refers to the activities that are recognized to be part of the job itself, thus contributing to the core of an organization (Borman and Motowidlo, 1997); these activities would be mentioned when an individual describes his/her job (Kaplan *et al.*, 2009a). More simply stated, task performance is the accomplishment of duties and responsibilities associated with a job (Murphy, 1989).

*Organizational citizenship behaviors* are those behaviors that are considered "above and beyond" an employee's formal job description to facilitate organizational functioning (Lee and Allen, 2002). Sometimes called citizenship performance, the construct refers to "the behavior that contributes to the organization by contributing to its social and psychological environment" (Rotundo (2002), p. 68).

In contrast, *counterproductive work behaviors* are voluntary behaviors that go against organizational values and norms and harm the well-being of the organization Rotundo (2002) and/or its members (Robinson and Bennett, 1995) . Examples reported by Kaplan *et al.* (2009a) include theft, harassment, sabotage, and physical aggression.

Lastly, *work withdrawal* refers to employees' attempts to avoid their work tasks, or disappear from the work environment (Hanisch and Hulin, 1991). Examples include absenteeism, tardiness, and turnover.

Problem solving is often employed to conceptualize task performance (Brief and Weiss, 2002), and it has been recently proposed as one of the main components of a general theory of software engineering (Johnson and Ekstedt, 2015). In section 2.2.3, we will show that developers value problem setting and solving as proxies for understanding their own performance; therefore, we explore the concept of problem solving performance in the next section.

## 2.2.2 Problem-solving performance

As software development is mainly characterized by intellectual and cognitive activities (Glass *et al.*, 1992; Darcy, 2005; Khan *et al.*, 2010), performance in software development can be conceptualized as a (series of) problem-solving activity. Problem solving, which is often employed to conceptualize performance (Brief and Weiss, 2002) refers to a state of desire for reaching a definite "goal" from the present condition that either is not directly moving toward the goal, is far from it, or needs more complex logic for finding a missing description of conditions or steps toward the goal (Robertson, 2001).

Goal settings has an established line of research in organizational behavior and psychology, especially in the works of Locke—one of the seminal works is Locke (1968). It involves the development of a plan, which in our case is internalized, designed to guide an individual toward a goal Clutterbuck (2010).

In a recent general theory of software engineering, Johnson and Ekstedt (2015) have proposed problem solving as the lens through which we should see the software engineering endeavor of developing computer programs that provide utility to stakeholders. Problem-solving can be classified as being analytical (or rational, mathematical) or creative (Simonton, 1975; Friedman and Förster, 2005).

Researchers have sometimes related creative and analytic problem solving to convergent and divergent thinking (Cropley, 2006; Csikszentmihalyi, 1997), which map roughly onto creativity and analytic problem solving studies, according to Csikszentmihalyi (1997). Divergent thinking leads to no agreed-upon solutions and involves the ability to generate a large quantity of ideas that are not necessarily correlated (Csikszentmihalyi, 1997). Convergent thinking involves solving well-defined, rational problems that often have a unique, correct answer and emphasizes speed and working from what is already known, which leaves little room for creativity because the answers are either right or wrong (Cropley, 2006; Csikszentmihalyi, 1997).

Although divergent and convergent thinking have not proven to be synonyms for creativity and analytical problem-solving capabilities—there are no clear definitions for these terms (Amabile, 1982)—these are the two dimensions that most studies analyze while still claiming to study creativity and analytical problem-solving Csikszentmihalyi (1997).

*Creativity* has been studied in psychology and cognitive science since for than 60 years, and it has been acknowledged as being necessary for technology (Simonton, 2000). Creativity is a multi-faceted, multi-disciplinary concept that is difficult to measure (Piffer, 2012). Over one hundred definitions exist for creativity, spanning several disciplines (Hocevar and Bachelor, 1989). Rhodes (1961) suggested that creativity can be an attribute of a process, a product, a person or press, so called four P's model of creativity. In research papers, however, the outcomes of creative performance are often conceptualized as performance itself, in terms of novelty and value (Davis, 2009a). This means that the conceptualization of creativity in terms of process, person, and the press is often tied to the evaluation of a creative product. This investigation adheres to the standard practice of concentrating on creativity in terms of generated intellectual products, namely ideas and strategies. Yet, it is useful to define creative ideas.

Dean *et al.* (2006) argued that, to define idea creativity, it is helpful to differentiate it from the concept of creativity itself. Drawing upon MacCrimmon *et al.* (1994), they defined "*a creative idea as a quality idea that is also novel. That is, it applies to the problem, is an effective and implementable solution, and is also novel*". Based on a literature review of 51 studies on quality, novel and creative ideas, they summarized a conceptual framework of idea creativity. *Novelty* is considered to be the main dimension of creativity (Dean *et al.*, 2006). A novel idea is rare, unusual, or uncommon (Magnusson, 1993). According to this definition, the most novel idea is an idea that is totally unique;

conversely, the least novel idea is the most common one (MacCrimmon *et al.*, 1994). Dean *et al.* (2006) warned that when applying the framework, the novelty of any idea must be judged in relation to how uncommon it is in the mind of the idea evaluator or how uncommon it is in the overall population of ideas. Novelty is broken down into two constructs: originality and paradigm relatedness. Ideas are considered *original* when they are rare, but that also have the characteristic of being ingenious, imaginative or surprising. Idea originality ranges from those that are common and mundane to those that are rare and imaginative. *Paradigm relatedness* describes the transformation potential of ideas. It is the degree to which an idea relates to the currently prevailing paradigm. The second dimension of idea creativity is the *quality*, which is further divided into workability and relevance. An idea is *workable* (or feasible) if it can be easily implemented and does not violate known constraints. An idea is *relevant* if it applies to the stated problem and will be effective at solving the problem. There is a third construct suggested by Dean *et al.* (2006), namely specificity. An idea is specific if it is clear and worked out in detail. The construct is optional and should be included when it is the main focus of a study.

Despite that creativity can be defined by some constructs, there is the issue that it is domain specific (Silvia *et al.*, 2009). Additionally, regardless of the constructs and the measurement instrument, creativity and its assessment are subjective (Amabile, 1982; Lobert *et al.*, 1995; Silvia *et al.*, 2009; Piffer, 2012). One way to mitigate the subjectivity of creative evaluation is the consensual assessment technique (Amabile, 1982; Kaufman *et al.*, 2007). Experts in the domain of creative outcomes are asked to rate the creativity of products. These judges are never asked to explain or defend their ratings, but to use their expert sense of what is creative in the domain in question to rate the creativity of the products in relation to one another. In the consensual assessment technique and other measurement frameworks for creative outcomes, the measurements are often represented by Likert items.

### 2.2.3 Performance in software engineering research

In software engineering research, performance is a widely debated, probably never-ending dispute. Performance as a term has been substituting *productivity* in recent studies, yet our stance is that performance and productivity are two interchangeable terms, in line with Kettunen (2013); Fagerholm *et al.* (2015); Petersen (2011); Meyer *et al.* (2014); Kaplan *et al.* (2009a) from both software engineering and psychology research. We see a reason for this tendency of sidelining *productivity* as a term. Assuming that software engineering as a discipline was born at the end of the ́70s (Naur *et al.*, 1968), Boehm (1972) was perhaps the first to address the effort and the costs required to build a software. Some "*unhappy* options" to decrease the effort required to build software, thus enhancing productivity, have been to "skimp on testing, integration, or documentation" (p. 7). That is, to reduce time. Yet, Boehm (1972) had already anticipated the frustration that a search for a productivity measurement would have incurred, by suggesting that instead of creating aptitude tests of arithmetic and logic reasoning, e.g., (Reinstedt, 1967), productivity should be defined "in terms of source instructions" (p. 9). This call was echoed by much of the research conducted on developers; productivity. After all, it would be difficult to blame the quest for a formula based on the economics known ratio *output/input*.

Perhaps Chen (1978) was the first to formalize the well known ratio of source statements over the time taken for producing them. This ratio and its variations have been widely criticized. Three of these variations are (1) the effort needed when developing software depends on the problem being solved and will vary with the complexity of each task (Collofello and Woodfield, 1983); (2) different developers will design different solutions for the same problem, yet using different numbers of lines of codes, thus providing different levels of non-comparable efficiency; (3) the number of lines of code can only be determined with confidence near the completion of the project, thus predicting the productivity is difficult with this approach.

In order to overcome some of the limitations of the classic productivity ratio, some enhancements have followed over time. There have been proposals of employing new variables such as the size of the problem and the quality of the output (Chrysler, 1978), normalizing the ratio across programming languages (Vosburgh *et al.*, 1984), and weighting factors in the formula (Pfleeger, 1991). Other attempts to determine the effort needed to produce software have been the use of function points (Albrecht and Gaffney, 1983) instead of lines of code, or abstracting the programming languages characteristics (Halstead, 1977).

There have also been several proposals not building upon the ratio formula. For example, it has been suggested to analyze the earned value as the percentage of progress towards the final product (Kadary, 1992). Other proposals have employed non-parametric analysis like data envelopment analysis (Pfleeger, 1991), where inputs and outputs are weighted and their ratio is maximized. There are also proposals to use bayesian belief nets, where productivity is represented as a directed graph in which the nodes are productivity-related variables connected by cause-effect relationships (Stamelos *et al.*, 2003).

The Institute of Electrical and Electronics Engineers (IEEE) has attempted to end up the debate by standardizing a way to measure productivity in 1992 (IEEE, 1992). The standard was withdrawn in 2008.

Numerous other proposals exist, and it is outside the scope of this dissertation to review all of them. Please see the work by Petersen (2011) for a systematic review. Besides of issues related to our lack of understanding of what productivity is, the previous studies also lack in their validity threats evaluations or have insufficient number of empirical studies behind (Petersen, 2011). Additionally, the studies of the productivity factors highlight too many variables (Scacchi, 1995). The point is that we still do not know what productivity is in software development (Dale and van der Zee, 1992).

Yet, there have been numerous attempts to address the *human aspects* in software productivity studies.

More than 30 years ago, programmer's productivity was believed to be influenced by different characteristics. Chrysler (1978) considered developers' productivity to be influenced by characteristics at the technical level, the knowledge level, and the developer level. However, only skills and experience, measured as the number of months, were taken into consideration as human factors.

Boehm and Papaccio (1988) identified factors influencing productivity by controlling the costs of producing software. The study suggested strategies to improve productivity such as writing less code and "getting the best from people". This study suggested ways for improving productivity such as assigning people private offices, creating a working environment to support creativity, and providing incentives to enhance the motivation of people and their commitment.

Scacchi (1995) observed which factors influenced software productivity and how productivity can be improved. His review focused on the creation of a framework to predict the productivity of large-scale software systems. The study criticized the previous research, because they failed to describe the variation in productivity among individual programmers. An important impact was attributed to human-related factors. Scacchi (1995) argued that organizational and social conditions can even dominate the productivity attributable to in-place software development technologies. The study called for improvement and alternative directions in software productivity measurements.

A recent review of productivity factors by Sampaio *et al.* (2010) identified three main areas in the body of knowledge: product, project, and people. The identified people-related factors consisted of the motivation of the team, and the individuals' skills, but also relationships, and the quality of management.

Despite the fact that we do not really know what performance is in software development, when we ask developers about their productivity while they develop software, they are able to quickly respond to our inquiry. They can tell us if they are currently productive, they can tell us if they are having a productive day, and they can tell us if they have been productive recently. For this reason, the self-assessment of productivity and performance has been employed in recent studies, e.g., in (Muller and Fritz, 2015). Self-assessed performance is commonly employed in psychology studies (Beal *et al.*, 2005), for example in the studies by Fisher and Noble (2004); Zelenski *et al.* (2008) that are included in the literature review of this thesis. Self-assessed performance rates are often necessary in real-time studies of affects and performance (Beal *et al.*, 2005); yet self-report are consistent to objective measurements of performance, although objective measures are still preferable (Dess and Robinson, 1984; Miner and Glomb, 2010). There is also the evidence that bias is not introduced by mood in self-reports of performance when individuals alone are being observed (Beal *et al.*, 2005; Miner and Glomb, 2010).

The perception of productivity in software developers has been explored by Meyer *et al.* (2014) in two studies. We review here the first study only, as it is the one relevant for this PhD dissertation. In the first study, a survey, the researchers collected 185 complete responses on how developers consider a day as a productive day, how they perceive work activities as productive activities, and how they measure and assess their productivity. More than half (53.2%) of the participants considers a productive day when they complete their tasks, achieve some planned goals, and perceive a progression on their goals. The second most mentioned (50.4%) reason for considering a day as a productive day is when they are able to enter into a *flow* mode of programming, where context-switching is limited and where distractions are almost absent. Other reasons include having no meetings, having clear goals, and having clear plans for a workday. Regarding how developers measure their productivity, Meyer *et al.* (2014) let the participants rate 23 possible measures. Overall, the metric with the highest rating was "The number of work items (tasks, bugs, ..) I close". In an open-ended field, where the developers were free to input what they would instead use as ways to measure performance, the developers responded with "Time spent per activity" (27.0%), and again "Achievement" (17.7%) in terms of the actual work done and the progress. Overall, the results suggest that developers see performance as the capacity to define problems and solve them, by progressing towards the solution of such problems.

> For the purposes of this investigation, we refer to **performance** as those actions, behavior and outcomes that individuals engage in or bring about that are linked with and/or contribute to some goals, which usually corresponds to those of an organization (Borman and Motowidlo, 1997; Viswesvaran and Ones, 2000; Murphy, 1989).

This definition, which roughly corresponds to the one of task performance in I/O research, is the one that has historically been identified mostly with job performance itself (Kaplan *et al.*, 2009a).

In addition to adhering to a definition of performance derived from psychology, we have chosen to assess the performance of software developers using two approaches. The first approach has been to employ behavioral metrics of problem solving performance that can be translated to software development (Paper IV). The second approach has been the employment of self-assessments of performance (Paper V; Paper VI), in line with several recent studies (Muller and Fritz, 2015; Meyer *et al.*, 2014).

## 2.3   Affects and performance

In Section 2.2.3, we provided the literature review about performance as studied in job performance and I/O psychology, problem-solving performance in general cognitive psychology, and performance in software engineering studies. This section reviews the literature about the relationship between affects and performance.

### 2.3.1 Affects and job performance

Some words should be spent on the more general studies on the relationship between affective states and job-related performance.

Fisher and Noble (2004) employed the Experience Sampling Method (Larson and Csik-szentmihalyi, 1983) to study correlates of real-time performance and affective states while working. The study recruited different workers (e.g., childcare worker, hairdresser, office worker); however, none of them was reported to be a software developer. The measurement instrument was a questionnaire with 5 points Likert items. The study analyzed self-assessed skills, task difficulty, affective states triggered by the working task, and task performance. Among the results of the study, it was shown that there is a strong positive correlation between positive affective states and task performance while there is a strong negative correlation between negative affective states and task performance. The authors encourage further research about real-time performance and emotions.

Along the same line were Miner and Glomb (2010), who performed a similar study with 67 individuals working in a call-center and sampled up to 5 times per day. Within-persons, periods of positive mood are associated with periods of improved task performance, in this case in terms of shorter support call-time.

Oswald *et al.* (2015) argued that research in economics and management has lacked studies on the relationship between happiness of workers and their productivity. They conducted a controlled experiment where 182 participants were divided into two groups. The first group received a treatment of positive affective states induction–i.e., a comedy clip, while the second group did not receive any treatment. The participants performed two mathematical tasks and their performance in the tasks represented their productivity. The results show that a rise in positive affective states leads to higher productivity. The effect was found to be equally significant in male and female sub-samples.

### 2.3.2 Affects and problem-solving

According to a recent meta-analysis on the impact of affective states on creativity (i.e., creative outcomes) in terms of the quality of generated ideas, positive affective states lead to higher creativity than neutral affective states, but there are no significant differences between negative and neutral affective states or between positive and negative affective states (Baas *et al.*, 2008). Another recent meta-analysis agreed that positive affective states have moderate effects on creativity with respect to neutral affective states. However, it showed that positive affective states also have small, non-zero effects on creativity with respect to negative affective states (Davis, 2009a). Lewis *et al.* (2011) provided evidence for higher creativity under induced positive and negative affective states, with respect to non-induced affective states. Forgeard (2011) showed that participants low in depression possess higher creativity with induced negative affective states, and no benefits were found for individuals with induced positive affective states. Sowden and Dawson (2011) found that the quantity of generated creative ideas is boosted under positive affective states, but no difference in terms of quality was found in their study. On the other hand, there are studies empirically demonstrating that negative affective states increase creativity (Kaufmann and Vosburg, 1997; George and Zhou, 2002). Based on the literature review, the author of this study has agreed with Fong (2006) that the nature of the relationship between affective states and creativity has not been completely understood and that more research is needed. No direction could be predicted on a difference between creativity and affective states of software developers.

In contrast to the case for creativity, fewer studies have investigated how affective states influence analytic problem solving performance. The understanding of the relationship is still limited even in psychology studies. In her literature review on affects and problem-solving skills, Abele-Brehm (1992) reported that there is evidence that negative affects foster critical and analytical thinking. Successive theoretical contributions have been in line with this suggestion. In their mood-as-information theoretical view, Schwarz and Clore (2003) argued that negative affects foster a systematic processing style characterized by bottom-up processing and attention to the details, and limited creativity. Spering *et al.* (2005) observed that negative affects promoted attention to the details to their participants, as well as analytical reasoning. It appears that analytical problem-solving skills—related to convergent thinking—are more influenced by negative affective states than by positive affective states. However, there are studies in conflict with this stance. Kaufmann and Vosburg (1997) reported no correlation between analytical problem-solving skills and the affective states of their participants. On the other hand, the processes of transferring and learning analytical problems have been reported to deteriorate when individuals are experiencing negative emotions (Brand *et al.*, 2007). Melton (1995) observed that individuals feeling positive affects performed significantly worse on a set of syllogisms (i.e., logical and analytical reasoning). Consequently, based on the limited studies, no clear prediction about the relationship between affective states and analytic problem solving skill could be made.

## 2.4 Psychology in software engineering research

There have been studies in software engineering research using psychology, especially in venues like the International Workshop on Cooperative and Human Aspects of Software Engineering (CHASE), the International Workshop on Social Software Engineering (SSE), and some special issues on the human aspects of software developers. However, there has never been a solid line of research on this avenue, nor guidelines before this PhD dissertation. This is surprising, as the psychology of programming has been dealt with in the 1970s already (Weinberg, 1971) but has never reached mainstream attention in research communities.

The quotation page of this dissertation states "Junior programmer's bookshelf: 90% APIs and programming languages; Senior programmer's bookshelf: 80% applied psychology". This appears to imply something that we give as obvious: we all *know* that psychological issues are major in software development. Yet, the majority of software engineering research has been on the technical side, not on the soft (or human) side (Lenberg *et al.*, 2015).

We see in (Feldt *et al.*, 2008; Lenberg *et al.*, 2014, 2015) the first proposals to employ validated research from behavioral science in software engineering. Feldt *et al.* (2008) presented the preliminary results of a study on the personality of software engineers, which was conducted using validated measurements and theory of psychology. The results of the study are interesting but, as stated earlier, are outside the scope of this dissertation. However, the authors have argued through the article that researchers in software engineering should put a larger focus on the humans involved in software development. This will be achieved by "collecting psychometric measurements". Feldt *et al.* (2008) continued by arguing that when introducing a change in a software engineering endeavor, the lack of understanding of the people interested by the change has the risk that the effects are dwarfed by how the persons accept and adapt the introduced change. In order to avoid the issue, the authors stress that future research in software engineering should adopt psychometric instruments such as questionnaires to measure personality, attitudes, motivations and emotions.

Lenberg *et al.* (2014, 2015) provided a systematic literature review of the research area on the human aspects of software engineering concerned with the area of "behavioral software engineering". The authors have defined behavioral software engineering as "the study of cognitive, behavioral and social aspects of software engineering performed by individuals, groups or organization" (p. 18). The authors found out that less than 5% of software engineering studies have dealt with *soft aspects* or human-related topics, which arguably implies that software engineering research has neglected the human aspects of programming. Throughout the paper, Lenberg *et al.* (2015) have laid down clear definitions for psychology and its sub-fields work and organizational psychology, behavioral economics, and they identified the psychology of programming interest group (PPIG) and CHASE as the venues more interested in psychological aspects of software engineering. In the systematic literature review, the authors identified 250 publications in the area of behavioral software engineering among more than 10,000 screened for relevance. Lenberg *et al.* (2015) identified 55 areas of behavioral software engineering, which include for example cognitive style, job satisfaction, motivation, commitment, positive psychology, self esteem, stress, work life balance, decision making, group dynamics, leadership, and organizational culture. The analysis of the literature has indicated that the research area is growing and considering an ever increasing number of psychological issues, but for more than half of the identified 55 areas, less than five publications were found, indicating that much research is still needed. As a general strategy for the research area, the authors have proposed to conduct exploratory research which considers several behavioral software engineering concepts at the same time, for creating an understanding of which concepts should be considered in more detail in future research.

Concurrently to Lenberg *et al.* (2014, 2015), we conducted a workshop at the 2014 annual meeting of the International Software Engineering Research Network (ISERN) called

"psychoempirical software engineering" (Graziotin *et al.*, 2014d), where we presented the opportunity of conducting empirical software engineering research with proper theory and measurement instruments from psychology, and we initiated the opportunities for collaboration on this avenue and discussion on terminology and how to effectively operationalize the research. As a result of the discussion, which continued in several venues, we proposed our guidelines for psychoempirical software engineering research in Paper II, which complement Lenberg *et al.* (2014, 2015) on the operational side. The guidelines are included in this dissertation in Appendix C, and are discussed in the implications for research in Section 8.5.1.

## 2.5 Affects in software development

In software engineering research, there is a limited number of studies on the affects of software developers, let alone those regarding the relationship between affects and the performance of developers. The most meaningful technique that could be employed for discovering the related work has been snowballing, also because it has been shown to be a very effective method for the systematic discovery of literature (Wohlin, 2014). Over the years, we actively searched in IEEE Xplore, ACM Digital library, AIS (Association for Information Systems) eLibrary, SpringerLink, Thomson Reuters' Web of Science, Scopus [4], and Google Scholar for articles mentioning stemmed words of emotions, moods, feelings, and affects in their titles and abstracts.

We distinguish between those articles dealing with the affects of developers in general, and those about understanding the relationship between affects and the performance of developers.

### 2.5.1 Affects of software developers

Requirements engineering and mining software engineering data are perhaps the two fields that have dealt with affects mostly, although recent articles from those fields (Miller *et al.*, 2015; Jurado and Rodriguez, 2015) have claimed that the related research is pretty much non-existing. Researchers in requirements engineering have long argued that the emotions and moods of the customers should be taken into account when eliciting requirements Ramos and Berry (2005). Yet, we are aware of only one study that has incorporated the affects of the requirements engineers, too Colomo-Palacios *et al.* (2010). A third field that has dealt with affects is the one of information systems, which has also neglected those who develop software and has mainly focused on emotions when adopting information systems, e.g., Hekkala and Newman (2011); Hekkala *et al.* (2011).

---

[4] For general purpose academic search databases, we added stemmed keywords such as "software" and "developer", "programmer", "engineer" for shrinking down the relevant results, as the principal keywords alone yield several thousands of results.

Generally speaking, we could not find studies that explicitly deal with affects in the context of software development prior to the works of Shaw (2004). By snowballing all found articles, we would conclude that affects in software engineering have been neglected till the last decade.

Shaw (2004) was perhaps the first to address the importance of studying the affects of developers. He observed that the role of emotions in the workplace has been the subject of management research, but information systems research has focused on job outcomes such as stress, turnover, burnout, and satisfaction. The study explored the emotions of information technology professionals and how these emotions can help explain their job outcomes. The paper employed the Affective Events Theory (Weiss and Cropanzano, 1996) as a framework for studying the fluctuation of the affective states of 12 senior-level undergraduate students who were engaged in a semester-long implementation project for an information systems course. The participants were asked to rate their affective states during or immediately after their episodes of work on their project. At four intervals during the project, they filled out a survey on stress, burnout, emotional labor, and identification with their teams. Shaw considered each student to be a single case study because a statistical analysis was not considered suitable. The study showed that the affective states of a software developer may dramatically change during a period of 48 hours and that the Affective Events Theory proved its usefulness in studying the affective states of software developers while they work. Shaw (2004) concluded by calling for additional research.

Syed-Abdullah *et al.* (2005) conducted a longitudinal study about how the extreme Programming (XP) methodology is capable to raise positive affects among developers. The authors conducted an experiment with students, who were divided into groups for developing a software project each. No information regarding the group compositions (how many groups, and how many students per group) and the project types were provided in the article. Half groups were trained to use the XP methodology, while the other were trained to use the Discovery Method. Their affects were measured before starting the projects and before delivering the projects, 10 weeks after the start of the course. The Positive and Negative Affect Schedule (PANAS) (Watson *et al.*, 1988a), reviewed in Section 2.1.3) was employed as a measurement instrument. However, only the measures of positive affects (the PA in PANAS) was retained to compare the groups. That is, the authors did not evaluate negative affects (the NA in PANAS) of the groups. The initial assessment of the positive affects of 50 students showed similar values for the two "treatments" of XP and Discovery, 31.85 and 30.28 respectively, although the authors did not report what these numbers actually mean and what is the range of the positive affects value. Additionally, no significance was provided. The second measurements with 47 students showed a difference between the two treatments, that were 32.72 for XP and 29.92 for Discovery. The authors claim significant difference at 0.10 confidence level. That is, the claim was that the students following the XP methodology became happier on average. Additionally, the authors studied the "relationship between XP practices and positive temperament [. . .] using the Pearson Correlation test. There was a significant relationship between the two variables for every project [r = 0.331, p = 0.030] indicating that the higher the number of XP practices used, the higher is the level of positive temperament experienced." (p. 220). The present author has decided to quote the previous sentences regarding the evaluation to underline the lack in details of this article, because that quoted sentence was the entire description of the second assessment. Overall, the study was interesting and very novel for the field. It was actually the only reviewed article employing a psychometric assessment of the two positive affects measurements. Unfortunately, the authors omitted very basic details such as the group numbers and compositions–leaving the reader to wonder whether the groups were balanced or not–the reason why three students dropped at the second measurement session, an explanation of the PANAS metrics, a description of the number of XP practices employed over time and how they were introduced, an explanation and the data-points for employing a coefficient measurement, and the reason why negative affects were not observed beyond the hollow "positive moods operated as a single construct indicating that the fluctuation of positive moods has no effect on the negative moods of a person". Finally, the claimed significance at $p < 0.10$ is not considered a value for significant results in research (Nuzzo, 2014).

Colomo-Palacios *et al.* (2010, 2011) considered requirements engineering a set of knowledge-intensive tasks. The study aimed to integrate the stakeholder's emotions into the process of requirements engineering. The authors conducted two empirical studies on two projects. The first project consisted in the maintenance of a legacy system, while the second project was the development of a touristic information system. In total, 65 user requirements were produced between the two projects, which lasted between six and seven months respectively. Each requirement faced different revisions, up to 97 for the first project and up to 115 in the second project. Each participant rated the affective state associated to each requirement version. Affective states in this study were represented with the dimensional state using the components of valence and arousal. The results showed that high arousal and low pleasure levels are predictors of high versioning requirements. Additionally, valence increased throughout versions (thus, over time), while the arousal decreased. The authors questioned what could be the role of time in the emotional rating of the participants, and called for more research on the role of affects in software engineering.

Bledow *et al.* (2011) built on top of self-regulation theories to construct a model of work engagement based on affects. By applying the experience sampling method on 55 software developers (Larson and Csikszentmihalyi, 1983) over the course of two weeks, the authors measured, twice per day, (1) the participants' affect with the PANAS measurement instrument and with a list of 12 adjectives describing free-floating affects (as in moods), (2) the work engagement using the Utrecht Work Engagement scale, (3) the rates of the occurrence of positive and negative work events (e.g., praise from the supervisors, conflicts with colleagues). The multilevel modeling revealed that an affective shift, that is negative affect followed by positive affect, is positively related to work engagement. Among the implications, the authors argued that it is beneficial to accept negative experiences such as crises, conflicts, and errors, as unavoidable aspects of daily work, and that in the absence of negative experiences, people will likely value positive experiences less; thus showing a lower level of work engagement.

Denning (2012, 2013) authored two viewpoint articles in the Communications of the ACM. In them, it has been argued that recognizing the moods of all the stakeholders involved in producing software is essential to an Information Technology (IT) company success. Denning (2012) warned the readers that our own mood is shared with others around, and that moods affect us. Denning (2012) continued by attempting to differentiate emotions and moods, by stating that they are not the same thing. He defined emotions as "feelings that individuals experience in response to various stimuli", and moods as "general pervasive states of interpretation about the world" (p. 34). While the present PhD candidate understands that the Communications of the ACM magazine attempts to communicate to practitioners by limiting the academic language and writing style, Denning (2012) definitions of mood could be more specific, as moods are not necessarily interpretation of the world—indeed, we showed earlier that is object-free. Furthermore, Denning (2012) claimed that "psychologist recognize eight basic emotions, with each positive balanced by a negative, as follows: love-hate, joy-sadness, peace-anger, curiosity-fear". The claim is incorrect and not backed up by any reference. This entire chapter has shown that it is not true that a unique dominant, accepted theory exist for affect, emotions, and moods. Researchers should recognize this issue when employing such delicate concepts for constructing research.

Guzman and Bruegge (2013); Guzman (2013) published two short papers that emphasize emotion mining and summary Guzman and Bruegge (2013) and emotion visualization during the software development process Guzman (2013). In the first work, the authors collected 1000 collaboration artifacts, in the forms of mailing list messages and web pages, from three development teams over a period of three months. Subsequently, they performed textual sentimental analysis on extracted topics from the artifact. A sentimental score was thus assigned to each topic and single artifacts. Guzman and Bruegge (2013) interviewed the project leaders by showing them the scores and the artifacts. Overall, the project leaders agreed that the proposed emotion summaries were representative of the perceived emotional state of the project; thus, these summaries were found as useful for improving the team's emotional awareness. On the other hand, the interviews indicated that the current state of the summaries was not detailed enough for recalling events, and further improvements were still needed. Guzman (2013) adopted a similar approach to develop a prototype for visualizing different emotional aspects of a software development endeavor. The prototype displayed a view of summaries of textual contents together with an emotion color cue and score, and a detailed view displayed non-summarized textual contents with emotions cue and score, too. By interviewing one of the above-mentioned teams, Guzman (2013) observed an effectiveness of the tool to represent *mood swings* of the team before and after certain events, e.g., a deadline or a positive feedback. By analyzing the interviews, Guzman (2013) offered three hypothesis to be tested in future works. The following are the three hypothesis, quoted from the paper. (1) As important deadlines approach written communication in software development teams is lengthier, more frequent and more emotionally diverse. (2) Visualizations with topical and emotional information can help developers and managers remember and reflect about past events in a project and learn from their past experiences. (3) Detailed views explaining how the emotion score of an artifact was computed can increase the trust of the visualization and thus its usefulness.

Kolakowska *et al.* (2013) published a position paper on emotion recognition and how it is useful in software engineering. The authors propose eight scenarios for software usability testing with emotions (first impression, task-based, free [app] interaction, and combinations), and for software process improvement (Integrated Development Environment [IDE] usability comparison, influence on productivity, influence on code quality, comparison between local and remote workers). The authors proposed to adopt PAD models for emotion study with discrete labeling, which is an acceptable compromised also employed in some dimensional affective models. For assessing the affects, Kolakowska *et al.* (2013) proposed the usage of physiological sensors, video and depth sensors, and standard input devices (e.g., keystroke patterns).

Colomo-Palacios *et al.* (2013) have explored the emotional consequences of making hard decisions when managing software development. The authors reported two studies. In both studies, Colomo-Palacios *et al.* (2013) interviewed thirty managers with broad experience in managing development projects, using a semi-structured questionnaire. In the first study, the authors employed the Delphi method for finding out which are the most hard decisions in software development process management. In the second study, the authors interviewed the participants for understanding which emotions are more often felt while performing such decisions. The authors claimed that "there are six basic emotions or universal emotions: anger, happiness, fear, [. . .]" (Colomo-Palacios *et al.* (2013), p. 1079), which the present PhD candidate has shown to be a hollow claim. Furthermore, the authors of the study have decided to exclude two of the basic emotions (disgust and surprise) because their translation in the participants' languages was not having any meaning for them. The analysis was performed using a qualitative analysis tool (NVIVO) but no analysis strategy was reported. The results show that, when taking hard decisions in IT project management, the negative emotions are prevalent over positive emotions. The following were the three most mentioned emotions per each activity, in decreasing order. For project prioritisation: resignation, frustration, and annoyance. For requirements prioritisation: annoyance, resignation, and frustration. For selecting internal personnel: resignation, anxiety, and frustration. For partner-supplier election: frustration, resignation, and anxiety and anger (tie). For the development strategy election: anxiety, resignation, and frustration. Among the practical implications, the authors suggested that managers should be trained to face negative emotions in order to take decisions without these having a big influence.

De Choudhury and Counts (2013) investigated the expression of affects through the analysis of 204000 microblogging posts from 22000 unique users of a Fortune 500 software corporation. The study investigated the roles of (1) exogenous and endogenous factors, (2) physical location including distributed workers, (3) organizational structure and job roles, on affect expression. The authors employed a psycholinguistic lexicon tool called Linguistic Inquiry and Word Count. De Choudhury and Counts (2013) categorized each post with the two measures of positive and negative affects, thus somehow referring to the dimensional model of affects. A positive (negative) affect score for a post was defined as the ratio of positive (negative) words in the post to the total number of words in the post. Although the tool has been validated by several psychometric studies, De Choudhury and Counts (2013) conducted a successful validation test of the tool involving human raters. Regarding endogenous factors, the sentimental analysis of the blogposts revealed that positive affects drop significantly in the evening with respect to the morning (but negative affects do also drop, although less significantly). Regarding exogenous factors, the analysis revealed that IT-related issues were found to be often sources of frustration, which is not surprising but the authors suggested that the methodology would offer almost real-time monitoring of the specific causes of frustration. Day-to-day demands, e.g., meetings, were also associated with negative affects. Regarding the employees' physical locations, the study found that geographically distributed employees tend to express affects through social media more than those centrally located. One reason could be that non-distributed employees have several other media for expressing their feelings, like face-to-face interaction. Regarding organizational structure, the study found that those that are central in the enterprise network tend to share and receive high positive affect, while those in individual contributor roles tend to express more negative affect. One reason could be that it is known that those who are in managerial role have the power to influence the moods of employees; thus they should express positive feelings.

Guzman *et al.* (2014) performed a sentiment analysis of Github's commit comments. In particular, the study aimed to investigate how emotions are related to a project's programming language, the commits' day of the week and time, and the approval of the projects. The analysis was performed over 29 projects coming from a dataset of top-starred Github repositories. The projects were implemented in 14 different programming languages, and the results showed Java to be the programming language most associated with negative affect. Most commits (78%) were created during work-days, and the most negative day was Monday, which was followed by Saturday and Friday. However, only Monday was found to be significantly more negative than the other days. No significant differences were found regarding the time of the day. The project approval was defined as the number of Github stars—i.e., the number of users *linking* the project. No correlation was found between the number of Github stars and the affectivity of the commit messages.

Murgia *et al.* (2014) performed an exploratory analysis of emotions in software artifacts. The authors aimed messages from a project's public discussion board toward the creation of an automatic tool for emotion mining. Murgia *et al.* (2014) aim was to understand the feasibility of building automatic emotion mining tools, rather then building and validating one. Therefore, they set two research questions: "can human raters agree on the presence (absence) of emotions in issue reports?"; and "does context improve the agreement of human raters on the presence of emotions in issue reports?". The authors randomly sampled 392 developers' comments coming from the issue system of the Apache projects. The authors recruited 16 participants, which were students and researchers in computer science. The participants were divided into two balanced groups. They were asked to code each developer comment with emotions from the Parrott's framework for discrete emotions. The results showed that agreement, which was measured using Cohen $k$, could be reached only for love, joy, and sadness; thus, the suggestion provided by the authors is to focus only on the mining of these three emotions. Adding a context to a comment—which for the authors was achieved by showing the preceding comments—did enhance the raters' agreement score, but not significantly.

Novielli *et al.* (2014) proposed to mine emotions in Stack Overflow as a way to understand how the emotional style of a question asked influence the probability to obtain a satisfying answer in a certain time. The authors proposed a design where a successful question is a question with an accepted answer—although the present PhD candidate would argue to employ the answer with the highest score if one answer was accepted. The authors propose to use Stanford CoreNLP and LIWC as tools to analyze the sentiment and the affective word classes of the questions. The article was a short paper and no continuation was yet performed.

Tourani *et al.* (2014) wanted to understand how accurate is automatic sentiment analysis with SentiStrenght when applied to the mailing lists of software projects, what types of sentiments are observable in software projects mailing lists, and whether developers and users express different sentiments or not. The authors investigated the mailing lists of two Apache Foundation projects, namely Tomcat and Ant. Tourani *et al.* (2014) collected almost 600000 emails from the two mailing lists, after data cleanup (e.g., spam, automated messages). They then sampled 800 random emails (400 for each project), and two independent raters scored the sentiment of each e-mail using a score of -1, 0, or 1 (negative, neutral, and positive). The two raters obtained an agreement of 76%, although no agreement measure was indicated. By dividing the datasets into positive and negative months, the authors concluded that the sentiment analysis tool has a precision of about 30% for positive months and 13% for negative months, while the tool advertised a precision over 60% for both cases. The authors found out that this difference was to be found in the misinterpretation of many emails with a neutral tone. Regarding the type of the sentiment, the two previously mentioned raters categorized the emails. The positive categorizations were satisfactory opinion, friendly interaction, explicit signals, announcement, socializing, and curiosity. The negative sentiment categories were unsatisfied opinion, aggression, uncomfortable situation, and sadness. Regarding the users' vs. the developers' sentiments, the authors concluded that they show different proportions of sentiment categories during the construction and maintenance of the software project. Developers mainly communicate with each other, while when communicating with users, there is usually supportive communication exchange. User mailing lists feature more curiosity and sadness, but less aggression, announcement, and socializing.

Ford and Parnin (2015) have been exploring the causes of frustration in software engineering. Building on top of psychology research, which found out that frustrating experiences contribute to negative learning outcomes and poor retention, Ford and Parnin (2015) presented a study toward a common framework for explaining why frustrating experiences occur. The authors asked 45 software developers about the severity of their frustration and to recall their most recent frustrating programming experience. 67% of the participants reported that frustration is a severe issue for them. As for the causes for such frustration, the authors categorized the responses as follows: not having a good mental model of the code (for the category "mapping behavior to cause"), learning curves of programming tools, too large task size, time required for adjusting to new projects, unavailability of resources (e.g., documentation, server availability, ..), perceived lack in programming experience, not fulfilling the estimated effort for perceived simple problems, fear of failure, internal hurdles and personal issues, limited time, and issues with peers. The paper was a short article and a continuation is currently under work, but the authors have offered 11 interesting areas to focus on for understanding significantly strong negative affect that pervades software development.

Jurado and Rodriguez (2015) have (yet again) proposed the use of sentiment analysis techniques applied to issue tickets in order to monitor the underlying health of developers and projects. The authors performed an exploratory case study by gathering the issues of nine famous software projects hosted on GitHub, such as Homebrew, IPython, and Ruby on Rails. Instead of employing commercial sentimental analysis software, the authors built their own tool through a lexicon for detecting emotions in the title and the body of the issues. Jurado and Rodriguez (2015) correctly state that there is not a commonly agreed classification system for emotions and that current research either classifies the sentiment of some content as positive (negative) with a certain value, or detecting the basic emotions proposed by Ekman (1992). By preparing and applying their own lexicon using NLTK and SnowBall stemmer, the authors performed a correlation analysis of the occurrence of Ekman (1992) basic emotions, their distributions among the projects, and the cumulated sum for each sentiment ratio to an issue. The results showed that developers leave underlying sentiments in the text, and that those sentiments could be used to analyze the development process. Among the findings, the authors discovered that in open source projects, sentiments expressed in the form of joy is pervading and present at almost one magnitude of order with respect to the other basic emotions. Still, the content not classified as having a high sentimental content was more than 80% for each project.

### 2.5.2 Affects and performance in software engineering

Limited research was found on the affective states of software developers and their relationship with performance. To the present author's knowledge, six publications exist that employed psychological tests to study the affective states of software developers, excluding ours.

Lesiuk (2005) performed a quasi-experimental field study with an interrupted time series with removed treatment. She recruited 56 software engineers, who were working in four software companies, to understand the effects of music listening on software design performance. Data was collected over a five-week period, where the performance was self-assessed twice per day by the participants, together with their affective states. For the first week of the study (the baseline) the participants were only observed in natural settings. During the second and third week, the participants were allowed to listen to their favorite music whenever they preferred. During the fourth week, the software engineers were not allowed to listen to any music, all day long. During the fifth week, the participants were allowed again to listen to music. The results indicated that positive affects are positively correlated with music listening (or better, with the allowance of music listening). Then, positive affects of the participants and self-assessed performance were lowest with no music, while time-on-task was longest when music was removed. However, the correlation was not statistically significant. Narrative responses revealed the value of music listening for positive mood change and enhanced perception on design while working.

Khan *et al.* (2010) provided links from psychology and cognitive science studies to software development studies. The authors constructed a theoretical two-dimensional mapping framework in two steps. In the first step, programming tasks were linked to cognitive tasks. For example, the process of constructing a program – e.g. modeling and implementation – was mapped to the cognitive tasks of memory, reasoning, and induction. In the second step, the same cognitive tasks were linked to affects. Two empirical studies on affects and software development were conducted, which related a developer's debugging performance to induced affects. In the first study, affects were induced to software developers, who were then asked to complete a quiz on software debugging. The second study was a controlled experiment. The participants were asked to write a trace of the execution of algorithms implemented in Java. The results suggest that (1) induced high valence condition alone does not coincide with high debugging performance, (2) induced high arousal condition alone coincides with high debugging performance, and (3) induced high arousal and valence conditions together are associated with high debugging performance. This study called for more research on the topic.

Wrobel (2013) conducted a survey with 49 developers. The questionnaire assessed the participants' job-related affects during programming, and which emotions were perceived to be those influencing their productivity. The results showed that developers feel a broad range of affects while programming—all the affects of the measurement instrument's spectrum. Positive emotions dominate in their work. The five most frequently occurring emotional states were happy, content, enthusiastic, optimistic, and frustrated. That is, the four most experienced emotions were positive. Positive affects were perceived to be those enhancing their productivity. It is interesting to note that 13% of the developers indicated a positive or very positive impact on productivity when they were angry. However, the result was not significant overall. Finally, frustration was perceived as the negative affect more often felt, as well as the one perceived as deteriorating productivity.

Garcia *et al.* (2013) analyzed 10 years of data coming from the bug tracker and the mailing list of the Gentoo GNU/Linux community. More than 700000 textual messages were analyzed through sentimental analysis. The authors performed two studies. In the first study, they grouped the data based on the activity of a single person, namely Alice, who became a star of the community, and the key player of Gentoo's bug tracking management. One dataset was formed before her activity, one during her activity, and one right after she suddenly left the community. Garcia *et al.* (2013) defined performance as the time lapsed between the opening of a new bug and its closure. The results of the sentimental analysis showed that after Alice's departure, the negative sentiment were significantly higher with respect to the other two previous periods. The authors also showed that after Alice's departure, the Gentoo community never reached the levels of performance as during or before her involvement. Thus, a strong impact of affects on community turnover and performance was shown in the study.

Crawford *et al.* (2014) argued that affects are an important human factor that has to be considered when developing software. In the short paper, the authors first motivate the emotion research in software engineering, by reporting psychology literature on the impact of positive and negative affect on creativity and problem-solving of workers. They then reported PANAS and the Affect Grid as useful tools to measure emotions in software engineering, although emotion theory and the tools placement were not mentioned. Finally, the authors ask themselves what could be the impact of emotions on the quality of the software produced.

Muller and Fritz (2015) performed a study with 17 participants, 6 of which were professional software developers and 11 were PhD students in computer science. The participants were asked to perform two change tasks, one for retrieving StackOverflow scores and the other to let users undo more than one command in the JHotDraw program. During the development, the participants were observed using three biometric sensors, namely an eye tracker, an electroencephalogram, and a wearable wireless multi-sensor for physiological signals (e.g., heart rate, temperature, skin conductance). After watching a relaxing video, the participants worked on both tasks in a randomly assigned order. They were then interrupted after 5 minutes of working or when they showed strong signs of emotions. During each interruption, the participants rated their affects using a psychology measurement instrument. After other 30 minutes of work, the participants repeated the experiment design using the second task. Finally, the participants were interviewed. Overall, the study found that (1) developers feel a broad range of affects, expressed using the two dimensional measures of valence and arousal instead of labeling the affects, (2) the affects expressed as valence and arousal dimensions are correlated with the perceived progress in the task (evaluated using a 1-5 Likert scale), (3) the most important aspects that affect positive emotions and progress are the ability to locate and understand relevant code parts, and the mere act of writing code instead of doing nothing. On the other hand, most negative affects and stuck situations were raised by not having clear goals and by being distracted.

The literature review has shown that the relationship between affects and software development performance has been mostly unexplored before the start of this PhD's research activities, as only the studies by Lesiuk (2005); Khan *et al.* (2010) had been published. Furthermore, no studies have attempted to understand the evolution of the linkage between affect and performance of software developers. That is, how developers experience affect and their performance over time. Only the study by Muller and Fritz (2015) has covered this concern. However, Muller and Fritz (2015) mentioned that they built their study upon the present writer's work in Paper V; Paper VI, which was conducted two years before. In brief, the literature review has shown that there was the need to understand the static relationship between pre-existing affects and performance of software developers, the dynamic, real-time relationship of affects and performance of developers, and the process explaining the dynamic relationship.

### 2.5.3 Common misconceptions of affects [5]

While presenting our work to conferences, meetings, seminars, reviewing other authors' papers, and discussing with practitioners, we have found that there are some common misconceptions in literature regarding affects. Given the rising number of recent SE articles that deal with the affects of developers, e.g., Muller and Fritz (2015); Ford and Parnin (2015); Haaranen *et al.* (2015); Dewan (2015), we believe that it should be important for researchers to adopt a critical view of the phenomenon under study, and that they do not fall into the several misconceptions when dealing with the affect of developers (Paper III).

We understand that we have placed ourselves in a "very confused and confusing field of study" (Ortony *et al.* (1990), p. 2). We experienced this confusion especially during our talks at ISERN 2014, where we chaired a workshop called psychoempirical SE (Graziotin *et al.*, 2014d), and during the CHASE 2015 workshop Begel *et al.* (2015), where we presented some common misconceptions and measurements of the affect of software developers (Paper III). Such misconceptions include confusing affect and the related constructs of emotions and moods with motivation or job satisfaction, which has happened even in articles directly dealing with misconceptions of motivation with respect to job satisfaction França *et al.* (2014), although affects were not the focus of the study.

Other issues lie in missing the opportunity of using validated measurement instruments for affect. An example is the use of the niko-niko calendar for assessing the mood of a software development team, e.g. Sato *et al.* (2006), or the so-called happiness index, e.g., Medinilla (2014). None of them have been validated by psychology procedures.

---

[5] This subsection is built upon Paper III—Graziotin, D., Wang, X., and Abrahamsson, P. The Affect of Software Developers: Common Misconceptions and Measurements. In *2015 IEEE/ACM 8th International Workshop on Cooperative and Human Aspects of Software Engineering*, pp. 123–124. IEEE, Firenze, Italy (2015c). ISBN 9781467370318. doi: 10.1109/CHASE.2015.23.

Another example of a missed opportunity is when a unique truth is assumed. E.g., a CACM positional article has claimed that "psychologist recognize eight basic emotions, with each positive balanced by a negative", e.g. "love-hate" (Denning (2012), p. 34). On the other hand, a paper on an empirical study has been claimed that "there are six basic emotions or universal emotions: anger, happiness, fear, [...]" (Colomo-Palacios *et al.* (2013), p. 1079). We have previously shown that it is not true that a unique dominant, accepted theory exists for affect, emotions, and moods. On the other hand, it is important to explain why we adhere to any of the established or conflicting theories of affect. Furthermore, the authors of the previous example continued with "We removed 'disgust' and 'surprise' [...] because their translation did not denote [...] everyday emotions." (ibid, p. 1079). The statement introduces two flaws in the study. First, the authors removed two items over six in a validated measurement instrument. Second, the authors translated the measurement instrument without conducting a psychometric study to show if the reliability and validity of the translated measurement instrument would still hold. Researchers should recognize these issues when employing such delicate concepts and measurement instruments for constructing research.

Besides that they are often not assessed by following validated theories and measurements instruments, affects are often confused with job satisfaction, motivation, and commitment. While all these concepts are important and have been subject of important research even in software engineering, they are not affects, although affects play a role and might be part of them. We present here some of these constructs as misinterpreted by software engineering researchers.

The first is *job satisfaction*, which is often confused with affects in the workplace. Perhaps, this misconception might be born with the seminal work by Locke (1969), who defined job satisfaction as "*the pleasurable emotional state resulting from the appraisal of one's job as achieving or facilitating the achievement of one's job values.*" (p. 316). Locke considered job satisfaction and dissatisfaction as emotions Locke (1970) and this assumption guided organizational research for more than 30 years. However, since mid '90s it has been established that this is not the case. Job satisfaction is an attitude, not affect Brief (1998). An attitude is an evaluative judgment made with regard to an attitudinal object, in this case one's job Weiss (2002). Still, many current definitions of job satisfaction "*have obscured the differences among three related but distinct constructs: evaluations of jobs, beliefs about jobs, and affective experiences on jobs.*"(Weiss (2002), p. 173). There has been much inconsistency in the literature when considering satisfaction as affect or as attitude (Weiss 2002). More precisely, job satisfaction is "*a positive (or negative) evaluative judgment one makes about one's job or job situation.*" (Weiss (2002), p. 175). However, affects of individuals are related to job satisfaction. The Affective Events Theory (Weiss and Cropanzano, 1996) , as written also for explaining the linkage between an individual's internal processes—including emotions and moods—and job satisfaction. The most basic assumption of Affective Events Theory is that job satisfaction should be conceptualized as an evaluative judgment about one's job Wegge *et al.* (2006). In particular, Affective Events Theory proposes that positive and negative emotional incidents at work have a significant impact upon a worker's job satisfaction. The theory demonstrates that employees react emotionally to anything happening to them at work. These emotional reactions of the individuals influence their job performance and their job satisfaction Hume (2008). This has been empirically demonstrated by several studies, for example the one by Ilies and Judge (2002), where it was found that mood positively influences job satisfaction.

Affects are not *motivation*, either. The Oxford English Dictionary defines motivation as "*The general desire or willingness of someone to do something; drive, enthusiasm.*" (OED, 2002). Mitchell (1982) reported that several theories have existed for motivation. However, all theories deal with the individuality of people, and perhaps the multitude of theories is a demonstration of the individuality of subjects. Mitchell (1982) defined motivation as "*those psychological processes that cause the arousal, direction, and persistence of voluntary actions that are goal directed.*". This definition already suggests that motivation is not an affect, however the two constructs appear to be related. Indeed, Weiner (1985) suggested that affects are motivational catalysts and influence subsequent behavior. Arnold (1981) reminded us that valence has been the base of several motivation theories. In particular, according to the expectancy-valence models (Vroom, 1964), the force with which an individual engages in an activity is a function of the sum of the valence of the outcomes and expectations that the activity will lead to the attainment of those outcomes, and that humans will choose to perform the activity having the strongest positive valence or the weakest negative valence. Much research has extended and confirmed the model, including Arnold (1981) study. According to Seifert (2004), when presented with a task, individuals perform evaluative judgments about the task itself, and they respond affectively based upon task and personal characteristics. These generated affects dictate successive motivation towards the task. The theoretical work by Seifert (2004) suggested that motivation is a series of patterns of behavior and affect. In particular, while other factors and affects influence motivation, the feelings of competence and control (i.e., dominance) are suggested to be the strongest drivers. In brief, the literature suggests that motivation is a multifaceted construct, which might have affects as part of its components (Mitchell, 1982) and as being influenced by affects.

*Commitment* as a construct is not extraneous to the software process improvement literature, e.g., (Abrahamsson, 2001). Commitment has been defined as a psychological state of attachment that defines the relationship between a person and an entity (e.g., an organization) (O'Reilly & Chatman, 1986) Similarly to motivation, commitment is multifaceted. According to Meyer and Allen (1991), commitment is conceptualized in the forms of affective, normative and continuance commitment. While normative and continuance commitment deals with perceived moral obligations and the awareness of the costs associated with leaving the organization respectively, affective commitment refers to an employee's attachment to, identification with, and involvement within an entity, e.g., an organization, a project, or a team. When employees are affectively committed to an organization, they identify with the organization itself and the brand. Employees develop a sense of belonging to the firm and its vision, which intensifies their involvement in the activities of organization, their inclination to pursue the firm's goals as if these goals were personal, and their wish to keep staying with the organization (Meyer and Allen, 1991).Affective commitment is fostered by emotionally satisfying experiences in the context of working, which may lead employees to identify the organization's well-being with their own, making them affectively bound to the organization (Rhoades *et al.*, 2001). Affective commitment can be easily observed in small teams of software development projects. Software engineers become emotionally attached to a product they are developing or under use, as documented by Wastell and Newman (1993).

To summarize, this section has briefly shown how important constructs and concerns such as job satisfaction, motivation, and commitment are not to be confused with affects. Although they are important topics of research in several disciplines including psychology, their distinction with affects make affects an even more important and fascinating subject to be studied. Affects are at the base of the most important human-based organizational constructs; therefore, they need to be understood deeply in any organizational settings, including software development.

Let us conclude this section with two brief elucidations. We often employ the term *happy* in the title of our publications. Happy refers to *happiness*, which is another interesting topic of research and has triggered some attention to our work. Vigilant researchers have warned us regarding the possible pitfalls tied to affects, happiness, and general well-being (e.g., (WIlkstrand *et al.*, 2014)). Similar to job-satisfaction, *well-being* is an attitude (Guttman and Levy, 1982). Therefore, subjective well-being consists of two interrelated components, which are life satisfaction and positive and negative affects, where life satisfaction refers to a cognitive sense of satisfaction with life (Diener, 1984). That is, like job-satisfaction, subjective well-being can be considered as one's self-evaluation of life, which is influenced by affects. Happiness, on the other hand, is a complex construct, which has different psychological and philosophical definitions.

Under an Aristotelian view, happiness is *eudaimonia*, (Haybron, 2005). A person is happy because (s)he conducts a satisfactory life full of quality. On the other hand, a blend of affects constitutes an individual's happiness under the hedonistic view of happiness (Haybron, 2001). According to this view, being happy coincides with the frequent experience of pleasure; that is, happiness reduces to a sequence of experiential episodes (Haybron, 2001). Frequent positive episodes lead to feeling frequent positive affects, and frequent positive affects lead to a positive *affect balance* (Diener *et al.*, 2009a). Similar to this stance are Lyubomirsky *et al.* (2005), who consider a person happy if the person's affect balance Diener *et al.* (2009a) is mainly positive (that is, characterized by frequent central positive affects). That is, happy people are those who often experience positive affects, and are in a positive condition of affects (affect balance).

As argued by the philosopher Haybron (2001), a quantitative view of happiness based solely on frequency of affects is psychologically superficial because some affects do not have distinct episodes or attributions (as in moods). Even more, Haybron (2005) has seen happiness as a matter of a person's affective condition where only *central affects* are concerned—for example, the pleasure of eating a cracker is not enduring and probably not affecting happiness; therefore, it is considered as a peripheral affect.

# Chapter 3

# Empirical research design

This chapter reports the strategy followed for executing this PhD's research activities. The chapter starts by briefly discussing what constitutes a research activity and the scientific method. Section 3.1 summarizes the different views on ontology, epistemology, and worldviews when conducting research activities, and it attempts to define the present author's worldview.

Section 3.2 describes the approaches for research, namely quantitative, qualitative, and mixed research. Section 3.3 summarizes the existing research approaches and the most common research strategies in software engineering research. Section 3.4 introduces and defines research methods as understood by the present author when writing this dissertation. After the general explanations about conducting research, Section 3.5 explains the approach adopted in this PhD's research activities, namely the research phases, the adopted research strategies, and the chosen data collection methodologies.

During this PhD cycle, the present candidate and his colleagues undertook several research activities. Research is commonly understood as an investigation toward a contribution of knowledge about phenomena (OED Online, 2015d), whereas a scientific research employs the scientific method as a body of techniques for investigating phenomena and producing knowledge. The present author is aware that much disagreement still occurs regarding what counts as science and what science is not (Godfrey-Smith, 2003), e.g. (Kuhn, 1970; Lakatos, 1977; Feyerabend, 1993), and it was not the intention of this thesis to enter the diatribe.

The general "textbook-like" definition of the *hypothetico-deductive method* reported by Godfrey-Smith (2003) fits fairly well a general idea of how science is conducted, which, according to the author, covers "a method of doing science and for a more abstract view about confirmation" (p. 236). According to Godfrey-Smith (2003), scientific methods comprise iterative phases of (1) observation gathering, (2) hypotheses formulation, (3) deduction of predictions from the hypotheses, and (4) verification of the prediction. Some views omit point 1, or consider a review of the existing literature and logical deduction as point 1. One should be well aware, however, that this view reduces the scientific method as the typical hypothesis testing approach that is usually taught in school (Gauch, 2003). Also, many prominent opponents, Feyerabend (1993) among them, have argued that the view is not really representative of how science is actually practiced.

Creswell (2009) has defined a framework for research, which was adopted for this PhD's research activities, and it is represented in Figure 3.1. The framework for research consists in an interconnection of the components of ontology and epistemology (Section 3.1, called worldviews by Creswell (2009)), approaches (Section 3.2), strategies (Section 3.3, called designs by Creswell (2009)), and methods. The successive four sections describe the components of the framework.

## 3.1 Ontology, epistemology, and worldviews

Perhaps more important than discussing what constitutes a scientific contribution or how a contribution is received by the scientific community is a discussion about what we accept as scientific knowledge, in terms of ontology and epistemology (Chalmers, 1999). According to Easterbrook *et al.* (2008), "philosophers make a distinction between epistemology (the nature of human knowledge, and how we obtain it) and ontology (the nature of the world irrespective of our attempts to understand it)" (p. 290). Understanding how we see the world is important because the "preference of each research methodology depends on philosophical issues related to the question of ontology and epistemology" (Tuli (2011), p. 99).

It is desirable to simplify the discussion of ontology and epistemology as *philosophical worldviews* (or paradigms), in line with Creswell (2009); Corbin and Strauss (2008) who have interchanged the terms or employed worldviews as a categorization of ontological and epistemological questions. Even more, Guba and Lincoln (1994) have defined worldviews as basic belief systems based on ontological, epistemological, and methodological assumptions.

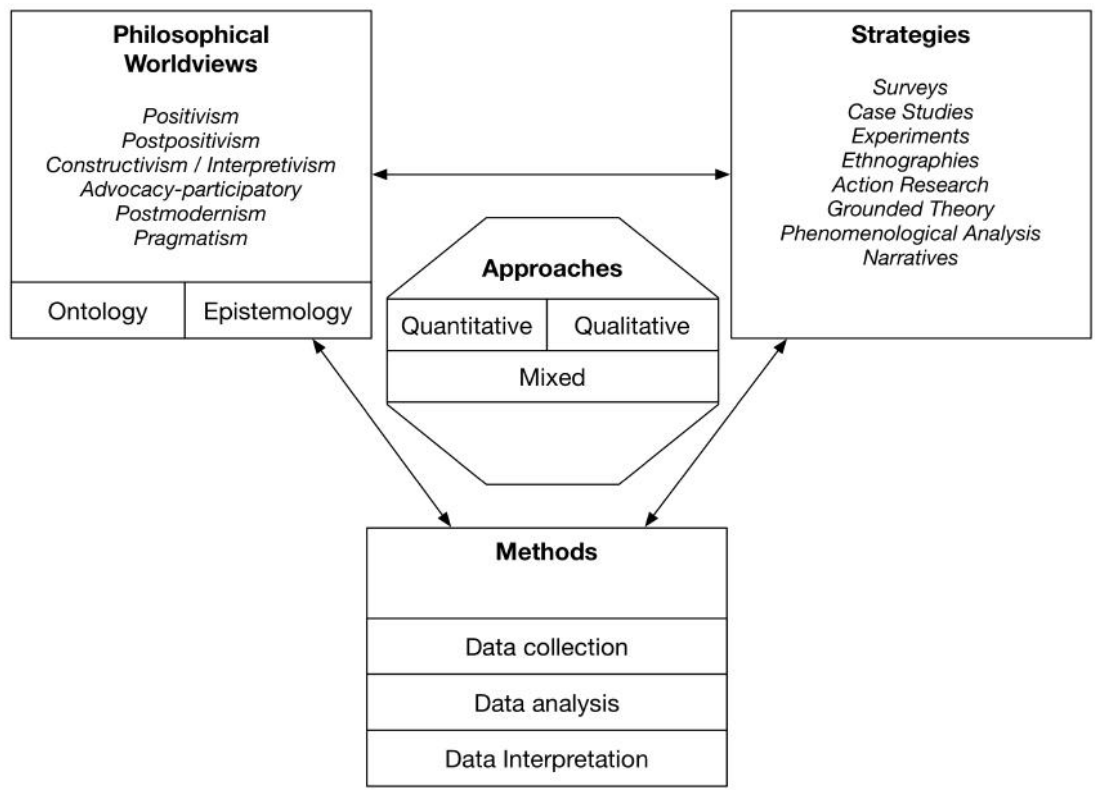

FIGURE 3.1: *Framework for research (from (Creswell, 2009))*

Worldviews are basic set of beliefs that guide action" ((Guba, 1990), p. 17) and describe reality ((Koltko-Rivera, 2004), p. 4). Human cognition and behavior are powerfully influenced by sets of beliefs and assumptions about life and reality (Koltko-Rivera, 2004). It is important to understand the worldviews of those who conduct research, because worldviews are orientation about the world, the way knowledge is acquired and represented and how truth are seen (Creswell, 2009). Several worldviews have been located over the years in the philosophy disciplines. According to Creswell (2009), the major worldviews are postpositivism, constructivism, pragmatism, and advocacy/participatory. According to Blaxter *et al.* (2010), the most prominent worldviews are positivism, postpositivism, interpretivism, critical paradigms, and postmodernism.

Positivism is the view that all sciences should mirror those of the natural science, thus making the researcher objective and detached from the object of research (Blaxter *et al.*, 2010). Positivism rejected Hegel's idealism and has had its roots on empiricism. Positivists have also adopted logical reasoning and a general theory of language (Godfrey-Smith, 2003) for their theory of knowledge, and they have focused on clarifying how a sentence could be stated in a meaningful way (Bendassolli, 2013). Positivism holds that the goal of knowledge is experienced phenomena, and science has to stick to what we can observe and measure, and nothing beyond that (Trochim and Donnelly, 2007). Of particular importance for positivists has been the objectivity of science as a rational surest activity to find out unique truths of the world (Okasha, 2002). Positivism is reductionist in the sense that ideas are reduced into a small set of variables and testable relationships that comprise hypotheses and research questions (Creswell, 2009). Positivists have believed that the world is governed by laws and theories, which need to be tested out, quantitatively most of the times, for finding them (Creswell, 2009). The world, according to positivists, is deterministic, controllable and predictable (Trochim and Donnelly, 2007). Positivism ontology is defined as naïve reality, as truth is out there and the objectivity of the researcher is required for finding the truth. The positivist epistemology is objectivitist, and the findings are considered to be "true" (Guba and Lincoln, 2005).

The postpositivist worldview has been the most pervasive in research, especially for quantitative research, to the point that it has been called *the scientific method* (Creswell, 2009). Postpositivism has had its roots in (logical) positivism. Postpositivism was born as a response to criticisms made to positivism (Blaxter *et al.*, 2010). Postpositivism takes into account the fallacy of the human being trying to understand the world. In this sense, postpositivists are critical realists (Trochim and Donnelly, 2007), who still believe that there is a unique reality that is independent from our ways of understanding it. However, objectivity is not achievable because we are biased when we try to understand the world through conjectures (Robson, 2011). Thus, we can only achieve knowledge of truths probabilistically (Blaxter *et al.*, 2010). Yet, this fallacy is not an issue as long as bias is discussed and critical discussions happen (Trochim and Donnelly, 2007; Creswell, 2009). The ontology of postpositivism is not purely realist, as an objective truth does exist but it can only be approximated (Robson, 2011). The ontology of postpositivism is critical realism, as a "real" reality does exist but it can only be apprehended imperfectly and probabilistically (Guba and Lincoln, 2005). The epistemology of postpositivism is similar to the one of positivism. However, the findings can aspire to become only *probably* true.

Social constructivism, which has often been united with the concept of interpretivism (Easterbrook *et al.*, 2008; Klein and Myers, 1999a), has been defined succinctly by Geertz (1973) as "really our own constructions of other people's constructions of what they and their compatriots are up to" (p. 9). Constructivists assume that individuals seek understanding of the world they live in, and they develop subjective meanings of their experience toward object, things, and phenomena (Creswell, 2009). Researchers are individuals, as well. Therefore, the construction of knowledge is always social and built through the interaction of humans and the sharing of constructed realities. Social constructivism is associated with qualitative inquiry (Creswell, 2009), but it is not limited to qualitative studies. Interpretivism is now established in information systems research (Walsham, 2006), but we see it still emerging in software engineering research. The constructivism epistemology is considered as embracing relativism, as realities are local to the phenomena under study and co-constructed (Guba and Lincoln, 2005). Consequently, constructivism has subjectivism as epistemological basis (Guba and Lincoln, 2005).

The advocacy-and-participatory worldview, which has often been called critical theory (Easterbrook *et al.*, 2008) or critical inquiry (Blaxter *et al.*, 2010), arose as a way to address how postpositivism and social constructivism could not fit marginalized individuals and issues of social justice (Creswell, 2009). The advocacy-and-participatory worldview holds that research inquiry needs to be intertwined with politics and political agenda, and it contains an action agenda through intervention for reform that may change the lives of the participants (Petersen and Gencel, 2013). The research activities often begin with an important social issue, such as empowerment, inequality, oppression, domination, suppression, and alienation, and the participants may help to shape the research questions (Creswell, 2009).

Postmodernism advocates have argued that "the era of big narratives and theories is over: locally, temporally and situationally limited narratives are now required" (Flick (2009), p. 12). Postmodernism embraces relativism in its extreme, to the point that any description of the world is an illusion (Godfrey-Smith, 2003). Postmodernism rejects social conventions and focus on deconstruction (Marvasti, 2010). Central to postmodernism is its rejection of the concept of truth. There is no such thing as true reality out there to discover, and what we accept as truth is merely community-based (Uduigwomen, 2005). Finally, postmodernism holds that there is a continual change of perspectives, without any underlying common frame of reference. Reality at once is multiple, local, temporal and without demonstrable foundation (Uduigwomen, 2005).

Pragmatism is based on actions, situations, and consequences rather than antecedent conditions as in postpositivism (Creswell, 2009). Pragmatists acknowledge that all knowledge is approximate and incomplete; therefore the value of knowledge is dependent on the methods by which it was obtained (Menand, 1997). That is, knowledge is judged by how useful it is at solving problems, and truth is whatever works at the time (Easterbrook *et al.*, 2008). Researchers embracing pragmatism will emphasize the research problem and will employ any possible method for understanding it. Finally, pragmatism is not committed to any particular system of philosophy or reality, and has stopped asking questions about reality (Creswell, 2009).

During these years of PhD research activities, the present author has struggled to understand what truly is his worldview. He has received a strong empirical, postpositivist education during the MSc studies, whereas no alternative had ever been offered. Therefore, the present candidate was *trained* postpositivist. However, he has never blindly believed that an objective, single truth does exist. Perhaps, we do construct our truths of the world that surrounds us. Therefore, the ontology of this candidate has been laid on a continuum between critical realism and relativism. That would make him a social constructivist with a background in postpositivism, unrealistically. Yet, the present candidate has agreed with DeWitt (2010) that worldviews and beliefs and what we take as facts lie somewhere on ontological and epistemological continua. This PhD study has been exploratory and explanatory in its nature. As the *what* and the *why/how* of the topics under investigation were still not understood, both quantitative and qualitative studies have been conducted. This research and this author's worldview conform to mixed method research and, consequently, if we were forced to declare a single, discrete worldview, pragmatism would be the safest choice for the present author (Creswell, 2009).

## 3.2 Research approaches

Research approaches are the high-level plans and procedures for research activities, which reflect the researchers' philosophical assumptions, designs, and methods for data collection and analysis (Creswell, 2009).

At the highest level of research, two main research approaches have been identified, that are quantitative and qualitative (Creswell, 2009). Blaxter *et al.* (2010) has warned us that such a distinction has often been presented for forming competing alternatives that spread up a political and contested nature of knowledge construction. Indeed, Creswell (2009) appears to fall into this limitation by stating that quantitative strategies have invoked the postpositivist worldview, while qualitative strategies correspond to social constructivism, advocacy-and-participatory worldview, and mixed methods strategies correspond to the pragmatism worldview. Perhaps along the same line of Blaxter *et al.* (2010) was Punch (2005), who stated that "quantitative research is empirical research where the data are in the form of numbers. Qualitative research is empirical research where the data are not in the form of numbers." (p. 3). Less polemically, Easterbrook *et al.* (2008) have stated "In quantitative research methods, the theoretical lens is used explicitly to decide which variables to isolate and measure, and which to ignore or exclude. In qualitative methods, the theoretical lens is often applied after data is collected, to focus the process of labeling and categorizing ("coding") the data." (p. 293). Finally, Blaxter *et al.* (2010) has attempted to clarify quantitative and qualitative research by means of comparison, as Figure 3.2 shows.

Selecting a research approach is influenced by several factors including our worldviews, and it is important to understand that no single approach is ideal. Any selection of research approach (and strategy) involves loss and gain as well (Tuli, 2011).

The mixed method approach recognizes that all methods have their limitations, thus it encourages triangulation through multiple data collection and analysis (Creswell, 2009). Mixed methods approaches inquiry by combining qualitative and quantitative forms of research in different degrees (Creswell, 2009).

| The differences between qualitative and quantitative research | |
| --- | --- |
| **Qualitative paradigms** | **Quantitative paradigms** |
| • Concerned with understanding behaviour from actors' own frames of reference | • Seeks the facts/causes of social phenomena |
| • Naturalistic and uncontrolled observation | • Obtrusive and controlled measurement |
| • Subjective | • Objective |
| • Close to the data: the 'insider' perspective | • Removed from the data: the 'outsider' perspective |
| • Grounded, discovery oriented, exploratory, expansionist, descriptive, inductive | • Ungrounded, verification oriented, reductionist, hypothetico-deductive |
| • Process oriented | • Outcome oriented |
| • Valid: real, rich, deep data | • Reliable: hard and replicable data |
| • Ungeneralizable: single case studies | • Generalizable: multiple case studies |
| • Holistic | • Particularistic |
| • Assumes a dynamic reality | • Assumes a stable reality |

**The similarities between qualitative and quantitative research**

- While quantitative research may be used mostly for testing theory, it can also be used for exploring an area and generating hypotheses and theory.
- Similarly, qualitative research can be used for testing hypotheses and theories, even though it is used mostly for theory generation.
- Qualitative data often includes quantification (e.g. statements such as more than, less than, most, as well as specific numbers).
- Quantitative approaches (e.g. large-scale surveys) can collect qualitative (non-numeric) data through open-ended questions.
- The underlying philosophical positions are not necessarily as distinct as the stereotypes suggest.

FIGURE 3.2: *Differences and similarities between quantitative and qualitative research (from (Blaxter et al., 2010), p. 66)*

## 3.3 Research strategies

Research strategies (also called research designs (Creswell, 2009)) are types of inquiry that fall within the approach categories of qualitative, quantitative, and mixed methods (Creswell, 2009). However, it is reasonable that qualitative and quantitative strategies are intertwined within a single major strategy.

Several research strategies exist. According to Wohlin *et al.* (2012), all studies in empirical software engineering fall into the three families of surveys, case studies, and experiments. Easterbrook *et al.* (2008) have added to the previous families those of ethnographies and action research. Furthermore, Creswell (2009) has included more qualitative strategies, namely grounded theory, phenomenological research, and narrative research. Some authors, e.g., Creswell (2009), have categorized the research strategies in terms of research approach. The present author has agreed that certain research strategies are better suited for certain research approaches, e.g., controlled experiments with quantitative research. However, research strategies can (and should) take advantage of both quantitative and qualitative approaches. Therefore, this dissertation will not categorize the research strategies according to any predefined research approach.

Surveys are a system for collecting information from or about people to describe, compare, or explain their knowledge, attitudes, and behavior ((Fink, 2003) as cited by Wohlin *et al.* (2012)). Surveys enable a broad understanding of the factors under study. In particular, the questionnaires have been used to enable the collection and the analysis of data in a standardized manner which, when gathered from a representative sample of a defined population, allows the inference of results to the population (Rattray and Jones, 2007). Surveys are administered to samples of a population to be queried for understanding a phenomena. According to Wohlin *et al.* (2012), surveys are employed to derive descriptive and explanatory conclusions. However, the present author is of the opinion that surveys, in the form of open-ended questions, are useful for exploratory purposes, as well.

Case studies, as defined by Thomas (2011), "are analyses of persons, events, decisions, periods, projects, policies, institutions, or other systems that are studied holistically by one or more methods. The case that is the subject of the inquiry will be an instance of a class of phenomena that provides an analytical frame–an object–within which the study is conducted and which the case illuminates and explicates" (p. 513). During a case study, researchers collect data using multiple data collection procedures over a sustained period of time (Creswell, 2009).

Experiments are inquiries that attempt to understand if a specific treatment—or characteristic—influences an outcome (Creswell, 2009). Wohlin *et al.* (2012) differentiate between controlled experiments and quasi-experiments. Controlled experiments are those where treatments are applied to or by different subjects while keeping other variables constant and measuring the effects on outcome variables. Treatments are administered (or not) to experimental groups according to different strategies. However, an important characteristic of controlled experiments is the randomization of the participants and/or the treatments. Quasi-experiments, on the other hand, are those experiments where the assignment of treatments cannot be based on randomization or when it emerges from the characteristics of the participants. The latter case is sometimes called natural experiment.

Ethnographies are strategies of inquiry where researchers study, by observing and interviewing, a cultural group in a natural setting over a prolonged period of time (Creswell, 2009). Researchers are immersed in the business of those being observed, and they perform conversations, attend meetings, read documents, and perform as much observational activities as possible (Sharp *et al.*, 2000).

Action research is characterized by an active attempt to solve a real-world problem while studying the experience of solving the problem (Davison *et al.*, 2004). Important aspects for action research are that the problem to be solved has to be authentic—that is, the problem has to be real and important to the point that its solving would have a real impact to the organization—and that the participants will benefit from authentic knowledge outcomes (Easterbrook *et al.*, 2008). Although several versions of action research have been developed, the essence of it is based upon iterations of problem identification, action planning, implementation, evaluation, and reflection Zuber-Skenitt (1993). The reflection phase inputs to the next iteration.

Grounded theory is a systematic methodology in social sciences involving the discovery of the theory through the analysis of qualitative but also quantitative data (Martin and Turner, 1986). Grounded theory is indicated to study human behavior in an iterative, explicit and systematic process (Easterbrook *et al.*, 2008). With grounded theory, researchers derive a general theory of phenomena (process, action, interaction, ..), which is grounded in the views of participants (Creswell, 2009). There are two main approaches for grounded theory, Glaser and Strauss (1967) approach and Corbin and Strauss (2008) approach, although the difference between the two approaches has been defined as a "*rhetorical wrestle*" (Heath and Cowley, 2004). What characterize grounded theory studies is a systematic categorization of chunks of data using coding techniques, a continuous comparison of data and its categorization, and a theoretical sampling strategy of participants (Creswell, 2009).

(Interpretive) Phenomenological research is an inquiry strategy where the researchers identify the essence of human experiences about a certain phenomenon (Creswell, 2009). The phenomenon is described by the participants, and the researchers should set aside their own experiences when describing the phenomenon. Phenomenological research combines psychological, interpretive, and idiographic components (Gill, 2014).

Narrative research is a strategy where the researchers ask the participants to provide stories about their lives and their experiencing of phenomena. The information is then re-told by the researchers into a narrative chronology, thus creating a shared view of experiences (Creswell, 2009).

## 3.4 Research methods

Research methods are the most concrete form in the research framework proposed by Creswell (2009). Research methods involve the actual forms of data collection, analysis, and interpretation that researchers propose for their studies (Creswell, 2009). Research methods are bounded within a study, and are usually what is described in a paper under the methodology section. A discussion regarding a research method will include several components, among which we are likely to find a description about the research participants, the required materials for running the research, the study procedure, and the required measures (if any) (Creswell, 2009). Other components included in the research method discussion depend on the research approach, strategy, and of course, the philosophical worldview. For example, the established guidelines for running experiments in software engineering by Wohlin *et al.* (2012) and by Jedlitschka and Ciolkowski (2008) advise to report the expected threats to validity in the discussion of the research method.

## 3.5 Research phases, strategies, and methods of this dissertation

The research reported in this dissertation was systematically organized into three main phases, namely Phase I—*Knowledge acquisition and theory translation*, Phase II—*Variance theory testing*, and Phase III—*Process theory development*. These three phases are the result of a posteriori sense-making process. In the present author's opinion, they help to explain to the reader how the work has been conducted from the ground up, when psychology research was consulted, to its end, when a theory of affect and performance in software engineering was produced. This section describes the research approach in terms of the three phases, connects the research questions to the phases, explains the chosen research strategies and the rationale behind the choices, and reports the data collection methods of each study.

Table 3.1 summarizes the research phases, and links the research questions with the conducted studies.

Table 3.1: *Research phases, research questions, and related papers*

| Research question | Research phase | Related papers |
|---|---|---|
| RQ1 | Phase I | Paper III; Paper II; Paper I |
| RQ2 | Phase II | Paper IV |
| RQ3 | Phase II | Paper V; Paper VI |
| RQ4 | Phase III | Paper VII |

### 3.5.1 Phase I—Knowledge acquisition and theory translation

The first phase of the research activities was an attempt to understand what are affects, emotions, moods, and performance in the context of software development (RQ1). In order to understand these complex constructs and their relationships, an extensive literature review in the domains of software engineering, information systems, psychology, and organizational behavior was performed. The review showed that the software engineering literature has neglected affects in software processes and methodologies, as well as in general research. Psychology, on the other hand, has offered several models, theories, and measurement systems, to the point that sense-making had been required in order to choose proper devices. This phase was characterized by a comprehensive literature review, which started with the input from researchers in human-computer interaction and trained psychologists to find first seminal articles. Then, the research was carried out using various academic search engines and by employing the snowballing technique. Please refer to Chapter 2 for further details.

The studies reported in Paper I; Paper III; Paper II, which according to Table 3.1 provide answers to RQ1, were literature reviews for the most part. A small exception is present in Paper I, where an inductive qualitative analysis of the practitioners' comments to one of our published articles was conducted, in order to understand how much the practitioners are interested in studies exploring their affects.

Section 7.1 summarizes the contribution of this phase.

### 3.5.2 Phase II—Variance theory testing

Based on the translation of the psychology theory to software engineering, the research questions *How do affects indicate problem-solving performance among developers?* and *How are affects correlated to in-foci task productivity of software developers?* arose. The majority of the discovered work relied on variance theory. Variance theories provide explanations for phenomena in terms of relationships among dependent and independent variables (Langley, 1999; Mohr, 1982). In variance theory, the precursor is both a necessary and sufficient condition to explain an outcome, and the time ordering among the independent variables is immaterial (Pfeffer, 1983; Mohr, 1982).

Thus, the second phase of this PhD's research activities was of variance theory development and testing. Given the background in quantitative methods, the chosen strategies for understanding the linkage between affects and performance while developing software were experiments, because they are the most immediate strategy for quantitative assessment of the relationship between constructs (Wohlin *et al.*, 2012). However, quasi- and natural experiments have been preferred to controlled experiments, as explained in the following subsection. Natural experiments are quasi-experiments where no or limited manipulation is performed. These experiments can happen in laboratory settings as well as in natural settings. Natural experiments do not manipulate any variable and condition, so they have the advantage to study the participants in their natural conditions.

The studies reported in Paper V; Paper VI; Paper IV, which according to Table 3.1 provide answers to RQ2 and RQ3, were experiments and resulted in variance theory.

The study reported in Paper IV was quantitative. It was a quasi-experiment with 42 participants. The participants, who were all computer science students, performed two psychology-validated tasks that can be generalized to software development. The tasks were designed to allow measuring the performance of the participants in terms of creativity and analytic problem-solving. The study compared the performance achieved in these tasks with the pre-existing affects of the participants. A validated measurement instrument (a questionnaire) was employed to measure the pre-existing affects of the participants. The affect measurement happened before each task. The participants were split into two groups according to a median split of the affects score. The two groups were compared according to the scores obtained in the two tasks. The experiment was conducted in a controlled environment. However, no treatment was assigned to the participants, which would be mood induction. For this reason, the study could not be classified as a controlled experiment. A complete description of the study can be found in Chapter 4. The study in Paper IV was reviewed by editor and reviewers from psychology research. We decided to submit our article to a psychology venue as a way to assess how much we were pointing into the right direction regarding our understanding of the psychology related knowledge.

The study reported in Paper V; Paper VI was quantitative with few qualitative elements. The study was a natural experiment with repeated measurements within-participant. The design was conducted with eight participants (four students and four professionals), conveniently sampled and studied individually over 90 min. of programming. Each ten minutes, the participants assessed the affects raised by the development task itself, using a validated measurement instrument from psychology, and self-assessed their productivity. The study design was novel for software engineering and, to our knowledge, novel for psychology research. The design was inspired by the Experience sampling method (Larson and Csikszentmihalyi, 1983). Experience sampling method is a repeated measurement method in which the participants are studied in natural settings, as they perform their daily activities. At certain time intervals, which are sometimes random and sometimes fixed according to the design choices, the participants fill out a questionnaire. The time-based correlates are then studied. In the case of our study, the participants were developing software for their own projects, being the project work-related or related to university courses. The participants were interviewed before and after their task. The pre-task interview was conducted mostly for gathering demographic alike information. The post-task interview was employed as data triangulation and for a broader understanding. The data points were converted as z-scores for the reasons described in Section 2.1.3. A complete description of the study can be found in Chapter 5.

**Issues with affects and controlled experiments**

There are two major reasons why natural experiments have been preferred instead of controlled experiments. Firstly, controlled experiments dealing with affects in psychology have a mood-induction phase. That is, the enrolled participants, right before performing the experiment task, face an affect induction procedure according to the experimental grouping (Westermann and Spies, 1996). Examples include imagining and re-experiencing emotion-ridden events, watching, reading and listening to narratives or descriptive materials, listening to music, and so on (Lewis *et al.*, 2011; Westermann and Spies, 1996). One could imagine that, assuming the success of these procedures, the effect would be small and perhaps too artificial with respect to how individuals feel naturally. Indeed, pre-existing affective states of participants mediate the mood induction techniques (Forgeard, 2011) to the point that mood induction techniques have been lately criticized. Finally, a meta analysis has found that the effect of mood induction techniques is significant but rather small (Westermann and Spies, 1996).

Secondly, mood induction techniques for a prolonged period (days, weeks, months before the running of the experiment) might be more effective but pose serious ethical questions. How to deal with inducing negative affects in the long run to subjects? How to cope with possibly depressed participants who are in the negative group? Perhaps, this is what has refrained ethical research in psychology to perform effective long-term mood-induction techniques. A recent example of questioning the ethics of long-term mood induction techniques has been the Facebook emotion contagion experiment (Kramer *et al.*, 2014), where 700.000 Facebook users unknowingly had their feeds tweaked to show mainly posts either with negative feelings or positive feelings for a week. In other words, their virtual worlds have been surrounded by negative or positive affects for a long week. The study has been criticized all around the world, by researchers in all fields. Unfortunately, the existing guidelines in ethical research in software engineering (Vinson and Singer, 2008) do not deal with these aspects, as the present author has shown recently (Graziotin, 2014).

Therefore, natural experiments have seemed better suited and ethical, so they have been chosen for this PhD research. Two natural experiments have been performed, which resulted in one conference article and two journal articles.

Section 7.2 summarizes the contribution of this phase.

### 3.5.3 Phase III—Process theory development

The results of the second phase of this PhD's research activities have offered a general understanding of the relationship between affects and the performance of software developers. Variance models excel in quantifying relationship; however, they lack in terms of explaining the process behind the relationship. For this reason, the third phase of this PhD's research activities has focused on process based theory development.

Originating from evolutionary ecology, process based research development evolved in the field of management as a quest to explain how and why organizations change (Van De Ven and Poole, 1995). Process based theory development was then generalized as those activities concerned with understanding *how* things evolve over time and *why* they evolve in the way we observe (Langley, 1999). A process theory is the conceptualization of the progression of events and actions related to some entities' existence over time (Van De Ven and Poole, 1995). Examples of such events and actions include decision-making techniques, work flows, communication paths, and so on. According to Langley (1999), process data consist mainly of "stories"—which are implemented using several different strategies—about what happened during observation of events, activities, choice, and people performing them, over time.

Mohr (1982) has contrasted process theory to variance theory by stating that the basis of explanation of things is a probabilistic rearrangement instead of clear causality, and the precursor in process theory is only a necessary condition for the outcome. In addition, in process based-theory we deal with discrete states and events rather than continuous variables ((Mohr, 1982; Pfeffer, 1983)).

Paper VII was executed through an interpretive, qualitative study in natural settings. Two participants were actively observed and interviewed during a development cycle lasting more than one month. The participants were indirectly asked to describe the development performance through their affect, starting with the question "How do you feel?". The design of Paper VII corresponds to an interpretive phenomenological research. A complete description of the study can be found in Chapter 6.

Section 7.3 summarizes the contribution of this phase.

At this point, the reader might be left wondering about the reasons of our choice to start with variance based theory development (thus, hypothesis-driven theory development and testing) and to turn to process theory afterwards. If any, the common approach would be the opposite, to develop a theory first, then test it. The reason for such strategy was dictated by the choice of acquiring knowledge and theory from psychology first, then to increasingly translate the acquired knowledge to software engineering. Figure 3.3 illustrates the process over time. The PhD research activities began by eliciting the theoretical foundations from psychology. Subsequently, we put into research practice the acquired knowledge. Indeed the study in Paper IV has often been discussed by the reviewers as being "not enough software engineering". The studies as they are presented in this dissertation are "less psychology" and "more software engineering" (with psychology theory) in terms of methodology and design as they appear. One direct consequence of this has been the initial development of the theory using variance based approaches. The reason is that psychology disciplines have been dominated by postpositivism until recently (Trochim and Donnelly, 2007; Michell, 2003). Therefore, quantitative methods have been preferred to qualitative methods, and experiments guided by statistical hypothesis testing have been the norm (Hubbard *et al.*, 1997), to the point of becoming integral part of basic psychology research (Wilkinson, 1999)[1].

The pervasiveness of postpositivist, quantitative oriented research toward variance theory, which was also found by the presented author while reviewing the literature, brought us to initially apply the acquired knowledge and theory as it is usually applied in psychology. However, over time, our gained knowledge and confidence let us conduct more software engineering-oriented study, also toward process theory.

---

[1] The present author is also aware that the situation is about to change as huge criticism to statistical inference has been presented, up to defining hypothesis testing misleading (Tryon, 1991) and invalid (Trafimow and Marks, 2015). Very recently, the journal *Basic and Applied Social Psychology* has decided to withdraw support to p-value based studies and to ban them (Trafimow, 2014; Trafimow and Marks, 2015; Woolston, 2015).

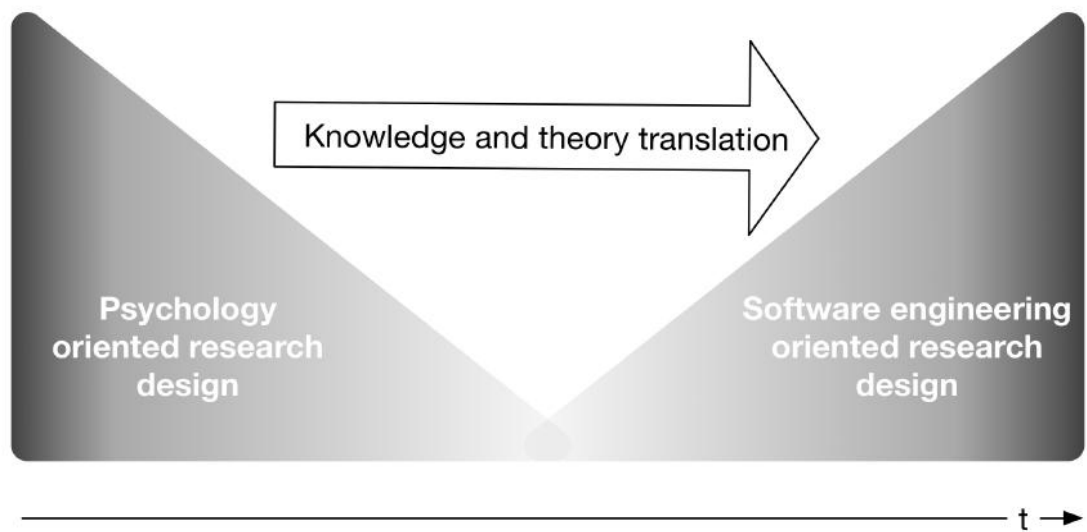

FIGURE 3.3: *Design choices over time*

### 3.5.4   Non-linearity of the process

The present candidate is willing to admit that the three research parts are necessary to make sense of the PhD process in retrospective. The process has not been perfectly linear. Figure 3.4 is a honest representation, alas still idealized for the purposes of representation, of this PhD's research activities.

As Figure 3.4 shows, each phase benefited from Phase I, the one where the literature from other domains was reviewed. Phase II was an input to Phase III. Even the published articles reflect this non-linearity of the process. For example, Paper I reviews some related work in software engineering and it reports evidence about how practitioners are deeply interested in our studies, thus motivating the research activities. However, Paper I was written after Paper IV; Paper V; Paper VI. Still, Paper I belongs to Phase I. The same can be stated for the literature review of this PhD dissertation, which indeed was the first activity ever to start. However, for the sake of completeness, the literature review has been continued throughout the entire PhD cycle, resulting in the publication of two papers, Paper III; Paper II.

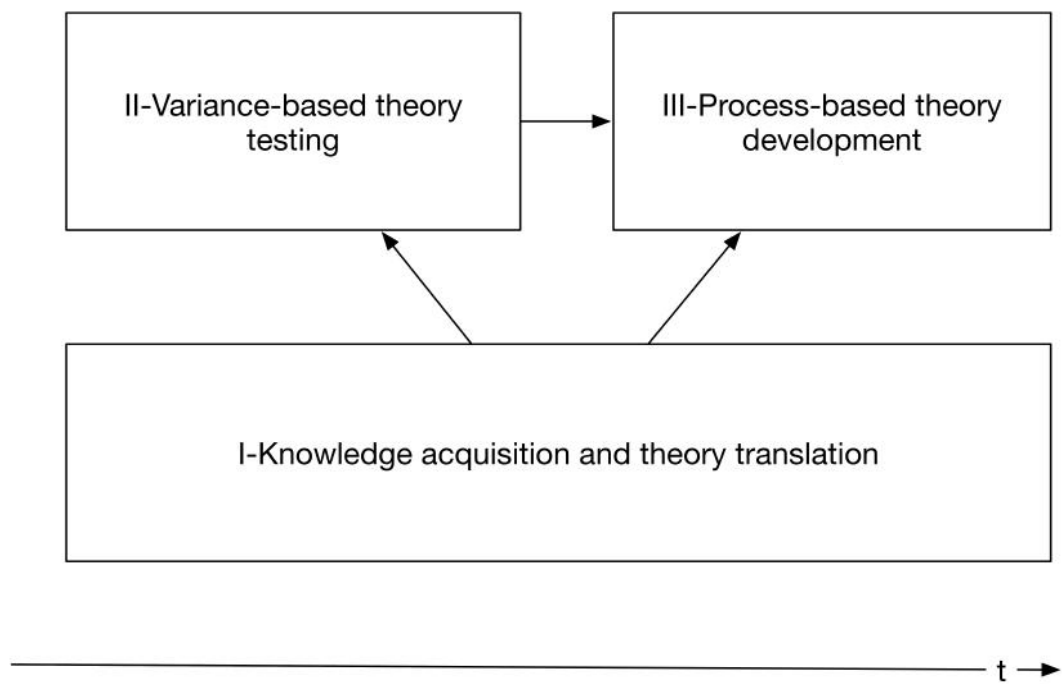

FIGURE 3.4: *PhD research phases*

This non-linearity of the process rarely appears on scientific papers and PhD dissertations, but it should not be considered an issue. According to the Nobel Prize winner Medawar (1964), scientific articles are fraudulent in the sense that they misrepresent the actual process of scientific discovery by giving a linear, rational-based yet misleading narrative of the process. This forced way of reporting scientific discovery brought to an implicit belief that the scientific method is a clearly defined series of steps, which causes confusion especially to students (Howitt and Wilson, 2014), who experience the (beautiful) chaos of doing science and yet are told to idealize the process when reporting it. Finally, the present author has agreed with Medawar (1964) that scientists should not be afraid or ashamed to admit the process is not linear, not error-proof, and not completely planned in advance.

### 3.5.5 Chapter summary

This chapter provided a meta-discussion on research activities in software engineering. First, it reported that this PhD work has adopted Creswell (2009) framework for research, which conceptualizes research into philosophical worldviews, approaches, methods, and strategies. The chapter then reiterated Creswell (2009) framework for describing phases and for stating the particular aspects of this PhD's research activities. In particular, the present candidate has a pragmatic worldview, which adheres to mixed method approaches for research. In fact, a mixed method approach was adopted for this PhD's research activities, which was divided into three main phases. Phase I was called *Knowledge acquisition and theory translation*, and it was performed in other to understand the theoretical foundations of this dissertation and translate them in the domain of software engineering. Phase II was called *Variance theory testing*, and it was performed as a direct consequence of translating much psychology knowledge in the domain of software engineering. It comprised of two quasi-experimental designs for developing and testing a variance theory of the impact of affects on developers' performance. Phase III was called *Process theory development*, and it was performed in order to develop a process theory of the impact of affects on developers' performance. It comprised a study, which was qualitative and interpretive, for constructing such theory.

# Chapter 4

# Affects and software developers' problem solving performance[1]

This chapter presents the first of the three main empirical studies, namely a quasi-experiment, which was conducted for testing a variance theory of affects and performance in software development. It is also the first study comprising Phase II of this PhD's research activities. This chapter is a contribution for answering RQ2—*How do affects indicate problem-solving performance among developers?*.

The chapter reports a quasi-experiment, which was designed for understanding if there are significant differences in terms of creative and analytic problem solving performance of software developers with respect to their pre-existing affects. Because of the lack of a clear relationship between affective states and problem-solving performance, we designed an experiment to test two related high-level hypotheses. We hypothesized that affective states would impact (1) the creative work produced by software developers and (2) their analytic problem-solving skills. 42 computer scientist students participated to the experiment which was comprised of two validated psychology tasks for creativity and analytic problem solving performance assessment. The two tasks were preceded by an affect measurement session, which was implemented by a validated measurement instrument, as well. The analysis was performed using a series of tests for group comparison,

---

as the participants were grouped into two groups according to a median split of their affect balance score. The two groups were compared in terms of the results obtained by each individual in the tasks. Overall, the results suggest that the happiest software developers outperform the other software developers in terms of analytic problem solving performance.

To our knowledge, the study in this chapter has been the first software engineering research activity using software development tasks that are suitable for measuring the creativity and analytical problem-solving skills of software developers. Moreover, the study in this chapter has been the first to ever compare the creative and analytic performance of developers with respect to their affects.

In section 4.1, the chapter extends in depth the description of the research design which was offered in section 3.5. First, the participants characteristics are provided, using descriptive statistics. Second, the section describes the required materials for running the experiment design, for example the Psychology Experiment Building Language. Third, the experiment procedure is explicated. Fourth, the employed validated measurement systems are carefully described. Section 4.2 reports the results of the executed experiment. Section 4.3 reports the limitations specific to this study.

## 4.1   Research design

To test the hypotheses we obtained various measures of creativity, and we developed a measure of analytical problem-solving. Often, a creative performance has been conceptualized in terms of the outcome of the process that leads to the creation of the creative results (Amabile, 1982; Davis, 2009b). A widely adopted task asks individuals to generate creative ideas for uncommon and bizarre problems (Forgeard, 2011; Kaufman *et al.*, 2007; Lewis *et al.*, 2011; Sowden and Dawson, 2011). For assessing the creativity of our participants, we used a "caption-generating" task. The quality of the creative outcome was assessed with subjective ratings by independent judges, and the quantity of the generated captions was recorded.

A common approach for testing analytical problem-solving is to assign points to the solution of analytical tasks (Abele-Brehm, 1992; Melton, 1995). We used the Tower of London test (Shallice, 1982), a game designed to assess planning and analytical problem-solving. The Tower of London game is a very high-level task that resembles algorithm design and execution. This task reduced the limitations that would have been imposed by employing a particular programming language. Furthermore, such a level of abstraction permits a higher level of generalization because the results are not bound to a particular programming language.

Although strict development tasks could be prepared, there would be several threats to validity. Participants with various backgrounds and skills are expected, and it is almost impossible to develop a software development task suitable and equally challenging for first year BSc students and second year MSc students. The present study remained at a higher level of abstraction. Consequently, creativity and analytical problem-solving skills were measured with validated tasks from psychology research.

### 4.1.1 Participant characteristics

Forty-two student participants were recruited from the Faculty of Computer Science at the Free University of Bozen-Bolzano. There were no restrictions in the gender, age, nationality, or level of studies of the participants. Participation was voluntary and given in exchange for research credits. The affective states of the participants were natural, i.e., random for the researchers. Of the 42 participants, 33 were male and nine were female. The participants had a mean age of 21.50 years old (standard deviation (SD) = 3.01 years) and were diverse in nationality: Italian 74%, Lithuanian 10%, German 5%, and Ghanaian, Nigerian, Moldavian, Peruvian, or American, with a 2.2% frequency for each of these latter nationalities. The participants' experience in terms of years of study was recorded (M = 2.26 years, SD = 1.38).

Institutional review board approval for conducting empirical studies on human participants was not required by the institution. However, written consent was obtained from all of the subjects. The participants were advised, both informally and on the consent form, about the data retained and that anonymity was fully ensured. No sensitive data were collected in this study. The participants were assigned a random participant code to link the gathered data. The code was in no way linked to any information that would reveal a participant's identity.

All of the students participated in the affective states measurement sessions and the two experimental tasks. However, the results of one participant from the creativity task and another from the analytical problem-solving task have been excluded; the two participants did not follow the instructions and submitted incomplete data. Therefore, the sample size for the two experiment tasks was N = 41. None of the participants reported previous experience with the tasks.

### 4.1.2   Materials

For the two affective states measurement sessions, the participants completed the Scale of Positive and Negative Experience (SPANE, (Diener *et al.*, 2009a)) questionnaire through a Web-based form, which included the related instructions. The SPANE questionnaire instructions that were provided to the participants are available in the article by Diener *et al.* (2009a) and are currently freely accessible on one of the author's academic website (Diener *et al.*, 2009b). Section 2.1.3 describes SPANE and the reason why it was selected as a valuable measurement instrument.

Six color photographs with ambiguous meanings were required for the creativity task. Figure 4.1 displays one of the six photographs. For legal reasons, the photographs are available from the authors upon request only.

For the analytical problem-solving task, a version of the Tower of London task implemented in the open source Psychology Experiment Building Language (PEBL; (Mueller and Piper, 2014; Mueller, 2012)) that has been used previously to examine age-related changes in planning and executive function (Piper *et al.*, 2011) was used to assess analytic problem solving. The PEBL instructed the participants, provided the task, and collected several metrics, including those of interest for our study. One computer per participant was required.

### 4.1.3   Procedure

The experimental procedure was composed of four activities: (1) the affective states measurement (SPANE), (2) the creativity task, (3) the affective states measurement (SPANE), and (4) the analytical problem-solving task. The second affective states measurement session was conducted to limit the threats to validity because the first task may provoke a change in the affective states of the participants.

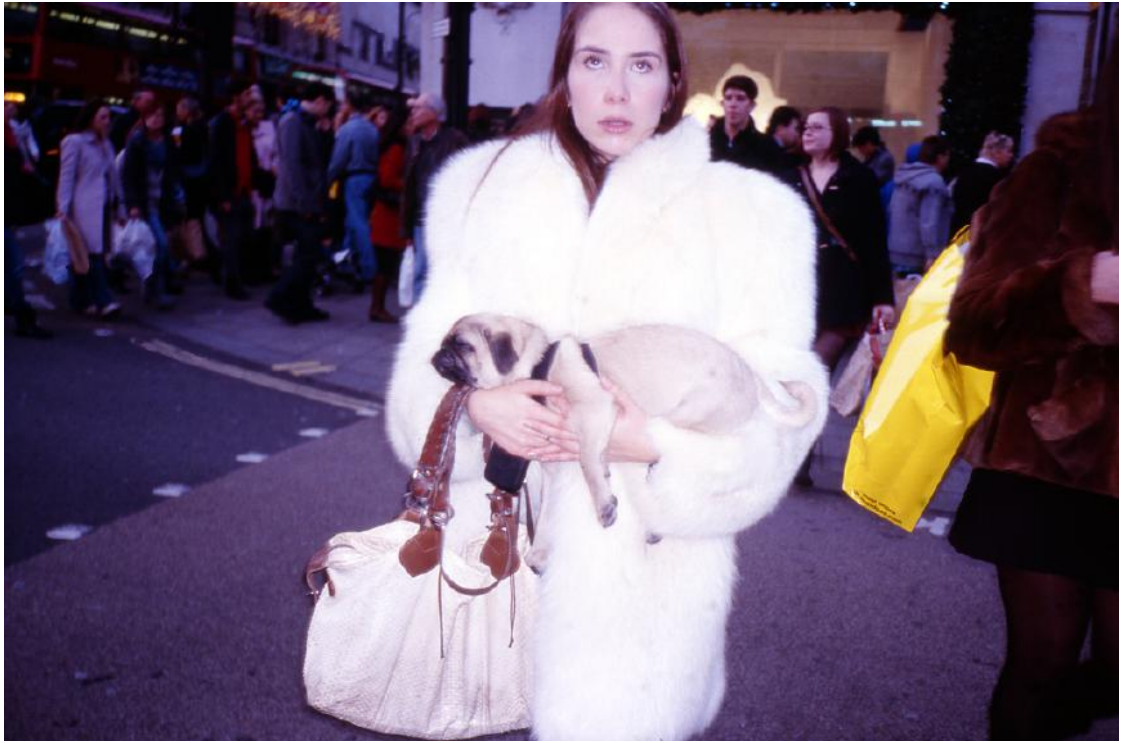

FIGURE 4.1: *A photograph for the creativity task.*

The participants arrived for the study knowing only that they would be participating in an experiment. As soon as they sat at their workstation, they read a reference sheet, which is included in Appendix A. The sheet provided a summary of all of the steps of the experiment. The researchers also assisted the participants during each stage of the experiment. The participants were not allowed to interact with each other.

During the creativity task, the participants received two random photographs from the set of the six available photographs, one at a time. The participants imagined participating in the *Best Caption of the Year* contest and tried to win the contest by writing the best captions possible for the two photographs. They wrote as many captions as they wanted for the pictures. The creativity task instructions are available as an appendix in the study by Forgeard (2011).

During the analytical problem-solving task, the participants opened the PEBL software. The software was set up to automatically display the Tower of London game, namely the *Shallice test* ([1,2,3] pile heights, 3 disks, and Shallice's 12 problems). The PEBL software displayed the instructions before the start of the task. The instructions stated how the game works and that the participants had to think about the solution before starting the task, i.e., making the first mouse click. Figure 4.2 provides a screenshot of the first level of the game. Because PEBL is open-source software, the reader is advised to obtain the PEBL software to read the instructions.

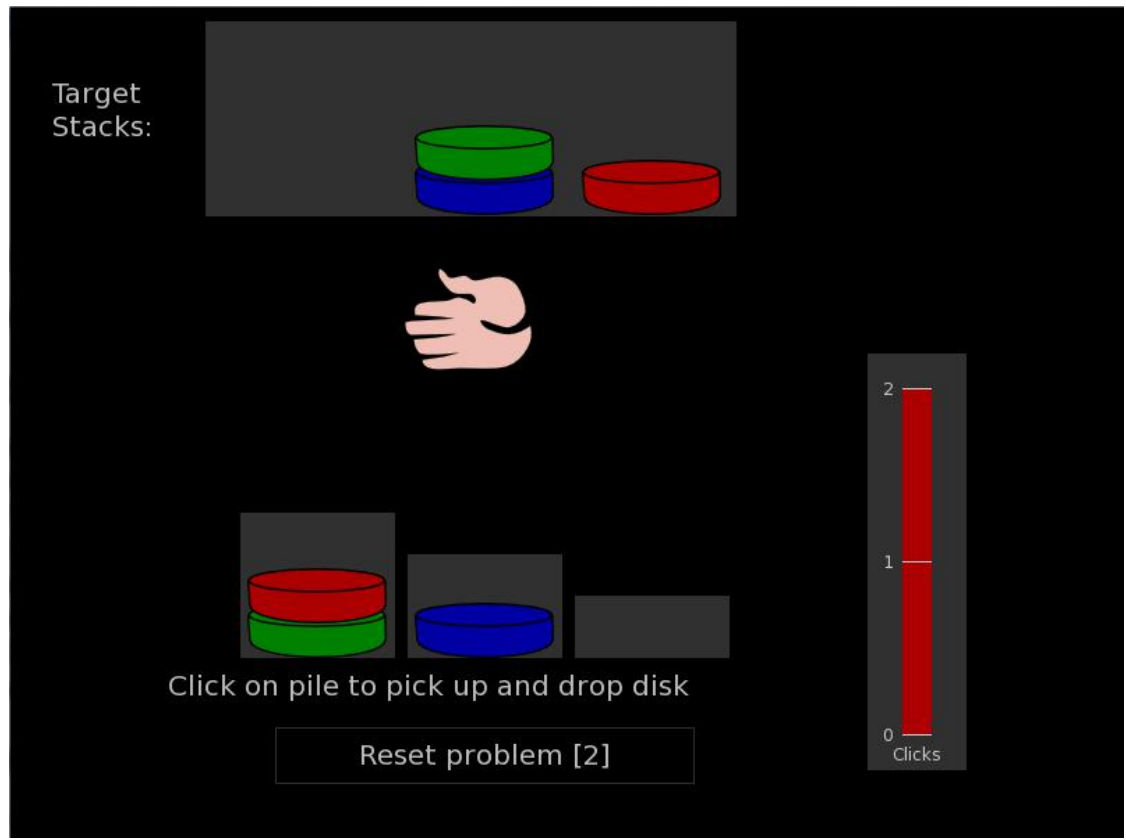

FIGURE 4.2: *The first level of the Tower of London game.*

Although the participants did not have strict time restrictions for completing the tasks, they were advised of the time usually required to complete each task and that the second task would begin only after every participant finished the first task.

The participants were not aware of any experimental settings nor of any purpose of the experiment.

Two supervisors were present during the experiment to check the progress of the participants and to answer their questions. All of the steps of the experiment were automatized with the use of a computer, except for the caption production in the creativity task. The captions were manually transcribed in a spreadsheet file. For this reason, a third person double-checked the spreadsheet containing the transcribed captions.

The study was conducted in January 2012. The designed data collection process was followed fully. No deviations occurred. Each of the tasks required 30 minutes to be completed, and the participants completed the two surveys in 10 minutes each. No participants dropped from the study.

### 4.1.4 Measures

To measure creativity according to the Consensual Assessment Technique (Amabile, 1982), independent judges who are experts in the field of creativity scored the captions using a Likert-item related to the creativity of the artifact to be evaluated. The judges had to use their own definition of creativity (Amabile, 1982; Kaufman *et al.*, 2007). The Likert-item is represented by the following sentence: *This caption is creative*. The value associated to the item ranges from one (*I strongly disagree*) to seven (*I strongly agree*). The judges were blind to the design and the scope of the experiment. That is, they received the six pictures with all of the participants' captions grouped per picture. The judges were not aware of the presence of other judges and rated the captions independently. Ten independent judges were contacted to rate the captions produced in the creativity task. Seven judges responded, and five of the judges completed the evaluation of the captions. These five judges included two professors of Design & Arts, two professors of humanistic studies, and one professor of creative writing.

The present study adopted measurements of quality and quantity for the assessment of creativity. The quality dimension of creativity was measured by two scores. The first quality score was the average of the scores assigned to all of the generated ideas of a participant (*ACR*). The second quality score was the best score obtained by each participant (*BCR*), as suggested by Forgeard (2011) because creators are often judged by their best work rather than the average of all of their works (Kaufman *et al.*, 2007). The quantity dimension was represented by the number of generated ideas (*NCR*), as suggested by Sowden and Dawson (2011).

Measuring analytical problem-solving skills is less problematic than measuring creativity. There is only one solution to a given problem (Cropley, 2006). The common approach in research has been to assign points to the solution of analytical tasks (Abele-Brehm, 1992; Melton, 1995). This study employed this approach to combine measures of quality and quantity by assigning points to the achievements of analytical tasks and by measuring the time spent on planning the solution. The Tower of London game (a.k.a. *Shallice's test*) is a game aimed to determine impairments in planning and executing solutions to analytical problems (Shallice, 1982). It is similar to the more famous Tower of Hanoi game in its execution. Figure 4.2 provides a screenshot of the game. The rationale for the employment of this task is straightforward.

The Analytical Problem Solving (*APS*) score is defined as the ratio between the progress score achieved in each trial of the Tower of London Game (TOLSS) and the number of seconds needed to plan the solution to solve each trial (PTS). The TOLSS scores range from 0 to 36 because there are 12 problems to be solved and each one can be solved in a maximum of three trials. PTS is the number of milliseconds that occurred between the presentation of the problem and the first mouse click in the program. To have comparable results, a function to map the *APS* ratio to a range from 0.00 to 1.00 was employed.

## 4.2   Results

The data were aggregated and analyzed using the open-source R software (R Core Team, 2013). The SPANE-B value obtained from this measurement session allowed us to estimate the SPANE-B population mean for software developers, $\mu_{SPANE-B-DEV} = 7.58$, 95% CI [5.29, 9.85]. The median value for the SPANE-B was nine. This result has consequences in the discussion of our results, which we offer in Chapter 8.

The multiple linear and polynomial regression analyses on the continuous values for the various SPANE scores and the task scores did not yield significant results. Therefore, the data analysis was performed by forming two groups via a median split of the SPANE-B score. The two groups were called *N-POS* (for *non-positive*) and *POS* (for *positive*). Before the creativity task, 20 students were classified as *N-POS* and 21 students were classified as *POS*.

The histograms related to the affective state distributions and the group compositions are in Figure 4.3. These data are not crucial for the purposes of this investigation. However, they have been attached to this article for the sake of completeness. The same holds for the boxplots (Figure 4.4) and the scatterplots representing non-significant data (Figure 4.5).

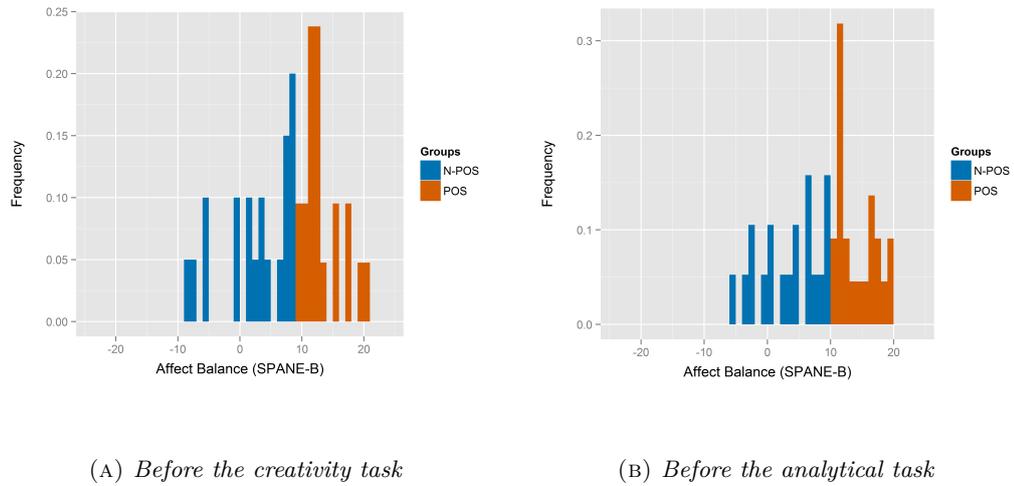

(A) *Before the creativity task*     (B) *Before the analytical task*

FIGURE 4.3: *Histograms of the affect balance (SPANE-B) before the two tasks*

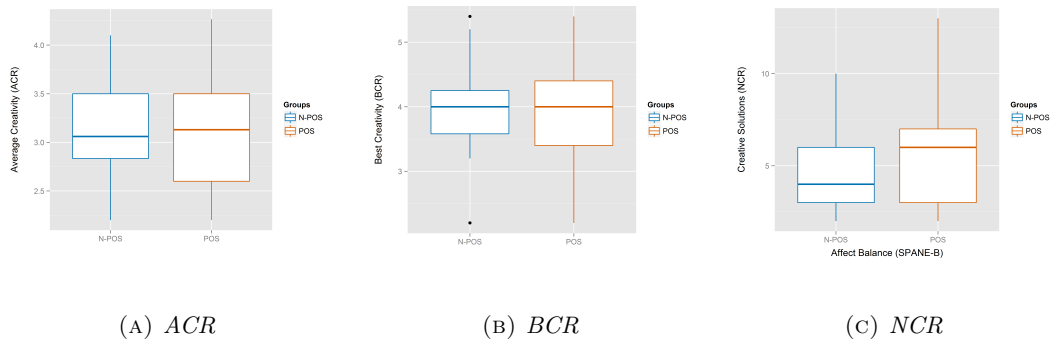

(A) *ACR*     (B) *BCR*     (C) *NCR*

FIGURE 4.4: *Boxplots for average and best creativity, and number of creative ideas between the N-POS and POS groups*

Table 4.1 summarizes the task scores of the two groups for the two tasks.

TABLE 4.1: *Mean and standard deviation of the task scores divided by the groups.*

|  | N-POS[1] | | POS[2] | |
|---|---|---|---|---|
| Variable | M (SD) | 95% CI | M (SD) | 95% CI |
| ACR[a] | 3.13 (0.45) | [2.92, 3.35] | 3.08 (0.58) | [2.81, 3.35] |
| BCR[b] | 4.02 (0.76) | [3.67, 4.38] | 3.98 (0.76) | [3.63, 4.32] |
| NCR[c] | 4.70 (2.34) | [3.60, 5.50] | 5.90 (3.46) | [4.00, 7.50] |
| APS[d] | 0.14 (0.04) | [0.12, 0.17] | 0.20 (0.08) | [0.17 0.25] |

[1] Non-positive group;
[2] Positive group;     [a] Average of the scores assigned to all of the generated ideas of a participant;     [b] Best score obtained by each participant;
[c] Number of generated ideas;     [d] Analytical problem-solving score.

The two creativity scores of *ACR* and *BCR* showed many commonalities. Visual inspections of the scatterplots of the *ACR* (Figure 4.5a) and *BCR* (Figure 4.5b) scores versus the SPANE-B score suggested a weak trend of higher creativity when the SPANE-B value tended to its extreme values (-24 and +24).

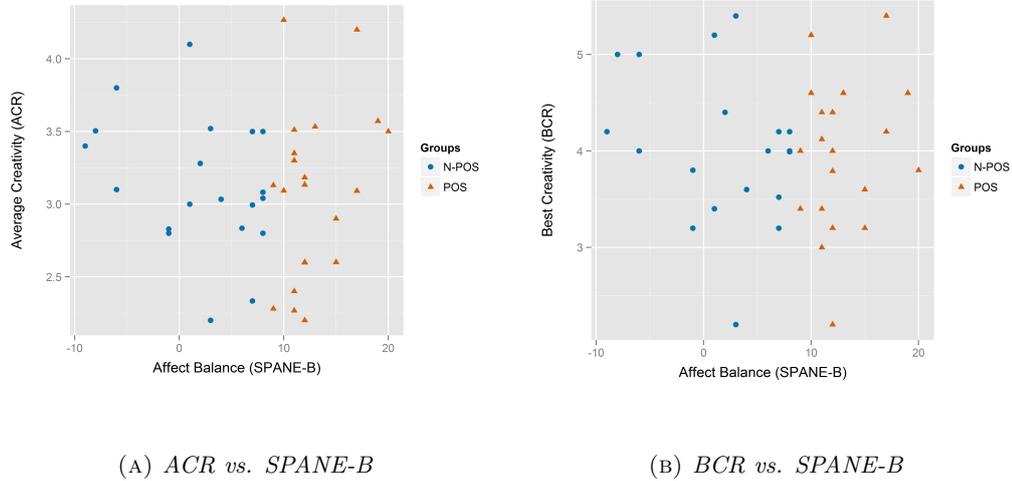

(A) *ACR vs. SPANE-B*                        (B) *BCR vs. SPANE-B*

FIGURE 4.5: *Scatterplots for average and best creativity vs. the affect balance (SPANE-B) between the N-POS and POS groups*

The median for the number of generated captions (*NCR*) was four for the *N-POS* group and six for the *POS* group. However, the lower quartiles of the two groups were almost the same, and there was a tiny difference between the two upper quartiles (Figure 4.4c).

We hypothesized that affective states would impact the creative work produced by software developers, without a direction of such impact. The hypothesis was tested using unpaired, two-tailed t-tests. There was no significant difference between the *N-POS* and *POS* groups on the *BCR* score (t(39) = 0.20, $p > .05$, d = 0.07, 95% CI [-0.43, 0.53]) or the *ACR* score (t(39) = 0.31, $p > .05$, d= 0.10, 95% CI [-0.28, 0.38]). The third test, which regarded the quantity of generated creative ideas (*NCR*), required a Mann-Whitney U test because the assumptions of normality were not met (Shapiro-Wilk test for normality, W = 0.89, $p = 0.02$ for *N-POS* and W = 0.87, $p = 0.01$ for *POS*). There was no significant difference between the *N-POS* and *POS* groups on the *NCR* score (W = 167.50, $p > .05$, d = -0.41, 95% CI [-2.00, 1.00]).

The second SPANE questionnaire session was performed immediately after the participants finished the creativity task. The average value of the SPANE-B was M = 8.70 (SD = 6.68), and the median value was 10. There was a significant increase in the SPANE-B value of 1.02 (t(39) = 3.00, $p < 0.01$, d = 0.96, 95% CI [0.34, 1.71]). Therefore, a slight change in the group composition occurred, with 19 students comprising the *N-POS* group and 22 students comprising the *POS* group. Cronbach (1951) developed the $\alpha$ (alpha) as a coefficient of internal consistency and interrelatedness especially designed for psychological tests. The value of Cronbach's $\alpha$ ranges from 0.00 to 1.00, where values near 1.00 indicate excellent consistency (Cortina, 1993; Cronbach, 1951). The Cronbach's $\alpha$ reliability measurement for the two SPANE questionnaire sessions was $\alpha = 0.97$ (95% CI [0.96, 0.98]), which indicates excellent consistency. We discuss the consequences of these results in Chapter 8.

We hypothesized that affective states would impact the analytic problem-solving skills of software developers. The boxplots for the *APS* score in Figure 4.6[2] suggest a difference between the two groups, and the relevant scatterplot in Figure 4.7 suggested that the *APS* points for the *N-POS* group may be linear and negatively correlated with the SPANE-B; excellent *APS* score were achieved only in the *POS* group. The hypothesis was tested using an unpaired, two tailed t-test with Welch's correction because a significant difference in the variances of the two groups was found (F-test for differences in variances, $F(21, 18) = 3.32$, $p = 0.01$, 95% CI [1.30, 8.17]). There was significant difference between the *N-POS* and *POS* groups on the *APS* score ($t(33.45) = -2.82$, $p = 0.008$, $d = -0.91$, 95% CI [-0.11, -0.02]). A two-sample permutation test confirmed the results ($t(168)$, $p = 0.01$, CI [-13.19, -1.91]).

A summary of the research output of this study is reported in Section 7.2. The discussion of the results obtained by this study are provided in Section 8.2. The implications for research brought by this study are available in Section 8.5.1. The implications for practice of this study are explained in Section 8.5.2.

The limitations of this study, and how they have been mitigated, are reported below as they are specific to this study only. The general limitations of this dissertation are offered in section 9.2.

## 4.3 Limitations

The primary limitation of the study in Paper IV, in Chapter 4, lies in the sample; the participants were all Computer Science students. Although there is diversity in the nationality and experience in years of study of the participants, they have limited software development experience compared with professionals. However, Kitchenham *et al.* (2002); Tichy (2000) argued that students are the next generation of software professionals. Thus, they are remarkably close to the population of interest and may even be more updated on the new technologies (Kitchenham *et al.*, 2002; Tichy, 2000). Höst *et al.* (2000) found non-significant differences in the performance of professional software developers and students on the factors affecting the lead-time of projects. There is an awareness that not all universities offer the same curricula and teaching methods and that students may have various levels of knowledge and skills (Berander, 2004). Still, given the high level of abstraction provided by the tasks in this study, a hypothetical difference between this study's participants and software professionals would likely be in the magnitude and not in the direction of the results (Tichy, 2000). Lastly, the employed affective states measurement instrument, SPANE, provided consistent data across full-time workers and students (Silva and Caetano, 2011).

---

[2]The color scheme for the graphs of this study have been generated by following the guidelines for producing colorblind-friendly graphics (Okabe & Ito, 2008).

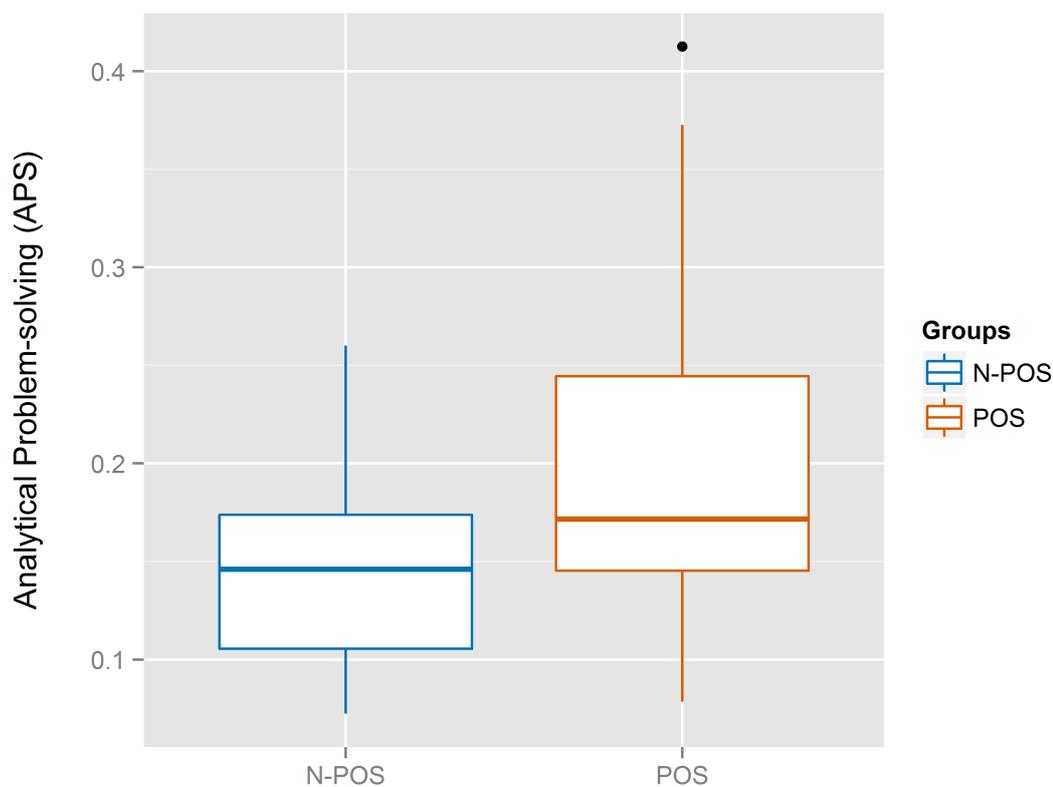

FIGURE 4.6: *Boxplots for the analytical problem-solving (APS) of the N-POS and POS groups.*

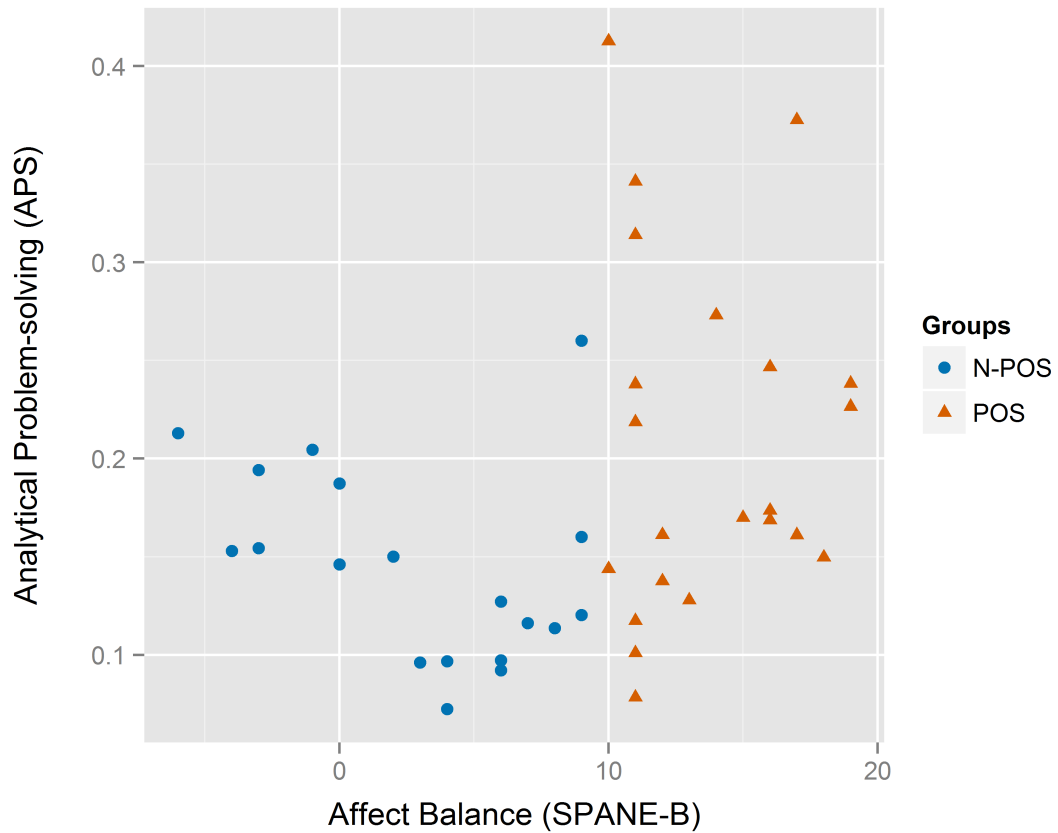

FIGURE 4.7: *Scatterplot for the analytical problem-solving (APS) vs. the affect balance (SPANE-B) between the N-POS and POS groups.*

Another limitation is that a full coverage of the SPANE-B range in the negative direction could not be obtained. Although 42 participants were recruited, the SPANE-B score did not fall below the value of minus nine, and its average value was always greater than +7 on a scale of [-24,+24]. Before the experiment, a more homogeneous distribution of participants was expected for the SPANE-B score. However, there is actually no evidence that the distribution of SPANE-B scores for the population of software developers should cover the full range of [-24, +24]. Additionally, studies estimating the SPANE-B mean for any population are not known. For this reason, an estimation of the affective states population mean for software developers was offered by this study: $\mu_{SPANE-B-DEV} = 7.58$, 95% CI [5.29, 9.85]. Thus, it may be that the population's true mean for the SPANE-B is above +7 and significantly different from the central value of the measurement instrument. This translated to a higher relativity when we discussed our results, especially for the comparison with related work. However, the results of this study are not affected by this discrepancy.

A third limitation lies in the employment of a median split to compose the groups. Employing a median split removed the precision that would have been available in a continuous measure of the SPANE-B[3]. Despite this, using a median split was necessary because no known regression technique could yield valid results; median splits on affective state measurements are not uncommon in similar research (Berna *et al.*, 2010; Forgeard, 2011; Hughes and Stoney, 2000).

## 4.4   Chapter summary

This chapter presented a quasi experiment, which was the first of the variance theory testing studies in Phase II of this PhD's research activities. The experiment was designed for understanding if there are significant differences in terms of creative and analytic problem solving performance of software developers with respect to their pre-existing affects. 42 computer scientist students participated to the experiment. The pre-existing affects of the developers were measured before each task using the SPANE (Diener *et al.*, 2009a) measurement instrument. The participant faced two validated psychology tasks for creativity and analytic problem solving assessment. The analysis of the results suggested that the happiest software developers outperform the other software developers in terms of analytic problem solving. The results of this study enabled us to develop the first facet of the theory of affects and performance in software developed, which was modeled in Figure 7.1a.

---

[3]The authors are thankful to an anonymous reviewer for pointing out this issue.

# Chapter 5

# Affects and software developers' productivity [1]

This chapter presents the second of the three main empirical studies, namely a natural experiment with repeated measurements design, which was conducted for testing a theory of affects and performance in software development. It is the second study comprising Phase II of this PhD's research activities. This chapter is a contribution for answering RQ3—*How are affects correlated to in-foci task productivity of software developers?*.

The chapter reports a repeated measures research on the correlation of the affects of software developers and their immediate self-assessed productivity. The research hypotheses of this study were on positive correlations between real-time affects and the self-assessed productivity of software developers. Figure 5.1 represents the formulation of the hypotheses, which are as follows:

**H1** The real-time valence affective state of software developers is positively correlated to their self-assessed productivity.

**H2** The real-time arousal affective state of software developers is positively correlated to their self-assessed productivity.

**H3** The real-time dominance affective state of software developers is positively correlated to their self-assessed productivity.

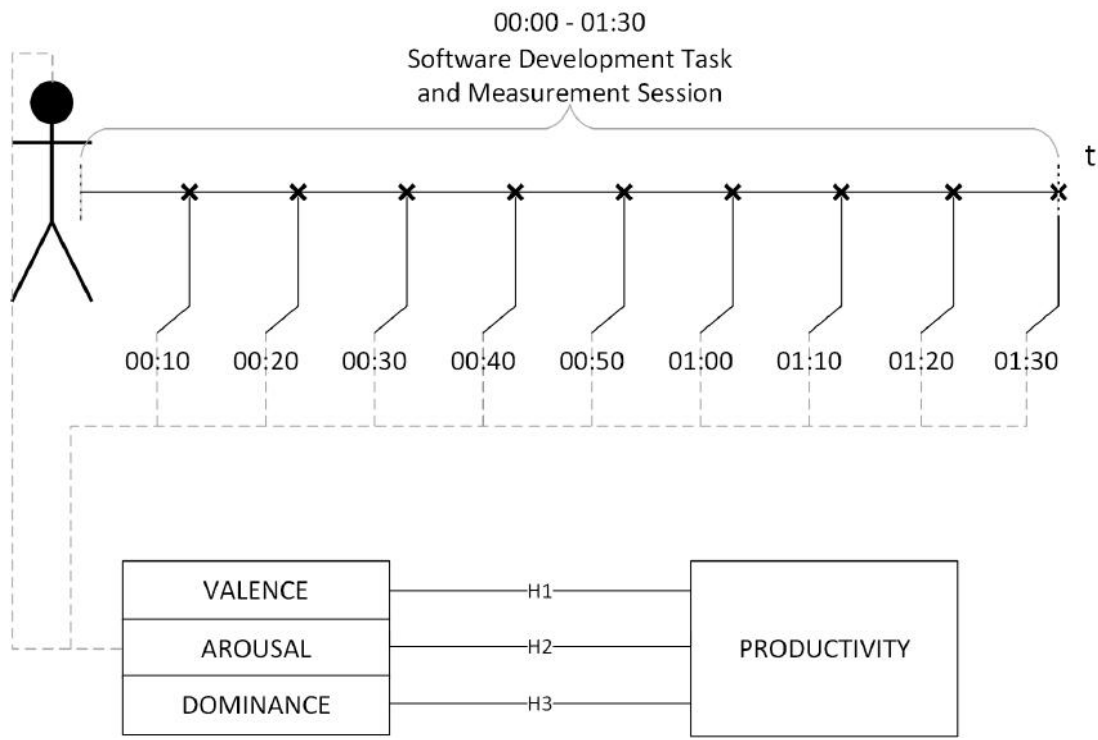

FIGURE 5.1: *Graphical representation of the hypotheses of this study, the fixed effects (valence, arousal, dominance), and the random effects (participant, time) in the context of the research design.*

As depicted by Figure 5.1, the three hypotheses are to be tested over the time dimension, and the easiest way to implement this is to verify the hypotheses at certain time intervals. Eight developers working on their individual projects have been observed for 90 minutes each. The developers were conducting a software development task on their own software project. Their affects and their self-assessed productivity were measured on intervals of ten minutes. A linear mixed-effects model was proposed in order to estimate the value of the correlation of the affects of valence, arousal, and dominance, as well as the productivity of developers. The model was able to express about the 38% of the deviance of the self-assessed productivity. Valence and dominance, or the attractiveness perceived towards the development task and the perception of possessing adequate skills, were able to provide nearly all the explanation power of the model.

To our knowledge, the study in this chapter has been the first software engineering research to study the real-time affects of software developers with respect to their self-assessed productivity over time. The study is novel in the field of software engineering to propose the use of linear mixed effects models as a way to cope with the complex net of data dependencies and z-scores when employing repeated sessions within subjects of psychological tests.

In section 5.1, the chapter extends in depth the description of the research design which was offered in section 3.5. First, the participants characteristics are provided. Second, the section describes the required materials for running the experiment design, for example the Self-Assessment Manikin. Third, the experiment procedure is explicated. Fourth, the employed validated measurement systems are carefully described, together with the data transformation measures and the analysis methods. Section 5.2 reports the results of the executed experiment. First, the preparation and the deviations of the experiment execution are described. Second, the descriptive statistics including the sample demographics, the pre-task interviews, the repeated measurements, and the post-task interviews are reported. Third, the hypotheses testing are reported. The development and the execution of this study has brought us several lessons learned, which are reported in Section 8.5.1. The discussions of the results, including the implications for research and for practice, and the conclusions of this study are presented in the related chapters 8 and 9.

## 5.1 Research design

### 5.1.1 Participants characteristics

The participants for this experiment are software developers. Professionals and students can both be taken into consideration for being participants. The only requirement is that the participants work on any software project, but can perform a task individually. There are no restrictions in gender, age, or nationality of the participants. Participation is voluntary and not rewarded. The rationale is that participants work on their project in their natural settings. Therefore, they only need to accept the presence of the researcher for a limited time. Confidentiality has to be assured to participants upfront when they are recruited. They are asked to participate in a study in which they are singularly observed for a limited amount of time and asked to fill a very short questionnaire at regular intervals of ten minutes during work. They are assured on the anonymity of the gathered data, as well. No personal information is retained.

### 5.1.2   Materials

On the participant side, there is no required material. Participants develop software in natural settings using their own equipments – i.e., a computer. The researcher, however, needs a suitable tablet device that implements the questionnaire to measure the affects of the participants and their self-assessed productivity. This study employed an ad-hoc Web-based survey, available upon request. Since the questionnaire is pictorial (see Appendix B), the effort required for a measurement session is reduced to four touches to the screen.

This experiment involves pre- and post-task interviews and the annotation of events during the task execution. Therefore, suitable devices for the recording of the observations are also required, i.e., a notepad and an audio recorder.

### 5.1.3   Procedure

Figure 5.2 summarizes the entire procedure for this study, as a timeline. The procedure is composed of three parts: a pre-task interview, a software development task, and a post-task interview.

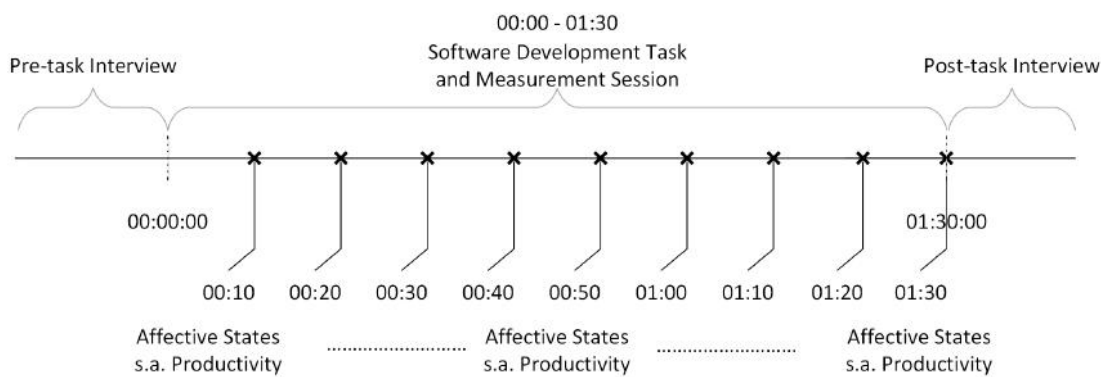

FIGURE 5.2: *Graphical representation (a timeline) of the research design.*

As indicated in the left part of Figure 5.2, the researcher and the participants first engage in a pre-task interview. The starting questions for the pre-task interview can be found in Appendix B. The participants are interviewed either in natural settings – e. g., their offices – or in any location that makes them comfortable, like a bar. During the pre-task interview, basic demographic data, information about the project, tasks, and the developers' skills are obtained. Descriptive data is collected on the role of the participant (either "professional" or "student"), as well as the perceived experience with the programming language, and the perceived experience with the domain of knowledge – including the task type - (low, medium, and high).

Right after the pre-task interview, as indicated by the central part of Figure 5.2, the participants and the researcher sit together in the working environment. Before the start of the task, the participants are informed about the mechanisms involved in the task session. The instructions given to each participant are available in Appendix B. The instructions for the Self-Assessment Manikin (SAM) questionnaire were written following the technical manual by Lang *et al.* (1999). For a period of 90 minutes, the participants work on software development tasks of real software projects. The researcher observes the behavior of the individuals while they are developing software. After each 10 minute session, the participants complete a short questionnaire on a tablet device. In this questionnaire, valence, arousal, and dominance are measured 9 times per participant. The same process holds for self-assessment of the worker's productivity. The researcher is present during the entire development period to observe the behavior of the participant without interfering.

After the completion of the working period, the researcher conducts a post-task interview represented by the third part of Figure 5.2. The questions are available in Appendix B and are related to the self-assessment of the productivity of the participants, the factors influencing the performance, and if and how the measurement system interrupts or annoys the participants. The interviews are structured and mostly close-ended questions are used. The purpose is to triangulate the findings obtained in the main experiment.

The reader is reminded that this experiment studies the participants one at a time. Each step of the experiment procedure is executed on an individual basis. Additionally, all steps of the experiment are automatized through electronic systems.

### 5.1.4 Measures

The affects dimensions - valence, arousal, dominance - describe differences in affective meanings among stimuli and are measured with the Self-Assessment Manikin (SAM, (Bradley and Lang, 1994; Lang *et al.*, 1999)) pictorial questionnaire. Section 2.1.3 describes SAM and the reason why it was selected as a suitable measurement instrument.

SAM's scores range from 1 to 5. However, as reported in Section 2.1.3, a participant's data is converted to the individual's *z-score* for the set of construct measurements, using the formula in Equation (2.1). As the variables are transformed to z-scores, their values will follow a normal distribution in their range. Therefore, the range of the variables, while theoretically infinite, is practically the interval [-3, +3].

Figure 5.2 shows that this experiment design is based on repeated measurements of multiple variables, with particular data dependencies at the participant's level and at the time level. Linear mixed-effects models are the most valuable tool to be employed in such cases. A linear mixed-effects model is a linear model that contains both fixed effects and random effects. The definition of a linear mixed-effects model by Robinson (1991) was given in Equation (2.2)

The estimation of the significance of the effects for mixed models is an open debate (Bates, 2006; R Community, 2006). One proposed way is to employ likelihood ratio tests (*ANOVA*) as a way to attain *p-values* (Winter, 2013). With this approach, a model is constructed by adding one factor (or interaction) at a time and performing a likelihood ratio test between the null model and the one-factor model. By adding one factor or interaction at a time, it would be possible to construct the best fitting possible model. However, this technique is time-consuming and prone to human errors.

Another proposed approach is to construct the full model instead. A way to express the significance of the parameters is to provide upper and lower bound *p-values*. Upper-bound *p-values* are computed by using as denominator degrees of freedom the number of data points minus the number of fixed effects and the intercept. Lower bound (or conservative) *p-values* are computed by using as denominator degrees of freedom the number of data points minus the within-participant slopes and intercepts multiplied per the number of participants. The reader is advised to read the technical manual (Tremblay and Ransijn, 2013) for additional details. This is the approach that was followed in this study.

The model has been implemented using the open-source statistical software *R* (R Core Team, 2013) and the *lme4.lmer* function for linear mixed-effects models (Douglas Bates, Martin Maechler, 2014). For the model construction, valence, arousal, dominance, and their interaction with time are modeled as fixed effects. The random effects are two: a scalar random effect for the participant-grouping factor (i.e., each participant) and a random slope for the measurement time, indexed by each participant. In this way, the dependency of the measurements within participants is taken into account at the participant's level and at the time level. The full model is given in 5.2 as a *lme4.lmer* formula.

$$productivity \sim (valence + arousal + dominance) * time + (1|participant) + (0 + time|participant)$$

$$(5.1)$$

where *productivity* is the dependent variable; *valence*, *arousal*, *dominance*, and *time* are fixed effects; (1 | *participant*) is the scalar random effect for each participant, and (0 + *time* | *participant*) is read as "no intercept and time by participant"[2]. It is a random slope for the measurement time, grouped by each participant.

## 5.2  Results

In this instance of the study, the participants were students of computer science at the Free University of Bozen-Bolzano and workers at local IT companies. The participants have been obtained using convenience sampling. However, the sample provides a fair balance in terms of knowledge and roles. As the participants work on their own software projects, no particular training was needed. However, participants were trained on the measurement instrument using the supplied reference sheet (see Appendix B). No deviations occurred. As no dropouts happened and no outliers could be identified, the only required data transformation was the formula in 2.2. The *z-score* transformation was implemented in a *R* script with the command *scale()*.

---

[2]The authors are thankful to an anonymous reviewer, who correctly suggested this slight change to the original model ending with (time|participant).

This study recruited eight participants for 72 measurements. The mean of the participants' age was 23.75 (standard deviation=3.29). Seven of them were male. Four participants were first year B.Sc. computer science students and four of them were professional software developers. The computer science students worked on course-related projects. The four professional software developers developed their work-related projects.

**Pre-task interviews**

The characteristics of the participants, gathered from the pre-task interviews, are summarized in Table 5.1. A first observation is that the roles did not always correspond to the experience. The professional participant P2 reported a LOW experience with the programming language while the student participant P8 reported a HIGH experience in both the programming language, the domain of knowledge, and task type. Table 5.1 also contains the characteristics of the projects and the implemented task. There were a high variety of project types and tasks. Five participants programmed using C++ while two of them with Java and the remaining one with Python. The participants' projects were non-trivial, regardless of their role. For example, participant P1 (a professional developer) was maintaining a complex software system to collect and analyze data from different sensors installed on hydrological defenses (e.g., dams). Participant P5 (a student) was implementing pictorial exports of music scores in an open-source software for composing music.

**Repeated measurements**

Figure 5.3, Figure 5.4, and Figure 5.5 provide the charts representing the changes of the self-assessed productivity (dashed line) over time, with respect to the valence, the arousal, and the dominance dimensions (solid line) respectively.

As seen in Figure 5.3, there were cases in which the valence score showed very close linkage to productivity (participants P2, P7, and P8). For the other participants, there were many matched intervals, e.g. P5 at interval 7, and P4 at intervals 4-7. Participant P1 was the only one for which the valence did not provide strong predictions. In few cases, the valence *z-score* was more than a standard deviation apart from the productivity *z-score*.

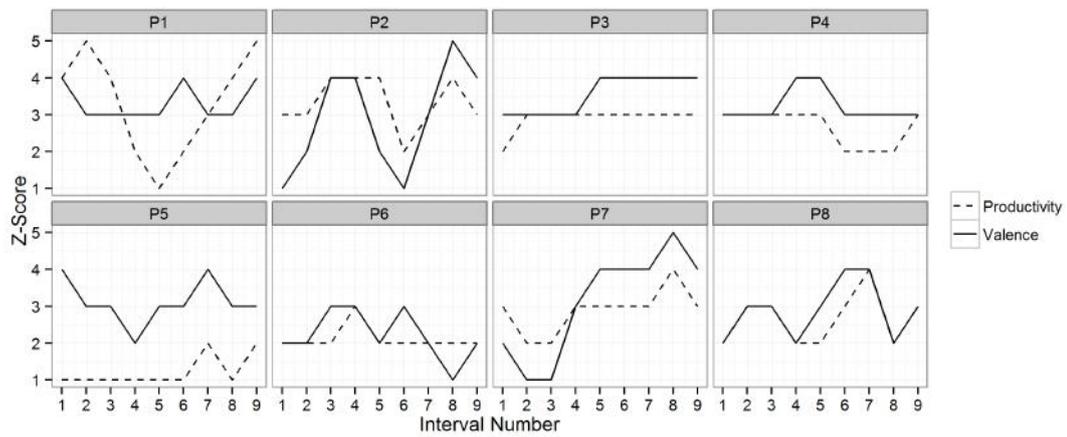

FIGURE 5.3: *Valence versus productivity over time, grouped per participant.*

TABLE 5.1: *Participants and project details*

| ID | Gender | Age | Role | Project | Task | Programming Language | Programming Language Experience | Domain Experience |
|---|---|---|---|---|---|---|---|---|
| P1 | M | 25 | PRO | Data collection for hydrological defense Research Data | Module for data displaying | Java | HIGH | HIGH |
| P2 | M | 26 | PRO | Collection & Analysis Human Resources | Script to analyze data | Python | LOW | HIGH |
| P3 | M | 28 | PRO | Manager for a School | Retrieval and display of DB data | Java | HIGH | HIGH |
| P4 | M | 28 | PRO | Metrics Analyzer | Retrieval and sending of metrics | C++ | HIGH | HIGH |
| P5 | F | 23 | STU | Music Editor | Conversion of music score to pictures | C++ | LOW | LOW |
| P6 | M | 20 | STU | Code Editor | Analysis of Cyclomatic Complexity | C++ | LOW | LOW |
| P7 | M | 20 | STU | CAD editor | Single-lined labels on objects | C++ | LOW | LOW |
| P8 | M | 20 | STU | SVG Image Editor | Multiple objects on a circle or ellipse | C++ | HIGH | HIGH |

The arousal dimension in Figure 5.4 looked less related to the productivity than the valence dimension. The behavior of the arousal line often deviated from the trend of the productivity line (e.g., all the points of participants P5 and P6). Nevertheless, there were intervals in which the arousal values were closely related to productivity, e.g., with participants P4 and P7.

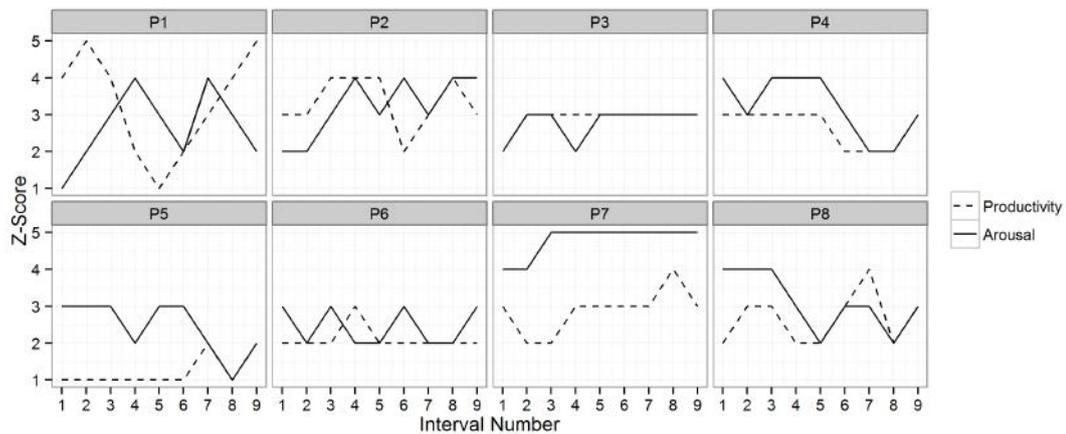

FIGURE 5.4: *Arousal versus productivity over time, grouped per participant.*

The dominance dimension in Figure 5.5 looked more correlated to the productivity than the arousal dimension. Participants P1, P5, and P7 provided close trends. For the other cases, there were intervals in which the correlation looked closer and stronger. However, it became weaker for the remaining intervals (e.g., with P4). The only exception was with participant P6 where a clear correlation between dominance and productivity could not be spotted.

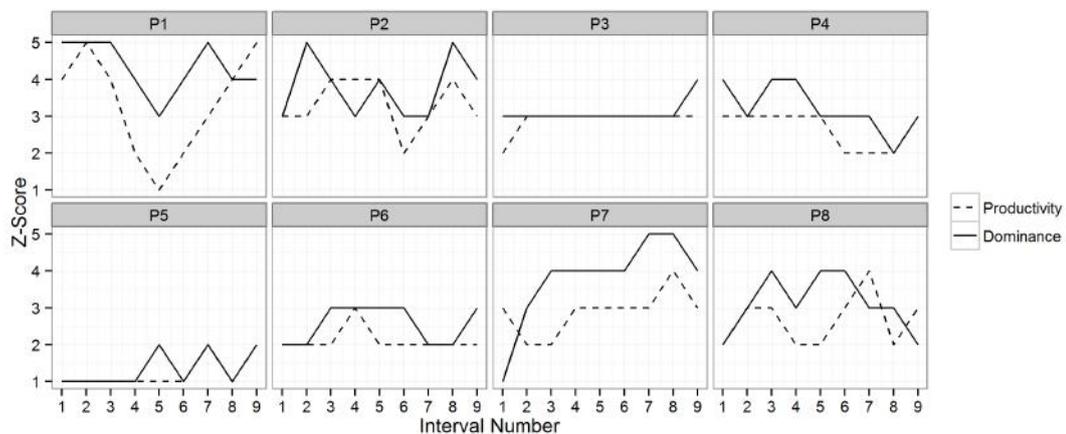

FIGURE 5.5: *Dominance versus productivity over time, grouped per participant.*

For all the participants, the *z-score* of the variables showed variations of about two units over time. That is, even for a working time of 90 minutes, there were strong variations of both the affects and the self-assessed productivity.

**Post-task interviews**

The post-task interviews were analyzed in light of the obtained results presented earlier. For the question regarding their satisfaction with their task performance, three participants (P4, P5, and P6) replied with a negative answer while the others answered with a nearly positive or completely positive answer.

All the participants had a clear idea about their productivity scores. None of them raised a question about how to self-rate their productivity. However, in the post-task interview, none of them was able to explain how they defined their own productivity. Six of them suggested that the self-assessment was related to the achievement of the expectation they set for the end of the 90 minutes of work. Their approach was to express the sequent productivity value with respect to the previous one – as in "worse", "equal", or "better" than before.

When questioned about difficulties and about what influenced their productivity, the participants found difficulties in answering in the beginning. Seven participants reported difficulties of a technical nature. For example, P2 was slowed down because of "difficulties in finding the right functions of Python for scraping data from non-standard representations in the files". P4 was slowed down because of "an obscure bug in the Secure Socket Layer library" which prevented a secure communication channel to be opened. Only P1 complained about non-technical factors. P1 was interrupted twice by two phone calls from a senior colleague, between interval 3 and interval 5. The phone calls were related to P1's task, as he was asked to perform "urgent maintenance on related stuff" that he was working on. This was reflected by P1's self-assessed productivity. It is noted here that no participants mentioned affective-related factors when answering this question. When directly inquired about the influence of their affects on the performance of their development task, six participants responded negatively: they were convinced that affects did not have an impact on their task performance.

Of the eight participants, none of them reported to be annoyed or disturbed at all by the way the measurements were obtained. This was probably achieved because a series of pilot studies with other participants had been conducted in order to reduce the invasiveness of the measurement sessions. However, two participants suggested that a wider measurement interval would have been more welcome. For the question "Would it annoy you if this system was employed during the whole working day?," all the participants agreed that a measurement interval of 10 minutes would most likely reduce their productivity.

**Hypotheses testing**

When fit with the gathered data, the proposed full model in 5.2 significantly performed better from its corresponding null model 5.2 in terms of likelihood ratio tests (*anova* in *R*; $\chi^2(7)$=30.88, *p*<0.01).

$$productivity \sim 1 + (1|username) + (0 + interval|username) \qquad (5.2)$$

The data has been checked for normality and homogeneity by visual inspections of a plot of the residuals against the fitted values. Additionally, there was no evidence for non-normality of the data (Shapiro-Wilk test; W=0.97, p>0.05). However, the likelihood ratio test only tells that there is statistical significance for the full model 5.2. It does not report meaningful results for its individual predictors and interactions. As reported in section 2.1.3, significance for parameter estimation is possible by providing lower- and upper-bound *p-values*.

Table 5.2 provides the parameter estimation for the fixed effects (expressed in *z-scores*), the significance of the parameter estimation, and the percentage of the deviance explained by each fixed effect. A single star (*) highlights the significant results (*p-value* less than 0.01). The values have been computed by the *pamer.fnc* function provided by *LMERConvenienceFunctions* (Tremblay and Ransijn, 2013). At the 0.01 significance level, valence and dominance are positively correlated with the self-assessed productivity of software developers. Therefore, there is significant evidence to support the hypotheses H1 and H3. Although there is no evidence to support H2, the hypothesis regarding arousal, a possible explanation is proposed in the discussions of this dissertation in Chapter 8. The random effects are reported in Table 5.3. The scalar random effects values for the participants belonged to the interval [-0.20, 0.15]; the random effects for the time were all estimated to be zero.

The linear mixed-effects model provided an explanation power of 38.17% in terms of percentage of the deviance explained. Valence was estimated to be 0.04 and dominance 0.55, in terms of *z-scores*. The percentage of the deviance explained by the two effects was 13.81 for valence and 22.42 for dominance, which is roughly the estimation power of the whole model.

## Lessons learned

There are several lessons learned that we would like to share.

First, the experiment design presented in the study is not suitable for continuous application in the industry. It would be counterproductive to ask software developers to self-assess their affects and their productivity each 10 minutes during their entire working days. The measurement interval employed in this study was the result of a pilot test where the participants agreed that an interval of 10 minutes is suitable for a single session of the experiment only. Future studies should aim to find suitable measurement intervals for longer sessions, e.g., the duration of an iteration of the software development lifecycle.

Second, it is important to explain clearly how the experimental task works to the participants. More importantly, some time should be spent in explaining them how the measurement instrument works. Appendix B reports clear instructions, which were derived from the literature in psychology research. The appendix reports that a paper-version of the survey should be available for enabling discussions and questions about the survey itself. The study showed that the participants do not understand the experiment aims, nor do they get that affects are measured. The authors of this article encourage re-use of the online appendix for organizing similar research. Finally, the authors encourage to perform pre- and post-task interviews with the participants, and to be physically present during the programming task and to keep a research diary.

TABLE 5.2: *Fixed effects estimation*

| Fixed Effect | Value | Sum of square | F-value | Upper p-value (64 d.f.) | Lower p-value (48 d.f.) | Deviance Explained (%) |
|---|---|---|---|---|---|---|
| valence | 0.04* | 8.65 | 22.09 | 0.000 | 0.000 | 13.81 |
| arousal | 0.07 | 0.19 | 0.49 | 0.487 | 0.489 | 0.30 |
| dominance | 0.55* | 14.04 | 35.86 | 0.000 | 0.000 | 22.42 |
| time | -0.03 | 0.09 | 0.24 | 0.626 | 0.626 | 0.15 |
| valence:time | 0.04 | 0.49 | 1.26 | 0.266 | 0.267 | 0.79 |
| arousal:time | -0.03 | 0.40 | 1.03 | 0.313 | 0.315 | 0.65 |
| dominance:time | 0.01 | 0.03 | 0.08 | 0.785 | 0.785 | 0.05 |

TABLE 5.3: *Random effects estimation*

| Random Effect | P1 | P2 | P3 | P4 | P5 | P6 | P7 | P8 |
|---|---|---|---|---|---|---|---|---|
| (1 | participant) | 0.01 | 0.15 | 0.04 | -0.02 | -0.20 | 0.00 | 0.00 | 0.02 |
| (0 + time | participant) | 0.00 | 0.00 | 0.00 | 0.00 | 0.00 | 0.00 | 0.00 | 0.00 |

Some lessons learned about linear mixed-effects models are shared, as well. These models are recent, and there is still remarkable discussion on how to employ them. An estimation of statistical significance for mixed effects models is possible via different methods, which will hold different numerical values but the same magnitude. At least in *R*, there is an older implementation of linear mixed-effects models called *nlme.lme*, which provides statistical significance out-of-the-box. However, *nlme.lme* does not handle unbalanced designs. The package has been superseded by *lme4.lmer*, which, on the other hand, does not provide *p-values* out-of-the-box. Likelihood ratio test is a good means to obtain significance testing for *lme4.lmer* model components. However, it should be employed for simple models with few parameters plus their interactions. The method provided by *LMERConvenienceFunctions* (Tremblay and Ransijn, 2013) is straightforward and should be preferable for more complex models. On the other hand, likelihood ratio tests are still useful for comparing two models in terms of fitting, for example by adding a single effect. When comparing two different models in terms of likelihood ratio test (*R anova* function), it is important to vary one effect only; attention should be given to random effects, as they should be kept consistent in the whole analysis.

Additionally, when comparing different models with likelihood ratio tests (*anova* function in *R*), it is important to set the *REML* parameter to *F* (false), to indicate that the maximum likelihood method (*ML*) is preferred to the restricted maximum likelihood method (*REML*). Despite the fact that *REML* estimates of standard deviations for the random effects are less biased than corresponding *ML* estimations (Bolker *et al.*, 2009) models with *REML* estimation do not work when comparing models using the likelihood ratio test (Winter, 2013).

Lastly, the article reported the full model 5.1 of which only two fixed effects were found to be significant (valence and dominance). The full model was kept as the final one because it shows a good example of fitting linear mixed-effects models. The model in 5.3 could have been the final output for this article, as it only reports the significant fixed effects and the random effects.

$$productivity \sim (valence + dominance) * time + (1|participant) + (0 + time|participant)$$
$$(5.3)$$

However, while choosing to report a reduced, significant model, it might be useful to check how much it differs from the full model initially chosen. In the case of this study, there was no significant difference between the reduced significant model and the initial full model (*anova* in *R*; $\chi^2(5)$=3.09, *p*>0.05). There was no advantage in reporting a reduced model.

A summary of the research output of this study is reported in Section 7.2. The discussion of the results obtained by this study are provided in Section 8.2. The implications for research brought by this study are available in Section 8.5.1. The implications for practice of this study are explained in Section 8.5.2.

The limitations of this study, and how they have been mitigated, are reported below as they are specific to this study only. The general limitations of this dissertation are offered in section 9.2.

## 5.3  Limitations

The limitations of the study in Paper V; Paper VI, in Chapter 5, have been reported while following the classification provided by Wohlin *et al.* (2012) as the article was written by following the guidelines for reporting experiments in software engineering.

*Conclusion validity* threats occur when the experimenters draw inaccurate inference from the data because of inadequate statistical tests. The employed linear mixed-effects models are robust to violations suffered by *ANOVA* methods caused by the multiple dependencies of data (see section 2.1). One threat lies in the limited number of participants (8) who worked for about 90 minutes each. However, the background and skills in the sample were balanced. Due to the peculiarity of the repeated measurements and the analysis method, all 72 measurements are valuable. The linear mixed-effects model is capable of addressing the variability due to individual differences and time effects: the obtained statistical results possessed degrees of freedom between 48 and 64, and F-values above the value of 20 when significance has been reached. It has been shown that repeated measures designs do not require more than seven measurements per individual (Vickers, 2003). Two more measurements have been added in order to be able to remove possible invalid data. Lastly, three hypotheses were tested on the same dataset.

While *p-value* adjustments techniques seem, to the authors' knowledge, not suitable for linear mixed-effects models, it can still be reported that the adjusted .05 *p-value* for this study would be 0.05 / 3 = 0.016, while the adjusted .01 *p-value* would be 0.01 / 3 = 0.003. As indicated in Table 5.2, the *p-values* obtained in this study are less than 0.001 (they are actually even less than 0.0001). Therefore, the results of this study have been obtained with an adjusted *p<0.01* (Bonferroni correction).

*Internal validity* threats are experimental issues that threaten the researcher's ability to draw inference from the data. Although the experiment was performed in natural settings, the fact the individuals were observed and the lack of knowledge about the experiment contents mitigated social threats to internal validity. A series of pilot studies with the measurement instrument showed that the minimum period to interrupt the participants was about 10 minutes if the case study was focused on a single task instead of longer periods of observations. The designed mitigation measures against internal validity threats were further confirmed through the post-interview data, which showed that the experiment design did not disturb nor negatively influence the productivity and the performance of software developers. However, more data is needed on how to extend the experimental session and the measurement intervals.

*Construct validity* refers to issues with the relation between theory and observation. A construct validity threat might come from the use of the self-assessed productivity. Given the difficulty in using traditional software metrics (the project, the task, and the programming language were random in this study), and that measuring software productivity is still an open problem, self-assessed performance is commonly employed in psychology studies (Beal *et al.*, 2005; Fisher and Noble, 2004; Zelenski *et al.*, 2008). Additionally, self-assessed performance is consistent (yet not preferred) to objective measurements of performance (Dess and Robinson, 1984; Miner and Glomb, 2010). There is also the evidence that bias is not introduced by mood in self-reports of performance when individuals alone are being observed Beal *et al.* (2005); Miner and Glomb (2010). The researchers carefully observed the participants during the programming task, and this further reduced the risk of bias. Post-task interviews included questions on their choices for productivity levels, which resulted in remarkably honest and reliable answers, as expected. The discussions at the post-interviews with the participants on their self-assessed productivity values reinforced the validity of the data.

*External validity* threats are issues related to improper inferences from the sample data to other persons, settings, and situations. The four professional developers and the four students were sampled using a convenient sampling method. Although this method limits the generalizability of the study, the research is applicable to any software development role with any programming language. Future studies should focus on restricted programming languages, project types, and programmers' experiences. Despite that half of the participants were students, it has been argued that students are the next generation of software professionals; they are remarkably close to the interested population if not even more updated on new technologies (Kitchenham *et al.*, 2002; Tichy, 2000). Secondly, it can be questioned why the present authors studied software developers working alone on their project. People working in a group interact and trigger a complex, powerful network of affects (Barsade and Gibson, 1998). Thus, to better control the measurements, individuals working alone have been chosen. However, no participant was forced to limit social connections while working, and the experiment took place in natural settings.

## 5.4   Chapter summary

This chapter reported a repeated measures research on the correlation of the affects of software developers and their self-assessed productivity. Eight developers working on their individual projects have been observed. Their affects and their self-assessed productivity were measured on intervals of ten minutes. A linear mixed-effects model was proposed in order to estimate the value of the correlation of the affects of valence, arousal, and dominance, as well as the productivity of developers. Valence was significantly positively correlated with the self-assessed productivity, and the correlation coefficient was estimated to 0.04 (standard deviations). Dominance was significantly positively correlated with the self-assessed productivity, and the correlation coefficient was estimated to 0.55 (standard deviations). The model was able to express about the 38% of the deviance of the self-assessed productivity. Valence and dominance, or the attractiveness perceived towards the development task and the perception of possessing adequate skills, were able to provide nearly all the explanation power of the model.

# Chapter 6

# An explanatory theory of the impact of affects on programming performance [1]

This chapter presents the last of the three main empirical studies, namely an interpretive phenomenological analysis, which was conducted toward a theory of affects and performance in software development. It is the study comprising Phase III of this PhD's research activities. This chapter is a contribution for answering RQ4—*How do software developers experience affects and their impact on their development performance?*.

We conducted an interpretive study of the software development performance through the affects of developers. By deeply observing and open interviewing two developers during a development cycle of about six weeks, for a total of 17 meetings, 657 hours of interviews, and uncounted direct observation for a total of 917 qualitative codes, we constructed an explanatory theory, called *Type II* theory by Gregor (2006), for explaining the impact of affects on development performance.

To our knowledge, the study in this chapter has been the first software engineering research to develop a process based theory of the relationship of the affects of software developers and their impact of the development performance. The results of the study are, to a certain degree, novel to the field of psychology as well and contribute to the bigger body of knowledge of affects and performance on the job.

In section 6.1, the chapter provides the theoretical framing of the reported study, which has its foundations in psychology research. In section 6.2, the chapter extends deeply the description of the research design which was offered in Section 3.5. First, the design of the study is explicated. Second, the data analysis technique is provided. Third, the employed measures to raise the reliability of the gathered data are discussed. Section 6.3 provides the results of the study, including the participants characteristics and the components of the constructed theory. The components of the theory and the relationships among the components are discussed and supported by interview snippets. While a brief discussion of the obtained results is provided as they are presented, in line with much qualitative research, the overall discussions of the results, including the implications for research and for practice, and the conclusions of this study are presented in the related chapters 8 and 9.

## 6.1 Theoretical framework

Our theoretical framework was primarily based upon the Affective Events Theory (AET) by Weiss and Cropanzano (1996) and the episodic process model of performance episodes by Beal *et al.* (2005). AET has been developed as a high-level structure to guide research on how affects influence job satisfaction and job-related performance.

In AET, the work environment settings (e.g., the workplace, the salary, promotion opportunities, etc.) mediate work events that cause affective reactions, which are interpreted according to the individuals' disposition. Affective reactions then influence work-related behaviors. Work-related behaviors are divided into affect-driven behaviors and judgment-driven behaviors. Affect-driven behaviors are behaviors, decisions, and judgments that have immediate consequences of being in particular emotions and moods. On example could be overreacting to a criticism. Judgment-driven behaviors are driven by the more enduring work attitudes about the job and the organization (Weiss and Beal, 2005). Examples are absenteeism and leaving.

As Weiss and Beal (2005) noted ten years after publishing AET, AET has often been erroneously employed as a theoretical model to explain affective experiences at work.

However, AET is a *macrostructure* for understanding affects, job satisfaction in the workplace, and to guide future research on what are their causes, consequences, and explanations. More specifically, AET is not a framework to explain the performance on the job, neither is it a model to explain the impact of all affects on job-related behaviors.

In their conceptual paper, Beal *et al.* (2005) provided a model that links the experiencing of affects to individual performance. Beal *et al.* (2005) model is centered around the conceptualization of performance episodes, which relies on self-regulation of attention regarding the on-task focus and the off-task focus. The cognitive resources towards the focus switch is limited. Affects, according to Beal *et al.* (2005), hinder the on-task performance regardless of them being positive or negative. The reason is that affective experiences create cognitive demand. Therefore, affective experiences, according to this model, influence the resource allocation towards off-task demand.

### 6.1.1 Theory construction and representation

Interpretive research is often conducted when producing theories for explaining phenomena (Klein and Myers, 1999b). Gregor (2006) examined the structural nature of theories in information systems research. Gregor (2006) proposed a taxonomy to classify theories with respect to how they address the four central goals of analysis and description, explanation, prediction, and prescription. We employed the widely established Gregor (2006) work as a framework for classifying and expressing our proposed theory.

A *type II*—or explanation—theory provides explanations but does not aim to predict with any precision. The structural components of a Type II theory are (1) the means of representation—e.g., words, diagrams, graphics, (2) the constructs—i.e., the phenomena of interests, (3) the statements of relationships—i.e., showing the relationships between the constructs, (4) the scope—the degree of generality of the statements of relationships (e.g., some, many, all, never) and statements of boundaries, and (5) the causal explanations which are usually included in the statements of relationship. While conducting this study, we ensured the constructed theory was composed of these elements.

Our study attempts to broaden our understanding of topics that are novel and unexplored in our field. Rindova (2008) warned us that "novelty, however, comes at a cost:

novel things are harder to understand and, especially, to appreciate" (p. 300). Therefore, we have to proceed carefully in the theory building process. The risk is to get lost in complex interrelated constructs in a confused and confusing field of study (Ortony *et al.*, 1990) brought in the complicated, creative domain that is software engineering. Furthermore, Barsade and Gibson (1998) advised researchers that, when understanding emotion dynamics, the bigger is the team under observation, the more complex and complicated are the team dynamics. Bigger teams have complicated, and even historical, reasons that are harder to grasp—triggering a complex, powerful network of affects (Barsade and Gibson, 1998). Therefore, there is the need to keep the phenomenon under study as simple as possible. For novel theory development, philosophers and economists often—but not always—draw from their own personal observation and reasoning, while still being able to offer a sound empirical basis (Yeager, 2011). Theorizing from the ivory tower can complement the scientific method by offering insights and discovering necessary truths (Yeager, 2011), to be further expanded by empirical research. Our empirical stance makes us eager to jump to data and start theorizing; yet, we need to take some precautionary measures before doing this.

When novel theories are to be developed in new domains, such as software engineering, a small sample should be considered (Järvinen, 2012). A small sample enables the development of an in-depth understanding of the new phenomena under study (Järvinen, 2012) and to avoid isolation in the ivory tower. Our research follows carefully Järvinen (2012) recommendations, which is reflected in our study design. Weick (1995) classic article is of the same stance by reporting that organizational study theories are approximations of complex interrelated constructs of human nature that often have small samples. Those works are often seen as substitutes of theory studies, but they often represent "struggles in which people intentionally inch toward stronger theories" (ibid, p. 1). Such struggles are needed when a phenomenon is too complex to be captured in detail (Weick, 1995). These issues were taken into account when we designed our study, which is demonstrated in the following section.

## 6.2 Research design

We describe our research as a qualitative interpretive study, which was based on face-to-face open-ended interviews, in-field observations, and e-mail exchanges. Given the aim of the study, there was the need to make sense of the developers' perceptions, experiences, interpretations, and feelings. We wanted to conduct open-ended interviews where the realities constructed by the participants are analyzed and reconstructed by the researcher.

Our epistemological stance for understanding these social constructs and interactions has been interpretivism, which we make coincide with social constructivism in line with other authors (Easterbrook *et al.*, 2008). Interpretive data analysis has been defined succinctly by Geertz (1973) as "really our own constructions of other people's constructions of what they and their compatriots are up to" (p. 9). Interpretivism is now established in information systems research (Walsham, 2006), but we see it still emerging in software engineering research.

As per our chosen design, the participants could be free to undergo the development of the system in any way, method, practice, and process they wished to employ. Our study comprised of regular scheduled face-to-face meetings with recorded interviews, impromptu meetings which could be called for by the participants themselves, e-mail exchanges, in-field observations, and a very short questionnaire right after each commit in the git system (explained in section *Reliability*). Therefore, the participants had to be aware of the design itself, although they were not informed about the aims of the study.

The participants' native language is Italian, but they have been certified as proficient English speakers. The first author of the present article employs Italian as first language, as well, and he was the reference person for the participants for the duration of the entire study. The other two authors of the present article have been certified as proficient and upper intermediate in Italian. The choice for the design of the study was therefore to conduct the interviews in Italian, as the native language let the participants express their opinion and feelings in the richest, unfiltered way (van Nes *et al.*, 2010). The interviews were subsequently transcribed in English as suggested by the common research practices (van Nes *et al.*, 2010; Squires, 2009), but the present case had the added value that the authors could validate the transcripts with the participants over the course of the study, given their advanced proficiency in English.

The in-field observations were performed by two of the present authors, and the personal communications such as e-mails or some impromptu meetings were exchanged between the first author of the study and the participants. The coding activities have been a collaborative effort among all the authors of this study.

In order to keep the study design and results as simple as possible and to provide precise answers to the research question, in line with what we stated in the section *Theory construction and representation*, we observed activities that produced code. Other artifacts such as requirements and design were not taken into consideration. Furthermore, our strategy to limit the complex network of triggered affects was to group and study them into the two well-known dimensions of positive and negative affects (Watson *et al.*, 1988b), which assign the affects—including those perceived as neutral—in a continuum within the two dimensions.

Our design took into account ethical issues, starting with a written consent to be obtained before starting any research activity. The consent form informed the participants of our study in terms of our presence, activities, data recordings, anonymity and data protection, and that their voluntary participation could be interrupted at any time without consequences. They were also informed that any report of the study had to be approved by them in terms of their privacy, dignity protection, and data reliability before it is disclosed to any third party. Furthermore, as an extra measure, any additional, personal data coming from e-mail exchanges and some impromptu meetings with a single author was approved by the participants before inclusion to the study data.

### 6.2.1 Data analysis

Grounded theory has been indicated to study human behavior (Easterbrook *et al.*, 2008), and it is suitable when the research has an explanatory and process-oriented focus (Eisenhardt, 1989). Qualitative data analysis techniques from grounded theory responded to our needs (Langley, 1999). We are aware that there has been some heated debate regarding which, between Glaser and Strauss (1967) or Corbin and Strauss (2008), is *the* grounded theory qualitative strategy (Creswell, 2009) or if it can be employed merely as a tool to analyze qualitative data (Kasurinen *et al.*, 2013). Heath and Cowley (2004) comparison study concludes that researchers should stop debating about grounded theory, select the method that best suits their cognitive style, and start doing research.

We agree with them and adopted Charmaz (2006) social constructivist grounded theory approach as a tool to analyze qualitative data coming from face-to-face open-ended interviews, in-field observations, and e-mail exchanges.

The adaption of grounded theory by Charmaz (2006) has merged and unified the major coding techniques into four major phases of coding, which are initial coding, focused coding, axial coding, and theoretical coding. The four coding phases have been adopted in the data analysis process of this study. Charmaz (2006) has often reminded her readers that no author on grounded theory methodology has ever really offered criteria for establishing what we should accept as a coding family, and that the coding phases are often overlapping, iterative and not strictly sequential within each iteration. This is true also for this study. An exemplar case of our coding activities is shown in Figure 6.1. The figure is divided into four columns. The first column provides an interview excerpt. The remaining columns show the intermediate results of the coding activities.

The *initial coding* phase should stick closely to the data instead of interpreting the data. The researchers should try to see the actions in each segment of data, and to avoid applying pre-existing categories to it. Therefore, Charmaz (2006) has suggested to code the data on a line-by-line approach so that the context is isolated as much as possible, and to code the data as actions. In order to help focusing on the data as actions, it has been suggested to use gerunds. For example, in Figure 6.1 the second column shows the initial codes assigned to a interview snippet.

The second coding phase is the *focused coding*. Focused code means that the most significant or frequent (or both) codes which appeared in the initial coding are employed to sift through larger amounts of data, like paragraphs, speeches, and incidents. This phase is about deciding which initial codes make the most analytic sense for categorizing the data. However, it is also possible to create umbrella codes as substitutes for other codes. During focused coding, the codes become more directed, selective, and conceptual. For example, as shown in Figure 6.1, the initial code "Improving productivity through the use of ST" was further abstracted as "Improving productivity through a tool".

The third coding phase is the *axial coding*. The axial coding phase has been proposed by Strauss and Corbin (1994). As synthesized by Charmaz (2006), the axial coding process follows the development of major categories, relates categories to subcategories, and relates them with each others. If during initial and focused coding the data is fractured into pieces, the axial coding phase brings the data back together again. In this phase, the properties and the dimensions of a category are specified. The fourth column of Figure 6.1 shows an iteration of axial coding.

| Interview snippet | Initial coding | Focused coding | Axial coding |
|---|---|---|---|
| [Interviewer: "Do you think that Sublime Text is enhancing your productivity then?"]<br><br>P2: "Absolutely. I was extremely excited by these features and they pushed me to do more and more." | Improving productivity through the use of ST; being motivated by ST to do more work; | Improving productivity through a tool; Feeling gratitude towards a tool; feeling motivated because of a tool | PERFORMANCE_positive; EVENT_using_useful_tool; AFFECT_gratitude; AFFECT_motivated; |
| [Interviewer: "Were you actually thinking about this while you were working?"]<br><br>P2: "Definitely. First, I turned the monitor towards P1 and showed him the magic. But I felt good for the rest of the day, and I accomplished more than what I hoped I could do." | Thinking about the improved performance brought by a tool; showing the features of a tool to a team mate; Feeling good during a workday because of tool functionality; accomplishing more than what planned; | Realizing positive performance; Sharing information; Feeling strongly good; Progressing strongly good on goal; | ATTRACTOR_good; PERFORMANCE_positive; FOCUS_positive; GOAL_progressing; |

FIGURE 6.1: *Example of coding phases for this study*

The fourth coding phase is the *theoretical coding*. Theoretical coding was introduced by Glaser (1978). As synthesized by Charmaz (2006), the theoretical coding phase specifies how the codes from the previous phases related to each other as hypotheses to be integrated into a theory.

It would be impractical to show the steps and complete examples of axial and theoretical coding as they would need several interview excerpts and resulting codes (Charmaz, 2006). What we could demonstrate in Figure 6.1 was that the interview excerpt was further coded in the later coding phases and became part of the evidence to support the key concepts, such as affect, and their components as shown in the fourth column. The overlapping of different categories over the same snippets indicated the potential linkage among them, which became the basis to develop the model proposed in this study.

### 6.2.2　Reliability

Here, we describe our procedures for enhancing the reliability of the gathered data and the results. The data was gathered using multiple sources. Each interview was accompanied by handwritten notes, recordings, and related subsequent transcriptions. All in-field observations were accompanied by audio recordings after obtaining permission of the participants. We wrote memos during the study. The transcriptions and the coding phases were conducted using *Atlas.ti 7.5*, which is a recognized instrument for such tasks.

In order to make the participants focus on their affects and recall how they felt during performance episodes, we asked them to fill out a very short questionnaire at each git commit. The questionnaire was the Self-Assessment Manikin (Bradley and Lang, 1994), which is a validated pictorial questionnaire to assess affects. We employed the questionnaire in a previous study (Graziotin *et al.*, 2015a) as it proved to be quick (three mouse clicks for completing one) and not invasive. We employed the gathered data to triangulate the observational data and the interview data during each interview. If there was disagreement between the qualitative data (e.g., several positive affective episodes but negative quantitative results), we asked for further clarification from the participants to solve the discrepancies.

As a further action to enhance reliability, but also ethicality of the study, we asked the participants to individually review the present paper in three different times. The first review session happened in the initial drafts of the paper when we solely laid down the results of the study. The second review session happened right before submitting the article. The third review session happened before submitting a revised version of the present article. For the reviews, we asked the participants to evaluate the results in terms of their own understanding of the phenomena under study and the protection of their identity and dignity. Because of their valuable help, the proposed theory is shared with them and further validated by them.

## 6.3 Results

The study was set in the context of a Web- and mobile-based health-care information systems development between July and September 2014. Two software developers, who were conducting a semester-long real-world project as a requirement for their BSc theses in Computer Science, were put in a company-like environment. Both developers, who we shall call *P1* and *P2* for anonymity reasons, were male. P1 was 22 years old and P2 was 26 years old. They both had about five years of experience developing Web and mobile systems. P1 and P2 had their own spacious office serving as an open space, their own desks and monitors, a fast Internet connection, flip-charts, a fridge, vending machines, and 24/7 access to the building. The developers accepted to work full time on the project as their sole activity. They were instructed to act as if they were in their own software company. Indeed, the developers were exposed to real-world customers and settings. The customers were the head of a hospital department, a nurse responsible for the project, and the entire nursing department. The development cycle began with a first meeting with the customer, and it ended with the delivery of a featureful first version of the working software.

It is beneficial to the reader to provide a brief summary of the main events, which have been extracted from our in-field memos. During the first week, P1 had to work on the project without P2. P2 failed to show up at work. During the first days, P2 gave brief explanations about the absence, e.g., housework or sickness. However, the explanations stopped quickly, and P2 stopped answering to text messages and phone calls. At the beginning of the second week, P2 showed up at work. P2 had some private issues, which brought some existential crisis. P1 was initially reluctant to welcome P2 in the development, as all the code so far was P1's creation. The first two days of collaboration brought some tension between the team members, crippled experimentation with the code, and a shared loss of project vision. On the third day of the second week, the team tensions exploded in a verbal fight regarding the data structures to be adopted. At that point, one of the present authors was involved in the discussion. The researcher invited the participants to express their opinion and acted as mediator. A decision was eventually made. The initial tensions between the developers began to vanish, and the work resumed at a fair pace. At the end of the second week, P1 and P2 had a further requirements elicitation session with the customer represented by the head nurse. The development appeared to be back at full speed, and a full reconciliation could be observed between the participants. The progresses succeeded one day after another, and the fully working prototype was demoed and tested during the sixth week.

Face-to-face open-ended interviews happened at the beginning of the project during 11 scheduled meetings and 5 impromptu shorter meetings called by the researchers or by the participants. The impromptu meetings were held mostly because of trivial issues, like casual chatting which turned into a proper interview. Only in one case an impromptu meeting was called by P2 when he finally came back to work. We also did not distinguish between the data coming from the scheduled meetings and the impromptu meetings. The interviews were open-ended and unstructured, but they all began with the question *How do you feel?*. In-field observations happened on an almost daily basis. The participants were informed if they were recorded. We recorded a total of 657 minutes of interviews. Finally, data was gathered via the exchange of thirteen emails.

The transcripts of the interviews were completed immediately after the interviews were concluded. The initial coding phase produced 917 unique codes. The focused coding phase was focused on the individual's experiences of the development process, and it produced 308 codes. Figure 6.1 provides an example of our coding activities. The axial coding and theoretical coding produced six themes, which are explained in this section. Inconsistencies between the qualitative data and the data from the Self-Assessment Manikin questionnaire happened three times during the entire study. All three discrepancies were minor, and they were immediately solved upon clarification from the participants. For example, in one case the participant P1 reported low values of valence and arousal, and a neutral value for dominance. During the interview, P1 often stated that he had a frustrating day, but there were no mentions of low-arousal negative affects. When asked to explain how the Self-Assessment Manikin values were representative of the work day, the participant added that he felt low esteem, which was caused by episodes of frustration. Overall, P1 was unexcited and lost over the day; thus the reported low value for arousal.

This section provides the proposed theory. The theory is represented in Figure 6.2. We describe the discovered themes and categories (boxes) and their relationships (arrows). While Type II theories are not expected to discuss causal explanations in terms of direction and magnitude (Gregor, 2006), we offer them as they were interpreted from the data. Each relationship is accompanied by a verb, which describes the nature of the relationship. Where possible, we precede the verb with some plus (+) or minus (−) signs. A plus (minus) sign indicates that we theorize a positive (negative) effect of one construct to another. A double plus (double minus) sign indicates that we theorize a strong positive (strong negative) effect of one construct to another with respect to a proposed weaker alternative. The reader should bear in mind that our theorized effects are not to be strongly interpreted quantitatively. That is, a double plus sign is not the double of a single plus sign or an order more of magnitude of a single plus sign. Every entity and relationship is supplied with interview quotes, codes, and related work.

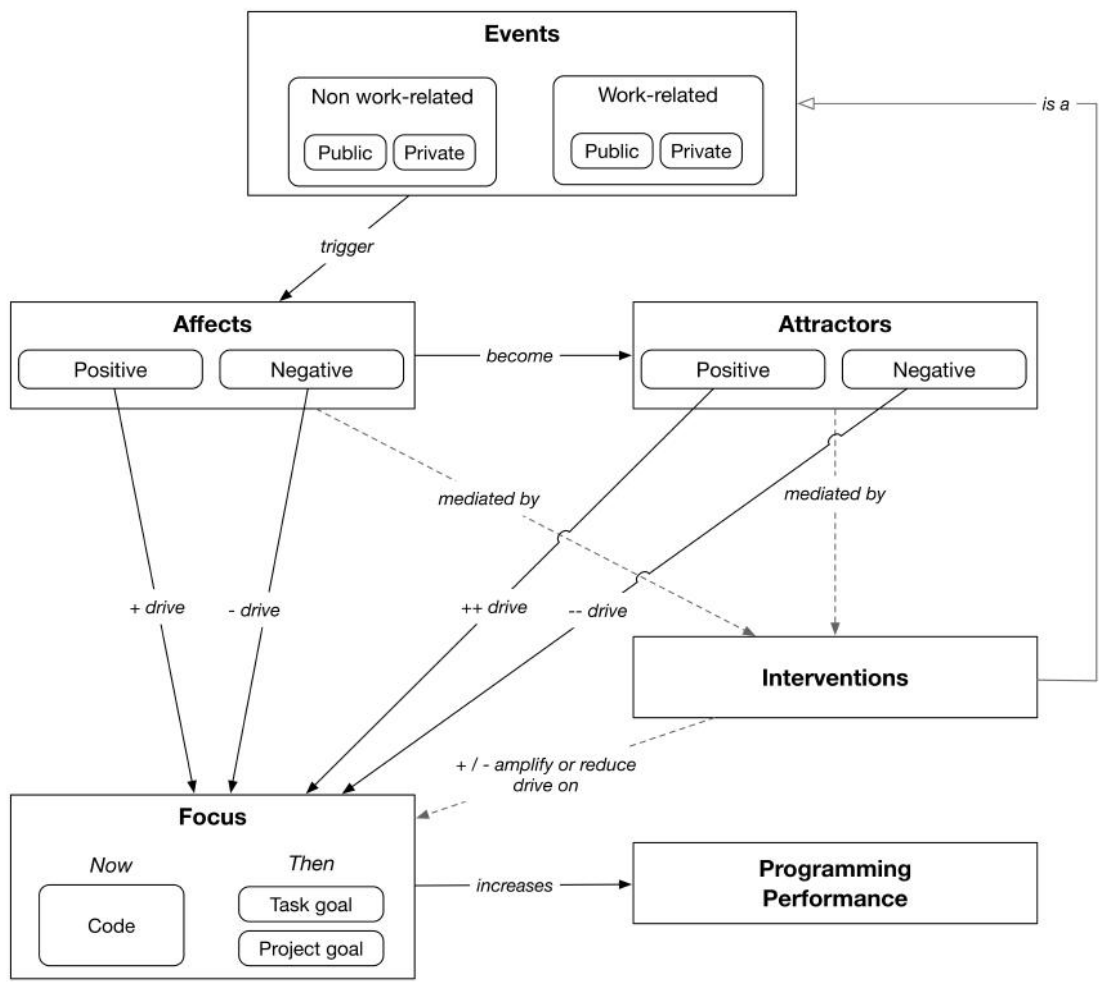

FIGURE 6.2: *A process theory of the relationship of affects and programming performance*

## Events

The *events* are perceived from the developer's point of view as something happening. Events resemble *psychological Objects*, which were defined by Russell (2003) as "the person, condition, thing, or event at which a mental state is directed" (p. 3) but also at which a mental state is attributed or misattributed.

Events may be *non work-related*—e.g., family, friends, house, hobbies—or they may be *work-related*—e.g., the working environment, the tools, and the team members. The interview quotes 1 and 2, and in-field memo 3 are examples of work-related events, while interview quote 4 is not related to work.

1. "*Suddenly, I discovered Google Plus Bootstrap, which is a Bootstrap theme resembling Google+. [I implemented it and] it was easy and looking good.*"—P1

2. "*I found a typo in the name of the key which keeps track of the nurse ID. The bug was preventing a correct visualization of patient-related measurements. Fixing the bug is very satisfying, because I can now see more results on the screen.*"—P2

3. P1, talking to P2 and visibly irritated "Again this? You still have not understood the concept! It is <component name> that is static, while the measurement changes!"

4. "*This morning I received a message with some bad news related to my mother. I immediately desired to abandon development in order to solve the possible issue. The focus was more on that issue than on any other issue at work.*"—P1

We further distinguish public events from private events. *Public events* are those that could be observed by a third person. The in-field memo 3 is an exemplar public event. *Private events* are known to oneself only, even if they are coming from the real world. For example, the event described in interview quote 4 was real and coming from the real world. However, it was not observable by a third person. Events have often an episodic nature, as P1 and P2 noted on several occasions. However, private events can also be reflections, realizations, memories, and situations as with psychological Objects.

5. Interviewer: "*Have you focused better on your programming task today?.*" P2: "*Yes, today went better [than usual]. It's probably..when you do that [programming]*"

> *alone that I am more.. it is more difficult, to write code. When I am working with somebody it goes better, you can work better."*

In the interview quote 5, P2 described the general situation, or a summary of the work day events with respect to usual situations. Situations can be causation chains or aggregation of previous events. The participants do not need to be aware of events as merely events or as situations as it does not make any difference to them. We are not representing situations in Figure 6.2 because we still consider them as events. The rest of the paper provides numerous other examples of events.

**Affects**

During the development process, several *affects* have been triggered by events and felt by the developers. We coded only affects which had been directly mentioned by P1 and P2.

The following are the detected positive and negative affects (respectively) being felt during the development cycle.

*accompanied, accomplished, attracted, contented, dominating, enjoyed, excited, fun, good, gratitude, happy, illuminated, motivated[2], optimistic, positive, satisfied, serene, stimulated, supported, teased, welcomed.*

*angry, anxious, bored, demoralized, demotivated, depressed, devastated, disinterested, dominated, frustrated, guilty, loneliness, lost, negative, pissed off, sad, stagnated, unexcited, unhappy, unsatisfied, unstimulated, unsupported, worried.*

Our qualitative results on the perceived affects agree with the quantitative results of Wrobel (2013); Muller and Fritz (2015), which indicated that developers do feel a very broad range of affects in the software development process.

---

[2] The careful readers might turn up their nose here. As we wrote in (Graziotin *et al.*, 2015c), affects are not motivation, as they are not job satisfaction, etc. Yet, affects are important components of these psychological constructs, and studying complex multifaceted constructs like motivation would require different approaches and different measurement instruments. For this reason, if the participants only stated that they felt motivated or satisfied, we considered them as affects, as it might well be the case that they were expressing emotional judgments about such constructs. In any case, the inclusion or exclusion of such terms as affects would not change the results of this study.

Examples of events that caused positive and negative affects (respectively), coded using the gerund principle of Charmaz (2006) method for analyzing qualitative data, are the following.

*'Feeling contented because a very low number of code changes caused big achievement in terms of quality [or functionality]', 'Feeling gratitude towards a tool', 'Feeling attracted by a junk of code because of anticipating its value for the end user', 'Feeling motivated because personal issues are now out clear', 'Feeling supported because of the brought automation of a framework', 'Feeling serene because of a low workload right after a high workload', 'Feeling happy because of sensing the presence of a team member after reconciliation'.*

*'Feeling alone [or unsupported] while working [or by a team member]', 'Feeling anxious because of a sudden, not localizable bug that ruined the day', 'Feeling anxious by not understanding the code behavior', 'Feeling bored by implementing necessary but too static details [e.g., aesthetic changes instead of functionalities]', 'Feeling frustrated by the different coding style of a team member', 'Feeling angry by failing to integrate [or extend] an external component', 'Feeling stagnated in life [or job, or studies]', 'Feeling unstimulated because of a too analytic task'.*

According to previous research, psychological Objects—sometimes in the form of events, sometimes as stimula—trigger affects all the time, and an individual is under a particular affect or a blend of affects all the time (Russell, 2003). Sometimes, these affects will be perceived strongly. Sometimes, they will not be perceived at all despite their presence. A failure to attribute an affect to an event does not demise the affect itself. This affect misattribution coincides with some theories of moods (Fisher, 2000; Weiss and Cropanzano, 1996), which consider affect as non attributed emotions or simply as free-floating, unattributed affect (Russell, 2003).

### Attractors

We observed that some events had a particular affective meaning to the participants. These affective experiences were assumed high importance to the participants with respect to other affective experiences; thus, we called them *attractors*.

Attractors are affects, which earn importance and priority to a developer's cognitive system. At a very basic instance, they gain the highest possible priority and emphasis to a developer's consciousness, to the point that behaviors associated to the attractor can be observed as it is experienced. An example can be offered by quote 6 below.

6. P2: "*I did a really good job and fixed things also due to Sublime Text (ST)*." Interviewer: "*What has ST done for you?*." P2: "*When you copy/paste code around and refactor, ST offers you at least three different ways for doing search and replace. It is really advanced*." Interviewer: "*Would another tool make a difference to your work instead?*." P2: "*With another editor or an IDE it would be another story, especially if an editor tries to do too much, like Eclipse. I think that the compromise between functionality and usability of ST is way better*." Interviewer: "*Do you think that ST is enhancing your productivity then?*." P2: "*Absolutely. I was extremely excited by these features and they pushed me to do more and more*." Interviewer: "*Were you actually thinking about this while you were working?*." P2: "*Definitely. First, I turned the monitor towards P1 and showed him the magic. But I felt good for the rest of the day, and I accomplished more than what I hoped I could do*."

In interview quote 6, P2 offered an insight regarding the affects triggered by a software development tool. The excitement toward the tool features was an attractor to P2. The attractor became central to the developer subjective conscious experience, not just an underlying affect. Moreover, the behavior caused by the experience of the attractor was directly observable. Interview quote 6 emphasizes that attractors are not necessarily concerns or negative in nature.

Interview quote 4 provides instead an example of a negative attractor. P1 realized that a non work-related event was not desirable, thus generating negative affects. What happened to his mother was important and demanded his attention. P1 was consciously experiencing the negative attractor, and the appraisal of such attractor had consequences to his way of working.

Attractors are not necessarily stronger than general affects for gaining a developer's subjective *conscious* experience. They might just *be there* and still have an impact. We can access them retrospectively. Interview quote 7 is an example of such occurrence.

7. "*I am not progressing.. in the working environment.. with my university career. With life. I feel behind everybody else and I do not progress. And I am not even sure about what I want to do with my life. I got no visual of this.*"—P2

Moreover, interview quote 7 shows that attractors are not always caused by single events. Attractors can become reflections on a series of events as a consequence of them and as a summation of them.

Another example of reflections of a series of events that have however an impact on a developer's subjective consciousness is shown in interview quote 8. P2 was having a life crisis which resulted in a loss of the vision of his own life.

8. "*When I was alone at home, I could not focus on my programming task. The thought of me not progressing with life did often come to my mind. There I realized that I was feeling depressed.*"—P2

In interview quote 8, the participant had a negative *depressed* attractor with the attached meaning *I am not progressing with life*. The rumination associated with this attractor was strong and pervaded P2 personal experience and his everyday life of that period.

Attractors are part of the personal sphere as much as affects are—indeed, they are special affects for us. In the software process improvement literature, the term *concern* has been used as commitment enabler (Abrahamsson, 2001). The commitments are formed in order to satisfy such concerns, i.e., needs (Flores, 1998). Attractors are not concerns as employed by Abrahamsson (2001). An important difference is that concerns are linked to actions, i.e., actions are driven by concerns. On the other hand, attractors are affects, and affects are not necessarily concerns, nor do they necessarily cause immediate actions.

Under our current theoretical framework, a blend of affects constitutes an individual's happiness, at least under the hedonic view of happiness (Haybron, 2001). According to this view, being happy coincides with the frequent experience of pleasure; that is, happiness is reduced to a sequence of experiential episodes (Haybron, 2001). Frequent positive episodes lead to feeling frequent positive affects, and frequent positive affects lead to a positive *affect balance* (Diener *et al.*, 2009a). Lyubomirsky *et al.* (2005) consider a person *happy* if the person's affect balance is mainly positive. However, we have just stated in this section that some developers' affects are more important than other affects. Let us now be more specific.

As argued by the philosopher Haybron (2001), a quantitative view of happiness based solely on frequency of affects is psychologically superficial because some affects do not have distinct episodes or attributions (as in moods). Even more, Haybron (2005) has seen happiness as a matter of a person's affective condition where only *central affects* are concerned. We see a similarity between attractors and Haybron (2005) central affects. As attractors are important affects, we agree that they are a strong constituent of the happiness of the individuals. However, non attractors could be central affects, as well. In our observations, we saw that attractors are also affects that are easily externalized by the participants, and we will show that their originating events are more visible to them. Furthermore, we will show that attractors are more linked to the focus and the developers' performance. Thus, we differentiate them from central affects.

The participants could sometimes realize the affective meaning of attractors by themselves, as in quote 8. There is often the need to externalize them in order for an observer to feel them. We found that sometimes, externalizing affects is alone beneficial, as seen in the next section.

### Interventions

While the presence of researchers has always an influence on the participant's behaviors (Franke and Kaul, 1978), it happened twice that our interaction with the participants had a clear effect on their feelings and behaviors. We call such events *interventions*. Interventions are events—as shown in Figure 6.2 by the UML-like grey arrow with a white arrowhead—that mediate the intensity of already existing negative attractors, thus reducing them as much as possible to normal affects. After externalizing his depressed state in interview quote 8, P2 continued as follows:

9. "*What we were doing was not 'in focus'. The result really didn't matter to me. To my eyes, we were losing time. However, once I've told you what I told you [the personal issues] you know that as well. It is not that I am hiding or that I am inventing things out..I now have no more the possibility to wriggle anymore. I told* *you why I was not there and I am feeling better already. I am now here for two days, and I feel way better than before.*"—P2.

The field memos provided more evidence on the effectiveness of interventions. For example, during the reconciliation, which happened at the beginning of week 2, the developers had frequent soft fights.

> P2 battles fiercely for his opinions and design strategies. However, he is listening to P1 opinions. On the other hand, P1 seems more interested to get stuff done, and he seems less prone to listen to P2. P2 is probably realizing this and responds using passive-aggressive modes. Some not-so-very nice words fly.

> P1 and P2 are less aggressive with each other. My proposal to let them express their opinions and to invite them to listen to each other seems to have a positive effect. A solution, albeit influenced by me, seems to have been reached.

A field memo six days after the reconciliation was much more positive.

> P1 and P2 have been working with an almost stable pace. There does not seem to be an elephant in the room anymore. Both of them smile often and joke with each other. You can feel them happier than before. I often see P1 and P2 showing their results to each other. The work seems way more productive than last week.

Even personal issues were having less impact on P2 as he revealed in a interview nine days after the reconciliation.

10. "*My personal issues are having a minor impact on my productivity, despite the fact that my mind wonders in different places. It is because we are now working well together and share a vision.*"—P2

Interventions in Figure 6.2 are reached by dashed arrows, which start from affects and attractors, and have a dashed arrow pointing to focus. The dashed arrows, together with the labels *mediated by* and *amplify (or reduce) drive on*, indicate alternative paths in the process. That is, affects and attractors are mediated by interventions, which amplify or reduce their drive on the focus.

These interventions suggest that a mediator is a useful figure in a software development team. The mediator should be able to gently push the team member to let out their opinions, views, and affects. A more concrete example could be an agile coach or a team leader Dybå *et al.* (2014) according to the team settings.

## Focus—progressing and goal setting

In this section, we explain the construct of focus, which is related to progressing toward goals and the setting of such goals. The *focus* often referred to a general mental focus, e.g., "*I was in focus after I could refactor all that code using Sublime Text search-and-replace capacity.*"—P2, which usually matched a focus on the current chunk of code. However, the focus on the current chunk of code was with respect to a goal. P2 mentioned focus in interview quote 8, where he told the interviewer that he could not focus on the programming task while at home, because of the realization of being depressed. A more tangible focus on the code at hand was portrayed by P1 in the following interview quote.

11. "*After our [between P1 and P2] reconciliation and after the meeting with [the head nurse], I often developed in full immersion. When I am in full immersion mode, nothing exists except what I am doing. I have a goal in mind and I work toward it. I don't think about anything else but my goal and my progress towards it.*"—P1

During the last interview, P1 was directly asked about the way he focuses while developing software and what he thinks about. Besides the full immersion mode that P1 described in quote 11, he described a "*lighter mode of immersion. I enter this mode when I am tired, when I write less functional aspects of the code.*" but also "*when I am interrupted by negative news or when I focus my attention more on some problems.*".

In quote 12, P2 shared his view on negative affects and how they hinder performance by changing the way he perceived events as attractors.

12. "*My negative thoughts have been the same lately—more or less–but I sometimes change the way I look at them. It is often positive, but it is often negative, too. Maybe I realize this more when I have a negative attitude towards them. It influences my work in a particular way: my concerns become quicksand.*"—P2

Our *focus* appears to be similar to the flow as depicted by Csikszentmihalyi (1997), and found in the related work by Meyer *et al.* (2014); Muller and Fritz (2015), which was described as an attention state of progressing and concentration.

Additionally, the participants often mentioned the term 'vision,' which was meant as the "*ability to conceive what might be attempted or achieved.*" (OED Online, 2015e). For this reason, we preferred using the term *goal setting*. The participants linked the focus and the capacity of setting goals. Goal settings has an established line of research in organizational behavior and psychology—one of the seminal works is by Locke (1968)— that would deserve its own space in a separate article. It involves the development of a plan, which in our case is internalized, designed to guide an individual toward a goal (Clutterbuck, 2010). Those goals found in our study were related to future achievements in the short and long run, i.e., the task and the project. One example of task goals lies in the interview quotes 13. Whenever the focus of attention was on the current code melted with the goal setting of task and project, the performance was reported and observed as positive. However, if something was preventing the focus on the current code—*now*—and the focus on the goal or the goal setting of the task or project—*then*— the performance was reported and observed as negative. P2 summarized these reflections concisely in quote 13.

13. "*It does not matter how much it is actually going well with the code, or how I actually start being focused. Then it [my thoughts about my personal issues] comes back into mind. It is like a mood. I cannot define it in any way. But it is this getting rid of a thought, focusing back to work and the task goal. Here [shows commit message] I wanted to add the deletion of messages in the nurses' log. But when it happens, I lose the task vision. What was I trying to accomplish? WHY was I trying to do this? It happens with the project vision, too. I don't know what I am doing anymore.*"—P2

The project goal setting is similar to the task goal setting. The difference is that project goal setting is the capacity of perceiving the completion of a project in the future and visualizing the final product before its existence as P1 outlined in interview quote 14.

14. "*After we talked to [the head nurse], we gathered so much information that we overlooked or just did not think about. [...] between that and the time you [the researcher] invited us to speak about our issues and mediated among our opinions, we had a new way to see how the project looked like. The product was not there still, but we could see it. It was how the final goal looked like.*"—P1

There is a link between focusing on the code and focusing on the task goal. Staying focused on the code meant staying focused on the *now* (and here). It is the awareness of the meaning of each written line of code towards the completion of a task. Focusing on the task and project goals meant staying focused on the *then* (and there). It was meant as the capacity of envisioning the goal at the shorter term (the task) and the overall goal of the project. At the same time, focusing on the task and the project meant the possibility to define a task completion criteria, the awareness of the distance towards the completion of such task, and to re-define the goal during the work day.

Our findings are in line with those of Meyer *et al.* (2014), where the participants in a survey perceived a productive day as a day where "they complete their tasks, achieve a planned goals or make progress on their goals" (p. 21). The number of closed work items, e.g. tasks and bugs, was the most valued productivity measurement among developers. The *full immersion mode* mentioned by P1 in interview quote 11 resembles the flow depicted by Csikszentmihalyi (1997) and mentioned in the related work by Meyer *et al.* (2014); Muller and Fritz (2015).

### Performance

The performance was generally understood by the participants as their perceived effectiveness in reaching a previously set expectation or goal. Or, whenever *then* became *now*.

15. "*Last week has been chaotic. We worked very little on the code. P2 played around with the programming framework. P2 tried to adapt an example program to fit our* needs. So, P2 studied the chosen framework. I can say that P2 was productive. I spent my time doing refactoring and little enhancements of what was already there. Little functionality was developed so far. In a sense, we still performed well. We did what we were expecting to do. Even if I did so little. I still laid down the basis for working on future aspects. So yeah, I am satisfied."—P1

16. Interviewer: "*What happened during this week?*" P2: '*Well, it happened that..I did not behave correctly in this week. I could not do a single commit.*"

We observed that the affects have an impact on the programming performance of the developers. This is achieved by driving the focus that developers have on the currently focused code, the ongoing task, or the project itself[3]. P2 suggested already, in interview quote 6, that the excitement caused by the discovery of the useful search-and-replace functionalities in his editor had pervaded his work day. This positive attractor caused him to be productive also when not using such functionalities. P2 could also offer cases of the opposite side, like the one in quote 17.

17. "*I was lost in my own issues. My desire to do stuff was vanishing because I felt very depressed. There was not point in what I was currently doing, to the point that I could not realize what I had to do.*"—P2

More precisely, positive affects have a positive impact on the programming performance—as they drive the focus positively—while negative affects have a negative impact on the programming performance—as they drive the focus negatively. While most of the previous quotes are examples on the negative side, quote 6 and the following quote are instances of the positive case.

18. P1: "*I now feel supported and accompanied by P2. We are a proper team.*". Interviewer: "*What has changed?*" P1: "*It's that now P2 is active in the project.*

---

[3]The aim of this study is to offer a theory of the impact of affects on performance while programming rather than proposing a performance or productivity theory. A plethora of factors influence the performance of developers—see (Wagner and Ruhe, 2008; Sampaio *et al.*, 2010) for a comprehensive review of the factors—and affects are one of them, although they are not yet part of any review paper. At the same time, software development performance is composed of several complex interrelated constructs—see (Petersen, 2011) for a review of productivity measurements—to which we add those driven by cognitive processes and *also* influenced by affects, e.g., creativity and analytic problem solving capability (Graziotin *et al.*, 2014b)

> *Before [the reconciliation] P2 was not here at all. [...] If he joined after our meeting with [the head nurse], there was the risk to see him as an impediment instead of a valid resource and team member. Now, I feel happier and more satisfied. We are working very well together and I am actually more focused and productive.*"

A positive focus has a positive effect on programming performance. But, a focus on the code toward a task or project goals (or a combination of them) have an even stronger positive impact on the programming performance.

We provide some codes related to the consequences of positive and negative affects (respectively) while programming.

*'Limiting the switch to personal issues because of feeling accompanied by a team member', 'Switching focus between the task and the positive feelings caused by a tool makes productive', 'Focusing better on code because of the positive feelings brought by reconciliation', 'Focusing less on personal issues [more on the code] because of a sense of being wanted at work', 'Focusing more on code because of feeling supported and in company', 'Committing code frequently if feeling in company of people'.*

*'Abandoning work because of negative feelings fostered by negative events', 'Avoiding coming to work because of lost vision [and depression]', 'Avoiding committing working code during day because of loneliness', 'Choosing an own path because of the loneliness', 'Switching focus between personal issues and work-related task prevents solving programming tasks', 'Losing focus often when feeling alone', 'Losing the project vision because of quicksanding in negative affects', 'Not reacting to team member input because of bad mood', 'Realizing the impediments brought by personal issues when they are the focus of attention', 'Trying to self-regulate affects related to negative events and thoughts lowers performance', 'Underestimating an achievement because of loneliness', 'Worrying continuously about life achievements and avoiding work'.*

A summary of the research output of this study is reported in Section 7.3. The discussion of the results obtained by this study are provided in Section 8.3. The implications for research brought by this study are available in Section 8.5.1. The implications for practice of this study are explained in Section 8.5.2.

The limitations of this study, and how they have been mitigated, are reported below as they are specific to this study only. The general limitations of this dissertation are offered in section 9.2.

## 6.4  Limitations

The most significant limitation of this research to be mentioned lies in its sample. Although it is very common for software engineering studies to recruit computer science students as participants to studies (Salman *et al.*, 2015), for some readers this might still be considered a limitation. First, it is true that our participants were enrolled to a BSc study in computer science, but they both had a working history as freelancers in companies developing websites and Web applications. While our developers did not have to be concerned about assets and salaries, they were paid in credit points and a final award in terms of a BSc thesis project. Tichy (2000); Kitchenham *et al.* (2002) argued that students are the next generation of software professionals as they are close to the interested population of workers, if not even more updated on new technologies. Indeed, the empirical studies comparing students in working settings with professionals did not find evidence for a difference between the groups (Svahnberg *et al.*, 2008; Berander, 2004; Runeson, 2003; Höst *et al.*, 2000; Salman *et al.*, 2015). The conclusions from the previous studies are that students are indeed representatives of professionals in software engineering studies.

The non-inclusion of female participants might be considered a further limitation of this study. There is a widespread popular conception that there are gender differences in emotionality (McRae *et al.*, 2008). Evidence has been found for gender differences at the neural level associated to reappraisal, emotional responding and reward processing (McRae *et al.*, 2008), and for a female having greater reactivity to negative stimuli (Gardener *et al.*, 2013) and adoption of different emotion regulation strategies (Nolen-Hoeksema and Aldao, 2011). While more studies on gender differences are needed as the produced evidence is not enough yet (Nolen-Hoeksema, 2012), it might be the case that the inclusion of a female developer would have made the dataset richer, and perhaps would have led to a more gender-balanced theory.

While we argued extensively about the choice of the sample size in section *Theory Construction and Representation*, we repeat here that there was the need to keep the phenomenon under study as simple as possible given its complex nature (Barsade and Gibson, 1998). Furthermore, when novel theories are to be developed in new domains, such as software engineering, a small sample should be considered (Järvinen, 2012). This strategy, while sometimes seen as limiting, pays off especially for setting out basic building blocks (Weick, 1995). As argued by Bendassolli (2013), even one observation could be sufficient for theorizing as so far as "phenomena should be directly explained by theory, and only indirectly supported by the data" (p. 15). Our choice of the small sample size was seen as a benefit for the purposes of this explanatory investigation. The reason is that in a real company, the source of events is vast and complex. There are team dynamics with complicated, and even historical, reasons that are harder to grasp—triggering a complex, powerful network of affects (Barsade and Gibson, 1998)—thus lifting the study's focus out from the programming itself.

## 6.5   Chapter summary

This chapter reported an interpretive study aimed to broaden our understanding of the psychology of programming in terms of affects perception and their impact while programming. We conducted a qualitative interpretive study based on face-to-face, open-ended interviews, in-field observations, and e-mail exchanges, which enabled us to construct a theory of the impact of affects on software developers with respect to their programming performance. As far as we know, this is the first study to observe and theorize a development process from the point of view of the affects of software developers. By echoing a call for theory building studies in software engineering, we offer first building blocks on the affects of software developers. The theory conceptualization portraits how the entities of events, attractors, affects, focus, goal settings, and performance interact with each other. In particular, we theorized a causal chain between the events and the programming performance, through affects or attractors. Positive (negative) affects have a positive (negative) impact on the programming task performance by acting on the focus on code, and task and project goals. We also provided evidence that fostering positive affects among developers boosts their performance and that the role of a mediator bringing reconciliations among the team members might be necessary for successful projects.

# Chapter 7

# Summary of research output

This chapter summarizes the results of the studies conducted during this PhD's research activities. The chapter is divided into three parts according to the three PhD research phases—as explained in Chapter 3, namely Phase I—*Knowledge acquisition and theory translation*, Phase II—*Variance theory testing*, and Phase III—*Process theory development*. For each phase, the chapter summarizes each paper's output.

## 7.1 Phase I—Knowledge acquisition and theory translation

During the phase *Knowledge acquisition and theory translation*, a comprehensive literature review was conducted. The aim of Phase I was to provide answers to the research question RQ1, namely *What are the theoretical foundations regarding the affects of software developers?*. Most of the output of this phase has been offered in Chapter 2. Paper I; Paper II; Paper III were published for this phase. What follows summarizes the research output and contribution of the published papers.

Paper I provided an initial literature review of the related work in software engineering, which were subsequently extended and reported in Section 2.5. The initial literature review highlighted that there had been a lack of understanding of the affect of software developers. The main contribution of the article, however, was a qualitative inductive analysis of the practitioners' reactions and comments to one of our published articles (Paper IV). Through the analysis of more than 200 comments in English, we have shown that practitioners are deeply interested in their affects while developing software, which causes them to engage in long and interesting discussions when reading related articles. Thus, we concluded that it is important to understand the role of affects in software development processes. The article is partially reproduced in Section 8.5.1

Paper II is a review article with a threefold contribution. First, it described the challenges to conduct proper affect-related studies with psychology. Second, it provided a comprehensive literature review of affect theories, which was mapped to Section 2.1. Third, it proposed guidelines for conducting what we have called psychoempirical software engineering, that is empirical software engineering with theory and measurement from psychology. The last part of the paper is enclosed in Appendix C.

Paper III is a research note, which summarized a part of the literature review. The note presented the common misconceptions of affects when dealing with job satisfaction, motivation, commitment, well-being, and happiness, the validated measurement instruments for affect measurement that were employed during this PhD's research activities, and our recommendations when analyzing the measurements of the affects of software developers, like employing mixed effects models. This article was mapped to Section 2.5.3 and Section 2.1.3.

## 7.2 Phase II—Variance theory testing

The second phase of this PhD's research activities was of variance theory development and testing. The articles Paper IV; Paper V; Paper VI were published as a result of this research phase.

Paper IV is a research article, which reported a quasi-experiment for comparing the performance of software developers—in terms of analytic problem solving and creativity—and their pre-existing affects. The study formulated four research hypotheses. The hypotheses predicted there is a differences in terms of problem solving and creative performance of software developers with respect to their pre-existing affects. Creativity was measured in three different, validated ways. Overall, there was evidence only to support the hypothesis that the happiest software developers were those with the highest analytic problem solving performance scores. Figure 7.1a summarizes the valence based theory. As the experiment was not a controlled experiment, we could claim only a correlation

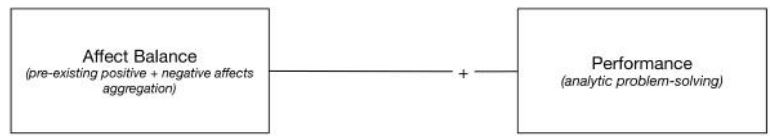

(A) *Variance theory of pre-existing affects and their relationship with problem-solving performance of developers*

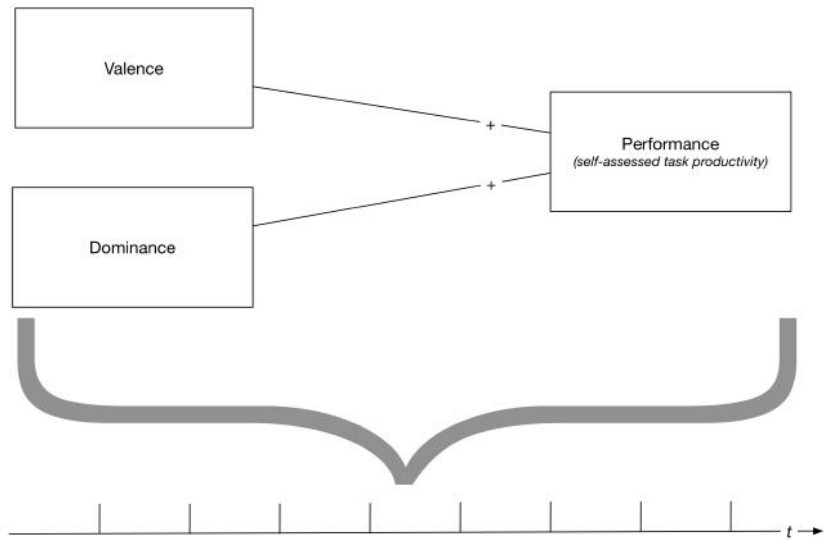

(B) *Variance theory of real-time affects and their relationship with software development performance*

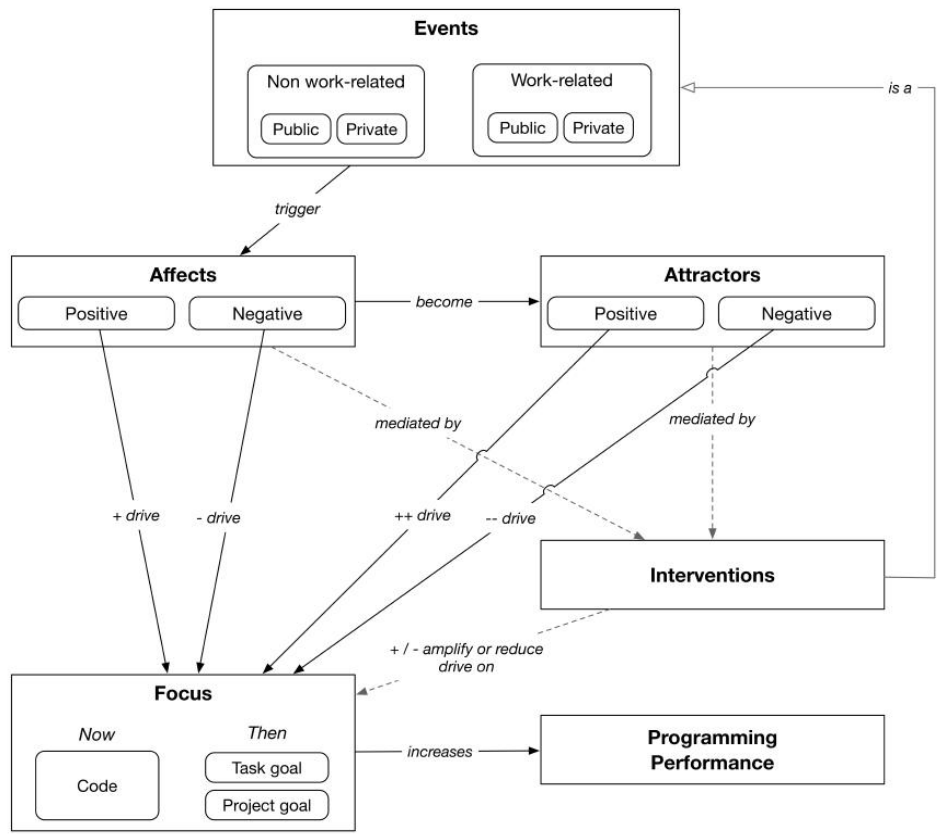

(C) *Process-based explanatory theory of the impact of affects on software development performance (repeated from 6.2)*

FIGURE 7.1: *A theory of affect and software developers' performance.*

and not a causality between the pre-existing affects of the software developers and their analytic problem-solving performance. Therefore, Figure 7.1a shows a line connecting the two constructs and not an arrow. Paper IV has offered the first iteration and facet of the model offered by this PhD research. The model was static, simple, variance based, and it was not able to depict the dynamics offered by studies over time.

The first facet of the theory, which was developed by testing hypotheses from psychology, offered an initial view of the relationship between the constructs. However, it lacked explanations about other two important components, which were the software development task itself and the time dimension. The next phase of the theory development was still variance based, because it derived from a research design from psychology, but it incorporated a dynamic component in it. The study in Paper V; Paper VI was hypothesis-based, like its predecessor. There was a central hypothesis, which predicted that real-time affect of software developers is positively correlated to their self-assessed development task productivity. The hypothesis was broken down into three directly testable hypotheses, which had the dimensions of valence, arousal, and dominance as sub-constructs of affect to be tested. The model was built using a series of linear mixed effects regressions. Overall, the study offered support for a positive correlation between the real-time affect, in terms of valence and dominance, and the self-assessed productivity of developers. The study had at its output the model in Figure 7.1b. As a natural experiment is a quasi-experiment, causality could not be claimed (see Section 3.5.2 for the reasons behind this design choice). So, the lines connecting the constructs of valence and dominance to performance are not arrows. This correlation holds over time, and time itself was not found to be significantly correlated with the self-assessed productivity of the participants.

## 7.3 Phase III—Process theory development

The research performed in Phase II added some understanding of the relationship of the affect of developers and their software development performance. The theory at that point offered a static part, where pre-existing affects and their relationship with performance of developers were studied and represented, and a dynamic part, where the immediate affects of the developers and their relationship with the software development task performance was studied and represented. However, the theory could be improved by qualitative studies towards process based theory. For this reason, the study in Paper VII was conducted.

The interpretive phenomenological analysis had as output the process based theory in Figure 7.1c. The theory, which was framed by the Affective Events Theory (Weiss and Cropanzano, 1996; Weiss and Beal, 2005) from psychology research, has offered explanations—thus, we classified it as an explanatory theory—about the impact of affects on development performance. It took an entire paper to describe the theory, which is an interpretive work, yet evidence-based. The remainder of this section offers a high-level description of the model. The reader is invited to read Chapter 6 for a complete description.

The theory as modeled in 7.1c has the concept of event as its entry point. The *events* are perceived from a developer's point of view as something happening. Events can be work-related and non work-related. Thus, they might be generated by working environment, co-workers, tasks, or from outside working environment, such as family. Under a psychology point of view, events resemble stimuli.

The events generate *affects* to the software developers, which could be either positive or negative, to different degrees.

Affects drive the *focus* on the *now* and the *then* of a software development task, which can be seen as a series of problem setting and problem-solving events. The *now* in the context of our research was the code, but it could be any software-related artifact. The *then* are goals, more specifically the task goal and the project goal. The focus has an important influence on the programming performance, which was generally understood by the participants as the progression from the *now* to the *then*.

We observed that some events had a particular affective meaning to the participants. These affective experiences were more important to the participants with respect to other affective experiences; thus, we called them *attractors*. Attractors are a novel concept introduced by the theory, although we found similarities in philosophy articles talking about central and peripheral affects (Haybron, 2005, 2001). Attractors have the capability of obtaining the attention of the developer and to guide the cognitive activities either positively or negatively. Focus is among those activities.

Finally, we observed that *interventions* were possible to mediate negative affects and attractors generated by events. Interventions reduce the effect that affects (attractors) have on the focus of developers, to the point of inverting the effect.

# Chapter 8

# Discussion

This chapter discusses the results obtained during the execution of research activities for this PhD dissertation, and the implications that these results imply. This PhD research was conceptualized into three phases, which were towards a theory of affects and their relationship with performance while developing software. The phases were Knowledge acquisition and theory translation; Variance theory testing; and Process theory development. The chapter presents the phases, and discusses the obtained results. Subsequently, the chapter discusses on performing research activities that matter to software developers. Finally, the chapter reports the major implications of this work.

## 8.1 Phase I—Knowledge acquisition and theory translation

The first phase of the research activities aimed to present, synthesize, and translate the theoretical foundations of affects, emotions, moods, performance, and their relationship in the context of software development (RQ1). In order to understand these complex constructs and their relationships, an extensive literature review was performed. It is not an easy task to discuss the results of literature review activities, as the results lie in the review itself.

The literature review showed that the software engineering research literature had neglected affects in software processes and methodologies as well as in general research. Psychology, on the other hand, has offered several models, theories, and measurement systems, to the point that sense-making had been required in order to choose proper devices.

The literature review also showed some common misconceptions when dealing with the affects of developers. The identification of the misconceptions are useful for defining the theoretical foundations of affect, because they explicate what affects are *not*. Research activities can build upon the article by designing studies that try to understand the affects of software developers and avoid accidentally studying related constructs. On the other hand, research activities on affect-related constructs can be built upon our results by adopting the understanding of affects achieved in it.

The results of Phase I had been our synthesis of the literature itself, which (1) laid down the basic building blocks for understanding and measuring the affect of software developers through the theories divided into the two competing frameworks of discrete and dimensional theories and the unifying theory of core affect (presented in Paper II), (2) presented the common misconceptions when dealing with the affect of software developers by clarifying that affects are not to be confused with related constructs such as motivation, commitment, and job satisfaction (presented in Paper III), (3) initiated the proposal of "psychoempirical software engineering" and its guidelines in the empirical software engineering sub-discipline. The latter will be discussed in the implications of this dissertation.

## 8.2 Phase II—Variance theory testing

The second phase of the research activities aimed to test some hypotheses derived form the literature review towards a variance based theory of affects and software development performance.

Based on the translation of psychology theory to software engineering, an assumption was made that there is a difference in analytic and creative performance among developers feeling positive affects and those feeling negative affects (RQ2), and that developers feeling positive affects are more productive in their development task than those feeling less positive (or negative) affects (RQ3).

We conducted a first study (also in (Paper IV)) to test a variance theory of affects—measured with the Scale of Positive and Negative Experience (SPANE, (Diener *et al.*, 2009a)) measurement instrument, which conceptualizes affect into the SPANE-B affect balance score—and their relationship with problem-solving performance—measured with validated psychological tasks such as the Tower of London (Shallice, 1982), the Psychology Experiment Building Language (PEBL, (Mueller and Piper, 2014)), a validated creativity generation task (Forgeard, 2011; Kaufman *et al.*, 2007), and the Consensual Assessment Technique for creativity evaluation (Amabile, 1982; Kaufman *et al.*, 2007). The study compared two balanced groups of 42 software developers, which were assigned to the groups based on their pre-existing affects, and their analytic and creative performance.

Our first affect measurement session, with SPANE (Diener *et al.*, 2009a), offered the estimation $\mu_{SPANE-B-DEV} = 7.58$ (95% CI [5.29, 9.85]) for the population's true mean over a possible score in the interval [-24, 24]. That is, it might be that the central value for the SPANE-B for software developers is above seven and significantly different from the central value of the measurement instrument, which is zero. While we further reflect on this in section 4.3, it should be noted that our discussion of the results takes this into account, especially when we compare our results with related work.

The empirical data did not support a difference in creativity with respect to the affective states of software developers in terms of any of the creativity measures we used. The results of this study agree with those of Sowden and Dawson (2011), who did not find a difference in the creativity of the generated ideas with respect to the affective states of the participants. We found no significant difference in the number of creative ideas generated, which is in contrast to Sowden and Dawson (2011), who found that participants in the positive condition produced more solutions than did those in the neutral and negative conditions. Our participants feeling the most positive affects did also generate more solutions with respect to other participants. However, the difference was not found to be statistically significant. Instead, the results of this study deviate from those in the study by Forgeard (2011), where non-depressed participants provided more creative captions under negative affective states. Nevertheless, it must be noted that the depression factor has not been controlled in this study. Overall, the results of this study contrast with past research that places affects–regardless of their polarity and intensity–as important contributors of the creative performance of individuals.

As we reported earlier, the second SPANE session was included for limiting the threats to validity because the first task could provoke a change in the affective states of the participants. During the execution of the creativity task, we observed how the participants enjoyed the task and how happily they committed to the task. This observation was mirrored by the data; the participants generated 220 captions, averaging 5.24 captions per participant. This enjoyment of the first task was reflected by the second SPANE measurement session, as there was a significant increase in the SPANE-B value of 1.02 (t(39) = 3.00, $p < 0.01$, d = 0.96, 95% CI [0.34, 1.71]). This further validates the capabilities of the adopted measurement instrument for the affective state measurements and shows that even simple and short activities may impact the affective states of software developers. The Cronbach's $\alpha$ value of 0.97 of the two SPANE measurement sessions present evidence that the participants provided stable and consistent data. The choice to include a second affective states measurement session in the design of the study is justified by the obtained results.

The empirical data supported a difference in the analytical problem-solving skills of software developers regarding their affective states. More specifically, the results suggest that the happiest software developers are more productive in analytical problem solving performance. The results of this study contrast with the past theoretical contributions indicating that negative affective states foster analytic problem-solving performance (Schwarz and Clore, 2003; Spering *et al.*, 2005; Abele-Brehm, 1992). The results of this study are in contradiction to those obtained by Melton (1995), who observed that individuals feeling positive affects performed significantly worse on a set of syllogisms (i.e., logical and analytical reasoning). Although we adopted rather different tasks, our participants feeling more positive affects performed significantly better than any other participants. Likewise, our results are in contradiction to those of Kaufmann and Vosburg (1997), where the performance on the analytic task was negatively related to anxiety (both trait and state) of the participants. However, there was no significant relationship between either positive or negative mood of the participants and their analytical problem-solving performance. Yet, our results tell that happiest software developers outperformed all the other participants in terms of analytic problem-solving.

We conducted a second study (also in Paper V; Paper VI) which tested a variance theory of affects—measured with the Self-Assessment Maniking (SAM, (Bradley and Lang, 1994)) measurement instrument that conceptualizes affect into the dimensions of valence, arousal, and dominance—and their relationship with self-assessed productivity over time. We studied eight software developers who worked for a real-world project, and assessed the affects raised by the task itself and the task productivity over time. The results were transformed into Z-scores, as explained in Section 2.1.3, and analyzed by fitting a linear mixed effects model, which are solutions that are robust and specifically designed for repeated, within-participant longitudinal data (Laird and Ware, 1982; Gueorguieva and Krystal, 2004; Baayen *et al.*, 2008).

The empirical results obtained in the study supported the hypothesis of a positive correlation between the affective state dimensions of valence and dominance with the self-assessed productivity of software developers. In other words, high happiness with the task and the sensation of having adequate skills are positively correlated with the self-assessed productivity.

No support was found for the correlation between arousal and self-assessed productivity. It is suspected that the participants might have misunderstood the correlation's role in the questionnaire. All participants raised questions about the arousal dimension during the questionnaire explanations. A possible explanation of no significant interactions between the affects dimensions and time is that the participants worked on different, independent projects. In addition, the random effects related to time were estimated to be zero, thus non-existing. The full model is still worth reporting with time as fixed and random effect because future experiments with a group of developers working on the same project will likely have significant interactions with time.

The results of this study are in line with the related work reported in section 2.5. Nevertheless, it must be noted that the results, methods, and context of this study are novel and can be compared with those of other studies only theoretically. The results of this study are in line with those of Fisher and Noble (2004) where real-time, positive affects (expressed with different constructs than those of this study) of different types of workers are found to be positively correlated with their productivity. However, the results of this study are not completely in line with those of Khan *et al.* (2010) which on the debugging written tests: (1) induced high valence condition alone did not coincide with high debugging performance, (2) induced high arousal condition alone did coincide with high debugging performance, and (3) induced high arousal and valence conditions together were associated with high debugging performance. Lastly, to the authors' knowledge, our study is the first research studying the correlates between software developers' performance and their self-assessed productivity in their natural, real-software working environments.

The post-task interviews acted as a triangulation to the quantitative findings. Events that visibly reduced the productivity of the participants were captured by the experiment data and were validated by the participants' post-task interviews. Although they were trained in employing the measurement instrument, the participants did not realize that the study was about the assessment of their affective states. On the other hand, the perceived uncertainty of the participants regarding arousal might threaten the validity of the study. However, there is no evidence that the misunderstanding happened, and the suspicion has been reported in this article for more transparency. Additionally, when they were asked about the factors that influenced their productivity, no participants mentioned affects or any synonym related to affects. This further enhances the reliability of the study.

Although not strictly related to this study, the uncertainty of the participants on how they defined their own productivity measurement while developing software suggests that alternative venues in measuring and defining development performance can be pursued. One alternative proposal is the newly discovered concept of "relative perceived productivity measure" built as incremental relative steps of previous estimations.

Overall, the results of the studies conducted during this phase indicate a clear direction for a positive correlation—if not a positive effect, although we are not in a position for strong argumentation—of the pre-existing affects and the task-related affects of software developers and their software development performance. Therefore, we may argue that happy software developers are more productive, and are more productive with respect to less happy and unhappy ones.

## 8.3 Phase III—Process theory development

The third phase of the research activities aimed to understand how affects are related to software development performance; therefore, the phase's aim was to develop a process theory of affects and software development performance.

We conducted a constructivist interpretive study over a software development cycle based on: face-to-face open-ended interviews, in-field observations, and e-mail exchanges. This approach enabled us to construct a novel a process based explanatory theory of the impact of affects on programming performance (also see Paper VII). The proposed theory builds upon the concepts of events, affects, attractors, focus, goals, and performance, and their linkage.

The proposed theory can be seen as a specialized version of Affective Events Theory (AET, (Weiss and Cropanzano, 1996)). It provides an affect-driven theory explaining how events, both work-related and not, impact the performance of developers through their focus and goal setting while programming. Therefore, our study produces evidence that AET is an effective macrostructure to guide research of affects on the job in the context of software development. At the same time, our proposed theory is reinforced by the existence of AET itself.

We also note that our theory is partially supported in Muller and Fritz (2015) independent study—built upon one of our previous studies (Paper VI)—which was conducted at about the same time of our study in Paper VII [1]. Among their findings, the self-assessed progressing with the task is correlated with the affects of developers; the most negative affects were correlated with less focus on clear goal settings and positive affects were linked with focusing and progressing toward the set goals.

Finally, our findings are in line with the general findings of goal settings research. That is, the task performance is positively influenced by shared, non conflicting goals, provided that there are fair individuals' skills (Locke and Latham, 2006).

The reader might have noted that the proposed theory provides a causal relation between the affects of software developers and their performance, while the studies for Phase II could only claim a positive correlation instead. Let us now explore this tension in the next section.

---

[1] Furthermore, at our submission time the work by Muller and Fritz (2015) had just been publicly accepted for inclusion in ICSE 2015 proceedings, but it is currently still not published formally. We obtained their work through an institutional repository of pre-prints.

**Happy, therefore productive or Productive, therefore happy?**

Considering the causality aspect between affects and performance, we note that the participants have always explicitly stated or suggested that the influence of affects on performance is of a causality type. Some researchers have warned us that there might instead be a correlation between the constructs, as well as a double causality (i.e., *I am more productive because I am more happy, and I am more happy because I am more productive*). Indeed, so far in our previous studies (Graziotin *et al.*, 2014b, 2015a) we have argued for correlation, not causation.

In the study presented in Paper VII, we could not find support in the data for a double causation, but for a causality chain *Happy, therefore productive*, in line also with related research (Wrobel, 2013). However, it seems reasonable that we are happier if we realize our positive performance.

We speculate here that a third, mediating option might exist. In the proposed theory, and in several other theories in psychology, being happy is reduced to frequent feeling of positive affects (Haybron, 2001). As argued by Haybron (2007), the centrality of affects might be relevant, as well. Haybron (2007) stated, as an example, that the pleasure of eating a cracker is not enduring and probably not affecting happiness; therefore, it is considered as peripheral affect. Peripheral affects arguably have smaller—if not unnoticeable—effects on cognitive activities. It might be the case that the positive (negative) affects triggered by being productive (unproductive) do exist but have a small to unnoticeable effect on future productivity. However, this is outside the scope of this study. We report our backed up speculation as causation for future work.

In Chapter 1, we wondered if practitioners would actually care about an added understanding about their affects, and if they would enjoy being studied from this point of view. The next section shows that the answer is strongly positive, to the point of having these studies called for by practitioners.

## 8.4   On research that matters to practitioners [2]

It has been argued that software engineering has to produce knowledge that matters to practitioners (Osterweil *et al.*, 2008).

Building upon (Paper I), we show here that practitioners are deeply interested in their affects while developing software, which causes them to engage in long and interesting discussions when reading related articles. The study in (Paper IV) was published in a *gold open access journal*. In short, gold open access journals are "those whose articles can be found in the peer-reviewed literature and are made freely available on the public Internet without any financial, legal or technical barriers" (Graziotin *et al.*, 2014a)[3].

The article was published in March 2014. It was immediately well received by practitioners via various social media platforms and the general Web. Four weeks after publication, the article had more than 5,000 views, gaining immediate attention from major news outlets all over the world. The article appeared on the most important social outlets for practitioners, including multiple appearances on Slashdot, Reddit, Hacker News, the Chinese Software Developer Network (CSDN), Digg, and Soylent News. Several news articles discussing our article appeared on more than 1,000 social networking shares. Examples include ITWorld (USA), Daily Mail (UK), Communications of the ACM (USA), Dotnetpro.de (Germany), hi-news.ru (Russia), roll.sohu.com (China), irorio.jp (Japan), Developpez.com (France), Quo (Mexico), Voice.fi (Finland), and the Independent Online (South Africa). The article was also linked[4] from the networks of Goldman Sachs, Oracle, Intel, Lexmark, Thomson Reuters, ChemAxon, and Treehouse Island, Inc.

Altmetric.com is a service to track the impact of articles in the social media, the news outlets, and the Web in general. It computes a score for the mentions. As of today, the article has a score of 303, positioning it in the high impact articles—in the top 5% of all articles ever tracked by Altmetric (3,643,699 across all journals in all disciplines). In one year, the article page was seen more than 15000 times by more than 11000 unique visitors.

---

[2] This section is built upon Paper I—Graziotin, D., Wang, X., and Abrahamsson, P. Software Developers, Moods, Emotions, and Performance. *IEEE Software*, 31(4):24–27 (2014c). doi: 10.1109/MS.2014.94.

[3]Please see Graziotin *et al.* (2014a) for a comprehensive review of the open access publishing business models. Our article was immediately accessible by any individual in the world through an Internet connection.

[4]We do not have access to the actual contents of the discussions, but we were able to track the links through the journal referrals.

Within two months of publication, readers had posted 304 comments on various sites, 210 of them in English. This "Voice of Practitioners" is worth listening to, and we did. Many comments stated, "*This is obvious,*" but quite a few stressed the importance of testing the obvious in science, "*Before you test something you don't know, you're guessing. [. . .] Research like this is important!*" Several comments show the practitioners understood why human aspects should have a major role in software engineering studies: "*[Happiness] has a bigger effect on software developers' productivity that it would in non-creative jobs [. . .] When a software developer is demoralized, you can get some truly awful code.*" That's obvious—or is it? Several practitioners correctly understood that "*the study isn't about whether music and a mini-bar make better programmers, it's about whether being happy makes one a better programmer.*"

A high number of developers reported how they were feeling down on their own companies and were "contemplating quitting over mistreatment." Others actually quit their jobs because they felt unhappy and unproductive: "*At that point, the motivating factor was making the project a success for another checkmark in my resume only... so I could leverage that for a position at a new company that would treat their developers better.*" While most of the unhappiness at work was related to interruption, the most surprising response was the dissatisfaction with open spaces and forced communication with other people: "*I got a lot more done in my quiet 6x6 cube than in a 16x32 open plan shared with eight other guys.*" The critical article by Skowronski (2004) supports these comments by criticizing overly strict application of agile without actually focusing on individual needs. Practitioners were willing to share what makes them feel happy at work in the comments—among the top cited being a quiet environment, limits on multitasking, frequent short breaks, and a comfort zone with coffee near the office.

According to practitioners, managers need to understand the uniqueness of people: "*For some people, music and a mini-bar make them happy, for others, it is silence and being free to develop. It comes down to 'When you get what makes you happy, you will work better.*'" In addition, "*The problem is that even if HR or PHB [5] understand this, they may try to apply a one-size-fits-all methodology to engender happiness.*" An experienced software engineer, referring to his work at a well-known semiconductor company, left what we think is an inspiring comment: "*I've never worked so hard, put in more hours, got more stuff done, cranked out more code, etc., as I have in my [Company Name]*

---

[5] Pointy-haired Boss, see http://en.wikipedia.org/wiki/Pointy-haired_Boss

*time. Why? In meetings, my ideas were listened to. I had a ton of freedom in my job to Get Things Done. I was recognized for Stuff I Got Done. I was not bogged down in daily staff meetings, weekly department meetings, etc. I had input on who to hire for my team. Most of all, I Had a Door I Could Close (but never did). Treat your employees like intelligent people, give them the tools they need, get out of the way, and they will not only be happy but productive as [censored]."*

The fact that our article "went viral", appearing in all the major social networks, in the most employed social-driven websites for practitioners, in the news outlets around the world, in the internal networks of tech giants and corporations, but also generating such a rich engagement from practitioners, has shown that practitioners are interested and keen to discuss them linkage between happy developers and performance.

This is notable because researchers often perceive that practitioners are not interested in their work. Practitioners, on the other hand, often feel that researchers are not listening to their needs. Instead, we learned that practitioners pose much attention to human-related research on software engineering. Therefore, we–as researchers–need to learn to communicate our interesting results and insights in venues and places where practitioners meet.

## 8.5 Implications

This section reports the major implications for research and for practice that are carried by the major research results of this PhD's research activities.

### 8.5.1 Implications for research

Three major implications for research were identified from the main results of this PhD's research activities. They are explained in this section.

The implications for research from the variance based studies lie in the tested theories themselves. First, positive affects of software developers are indicators of higher analytical problem-solving skills. Second, our work empirically demonstrates the linkage of affects and performance over time. The real-time affects related to a software development task are positively correlated with a programmer's self-assessed productivity.

These results have important consequences for successive studies in the field of human aspects in software engineering, because they lay down basic building blocks—i.e., future assumptions—for conducting research on affects of developers and any other related construct such as motivation, commitment, and job satisfaction.

The implications for research of the process based study are in the theory itself, as well. The theory incorporates the impact of affects on performance through an influence on the focus of developer's consciousness on coding and on several aspects of goal settings (task, project). In addition, we introduced the concept of attractors for developers, which is a novel construct based on affects and events at different spheres (work-related and not, private or public). The theory is proposed as part of basic science of software engineering, and it is open to falsification and extension. The theory can be employed to frame much future research on affects of software developers and act as a base to understand the complex team dynamics.

The third major implication for research is that our studies have shown, for empirical software engineering research, the benefits of employing validated psychological tests—e.g., the test for creative performance, the consensual assessment technique for evaluating creativity, and the psychology experiment building language—and how to conduct and analyze data with multiple dependencies in the context of repeated within-participant measurements. The positive feedback received by the lessons that we learned and shared caused us to the more active on this avenue. Therefore, we highlighted and presented a substantial body of knowledge in psychology and management research on the affects and their impact on cognitive processes. We presented the most important theories behind affects, their classification, and their measurements, and on the best practices to perform psychological measurements in the context of empirical software engineering. The theoretical foundations of affect have been presented to a software engineering research audience, and act as basic building blocks upon which any research in the domain of the affect of developers can build. As a result of these research activities for the community, we proposed the initial guidelines to conduct research in empirical software engineering with psychological measurements and theory, namely "psychoempirical software engineering".

The "psychoempirical software engineering" initiative is not within the principal focus of this dissertation—indeed, it was included as Appendix C. However, it might perhaps be the most important outcome of the research activities conducted during this PhD cycle. Together with the concurrent proposal of "behavioral software engineering" (Lenberg *et al.*, 2014, 2015), which itself has had its roots in "individualized software engineering" (Feldt *et al.*, 2008), "psychoempirical software engineering" research has the potential to reveal new avenues of research and a new understanding on how software is produced and how process can be improved. "Psychoempirical software engineering" in its current state has a more operational angle and it leans towards the empirical side of software engineering research, while "behavioral software engineering" has proposed a definition and categorized existing research. It is intended by the present author to initiate a discussion with the authors of "behavioral software engineering" towards unified terms, guidelines, and theoretical foundations, and to advocate the study of software engineering with psychology. The potential benefits for the research and practice are limited only by the willingness of the authors to adopt the guidelines and psychology as a way to better understand software construction.

### 8.5.2 Implications for practice

Three major implications for practice were identified from the main results of this PhD's research activities. They are explained in this section.

Our results have implications for management styles. The results have offered an initial support for the claim that an increase in developers performance is expected by making them happier. The results may partially justify the workplace settings of currently successful and notable Silicon Valley ventures, which provide several incentives to entertain their software developers. Secondly, our results suggested that increases in performance are explained by increases in the affective states of developers. Therefore, managers, team leaders, and leaders in general should expect higher performance when the affects triggered by a development task are positive. Moreover, the strong changes in affects and performance in short intervals of time have the implication that managers should care about their developers all day long and continuously.

Finally, the implications for practice are to be found in the evidence for a positive attitude of practitioners towards studies of their affect and the working conditions. Practitioners are prone to share their feelings and their attitude toward their working conditions. Managers and team leaders are thus encouraged to let developers express their affective states in order to provide new insights on their performance. Especially when negative affects are raised on the job, developers should be listened to and let to express their own individuality.

Overall, the theory proposed in this PhD dissertation has the implication that, despite the idea among managers that pressure and some negative feelings help in getting the best results out, there is growing evidence that fostering (hindering) positive (negative) affects of software developers has a positive effect on the focus on code, and task and project goal settings, and, consequently, on their performance. Additionally, we found evidence that a mediator role to reconcile developers' issues and conflicts is a way to foster positive affects and mediate negative attractors. The theory can be employed as a guideline to understand the affective dynamics in a software development process. The theory can be used to build a better environment in a software development team and to guide managers and team leaders to improve developers' performance by making the developers feel better. On the other hand, our conceptualized theory can guide the team leaders to understand the dynamics of negative performance when it is linked to negative affects.

# Chapter 9

# Conclusion

For years, it has been claimed that a way to improve software developers' productivity and software quality is to focus on people and to make software developers satisfied and happy. Several Silicon Valley companies and software startups are following this advice, by providing incentives and *perks*, to make developers happy. However, limited research has supported such claim.

A proposal to study human aspects in empirical software engineering research has been to adopt psychological measurements. By studying the reference fields—primarily psychology and organizational research—this thesis has supported the advocates of studying the human and social aspects of software engineering research by providing the basic building blocks for understanding software developers and their performance under the lens of how they feel. That is, this PhD's research activities aimed to theorize the impact of affects on software development performance.

Through the research activities published in 7 articles in international venues, we provided a new understanding of the complex relationship between the affects of developers and their performance while they develop software. This PhD's contribution lies in the theoretical foundations of the affects of software developers and the validated ways to assess these affects, the guidelines for conducting studies in software engineering with psychology, and a three-faceted theory of the relationship of affects and the performance of software developers. The latter is the main contribution of this dissertation.

The theory is expressed into two variance facets and one process based facet. Overall, the facets express what is the relationship of the pre-existing affects of software developers and their problem solving performance, the relationship of the immediate, real-time affects and the productivity on a software development task, and the process that explains how affects impact the performance while developing software.

Software developers are unique human beings. By embracing a multidisciplinary view, human factors in software engineering can be effectively studied. By inspecting how cognitive activities influence the performance of software engineers, research will open up a completely new angle and a better understanding of the creative activity of the software construction process.

## 9.1 Answers to the research questions

The research question of this dissertation, *How are affects of software developers related to their performance?*, was introduced in Section 1.2. Answering the research question of this dissertation requires a deep understanding of the theoretical foundations of the constructs to be studied, namely affects and performance, and how these constructs are related in terms of impact, over time, and the process behind their relationship.

For this reason, the main research question of this dissertation was divided into four research sub-questions. This section provides answers to the research questions with a summary of the results.

### RQ1—What are the theoretical foundations regarding the affects of software developers?

The answer to RQ1 is to acquire an understanding of the constructs under study, namely affects, emotions, moods, and performance. Therefore, Chapter 2 through the publications Paper II; Paper III answers the research question. The scientific knowledge provided by Chapter 2 is a contribution of this dissertation, which answers RQ1 by translating theory from the psychology disciplines to the software engineering domain.

The chapter (1) conceptualizes the system parts of the mind by emphasizing that affects are considered an important component of the mind, which influences the cognitive and conative parts; (2) extensively explores the two major frameworks from psychology and organizational behavior for categorizing affects, and a recent unifying framework and theory; (3) provides the common ways to measure the affects and how to cope with the several issues that arise when employing psychological measurements in software engineering; (4) highlights the common misconceptions when dealing with the affects of software developers. (5) reviews the theory about performance in psychology and organizational behavior research, and in software engineering research; (6) provides the theoretical foundations of the relationship between affects and performance in psychology and organizational behavior.

## RQ2—How do affects indicate problem-solving performance among developers?

We conducted the study in Paper IV for testing a variance theory of the relationship of pre-existing affects, in the form of affect balance, and the analytic problem solving performance, in the form of the Tower of London game. The results have shown that the happiest software developers outperform the other software developers in terms of analytic problem solving. The results reached statistical significance. Thus, with the limitations described in Section 4.3 and in the next section, we claim that the happiest software developers are those with the highest analytic problem-solving performance, and that positive affect is an indicator of higher analytic problem solving performance.

The study in Paper IV was also conducted for testing a variance theory of the relationship of pre-existing affects, in the form of affect balance, and the creative problem solving performance, in the form of a creative generating task later analyzed with the Consensual Assessment Technique. The results could not provide any support that happy software developers are more creative, and we report a so called *null result* (or negative result) for the creativity part of RQ2. Still, we encourage more research towards answering this question because our study was the first ever in the field, and more evidence should be collected.

## RQ3—How are affects correlated to in-foci task productivity of software developers?

We conducted the study in Paper V; Paper VI for testing a variance theory of the relationship of affects, in the form of the valence, arousal, and dominance dimensions, and the real time, self-assessed software development task productivity. The tested theory is variance based, yet it incorporates the dynamic component of time. By analyzing the data coming from developers working on their software development task using a linear mixed effects model, we found a significant, positive correlation between affect in the terms of valence—the attractiveness perceived towards the development task—and dominance—the perception of possessing adequate skills for facing the task—and the productivity of developers. Thus, with the limitations described in Section 5.3 and in the next section, we claim that there is a positive correlation of affects and the real-time task productivity of software developers. The claim is however limited in terms of the arousal components of affects, as no significant correlation could be found.

## RQ4—How do software developers experience affects and their impact on their development performance?

We conducted the study in Paper VII for constructing a process theory of the experience of affects while developing software and how affects impact the software development performance.

As a result of our observation activities, which were based on face-to-face, open-ended interviews, in-field observations, and e-mail exchanges, we constructed a novel explanatory theory of the impact of affects on development performance. The proposed theory builds upon the concepts of events, affects, attractors, focus, goal, and performance, and how they interact with each other. In particular, as an answer to RQ4, we theorized a causal chain between the events and the programming performance, through affects or attractors. We theorized that positive affects (negative affects) have a positive (negative) impact on the programming task performance by acting on the focus on code, and task and project goals.

The theory introduces the concept of attractors, which are affects that obtain importance and priority to a developer's cognitive system and, often, to their conscious experience. Attractors have an even higher impact on programming performance than ordinary affects.

Finally, we also theorized that fostering positive affects among developers boosts their performance and that the role of a mediator bringing reconciliations to team members might be necessary for successful projects.

## 9.2 Limitations

As with all studies, this work has limitations. The limitations related to the single studies, and how they have been mitigated, were reported at the end of the respective chapters. This section identifies the three major limitations of the overall research, which builds on top of those of the single studies.

The first major limitation of this work lies in the samples obtained in the studies, in terms of size and characteristics. While a bigger sample could offer an appreciably higher variation of participants' characteristics towards a higher generalizability of the results, this threat has been mitigated in all studies. Second, all three major empirical studies have had participants who were computer science students, with the exception of the study in Paper V; Paper VI which had half of the participants as software professionals. Although it is very common for software engineering studies to recruit computer science students as participants to studies (Salman *et al.*, 2015), for some readers this might still be considered a limitation. Tichy (2000); Kitchenham *et al.* (2002) argued that students are the next generation of software professionals as they are close to the interested population of workers, if not even more updated on new technologies. Indeed, the empirical studies comparing students in working settings with professionals did not find evidence for a difference between the groups (Svahnberg *et al.*, 2008; Berander, 2004; Runeson, 2003; Höst *et al.*, 2000; Salman *et al.*, 2015). The conclusions from the previous studies are that students are indeed representatives of professionals in software engineering studies. Third, one could argue that this work did not control factors for individual differences such as personality. Kosti *et al.* (2014, 2015) have recently studied personality clusters of software engineers, as well as some of their attitudes. Some personality archetypes were offered by the latest research. While this dissertation has excluded such avenues from its scope (see Section 1.3) as there was the need to concentrate on laying down basic building blocks, the present candidate recognizes that human aspects are variegated, complex, and often dependent on each other. More studies are needed on the specific constructs touched upon this section and Section 1.3, and future work should understand the relationship that such psychological constructs have on each other in the context of software engineering.

The second major limitation of this work lies in the issues with the relationship between theory and observation (or construct validity). One of our studies did not adopt strict coding-related tasks for assessing the software development performance. To our knowledge, there have been no studies in software engineering research using software development tasks that are suitable for measuring the creativity and analytical problem-solving skills of software developers. Although strict development tasks could be prepared, there would be several threats to validity. Participants with various backgrounds and skills are expected, and it is almost impossible to develop a software development task suitable and equally challenging for first year BSc students and second year MSc students. The study remained at a higher level of abstraction. Consequently, creativity and analytical problem-solving skills were measured with validated tasks from psychology research. Similarly, a validity threat might come from the use of the self-assessed productivity in the second study. Given the difficulty in using traditional software metrics (the project, the task, and the programming language were random in this study), and that measuring software productivity is still an open problem, self-assessed performance is commonly employed in psychology studies (Beal *et al.*, 2005; Fisher and Noble, 2004; Zelenski *et al.*, 2008). Additionally, self-assessed performance is consistent (yet not preferred) to objective measurements of performance (Dess and Robinson, 1984; Miner and Glomb, 2010). There is also the evidence that bias is not introduced by mood in self-reports of performance when individuals alone are being observed Beal *et al.* (2005); Miner and Glomb (2010). We carefully observed the participants during the programming task, and this further reduced the risk of bias. Post-task interviews included questions on their choices for productivity levels, which resulted in remarkably honest and reliable answers, as expected. The discussions at the post-interviews with the participants on their self-assessed productivity values reinforced the validity of the data.

One may argue that a limitation is an absence of a psychology degree in the present candidate. The present author has a BSc and MSc in computer science. There had been the risk that several psychological aspects would have been overlooked or misinterpreted. However, the present author has taken the challenge very seriously. The study of the literature and basic textbooks in psychology and organizational behavior took the majority of the research efforts in the first half of the PhD, and was carried on throughout the entire PhD cycle. The author has also spent the initial months of research by networking and actively discussing with researchers in the affective aspects of human-computer interaction and trained psychologists. While the present author would never claim to be able to act as a trained psychologist, a fair confidence has been reached on studying the topic of affects—and not the treatment of affect-related pathologies—and their placement in organization studies. This gained confidence was put into test by submitting the study in Paper IV to a psychology venue. The study was reviewed by researchers and editor from psychology, and it was accepted and published [1]. This has shown that the gained knowledge in this particular topic has put the present author in the position to sustain an academic discourse with peers from both the disciplines of psychology and software engineering.

## 9.3 Future research

The scope of this dissertation was to lay down the basic building blocks for understanding the affects of software developers and the relationship between the affects of developers and the programming performance. The basic building blocks are required in order to perform any kind of research activity, because they open up avenues for research. Yet, four major directions for future research are provided here.

An immediate indication for future research is a consequence of the intentional narrow scope of this work. This dissertation has focused on individual developers as unit of analysis—also when studying a team—and mostly on programming activities. Future work should raise our results at the team level, at the organizational level, and broaden the focus on artifacts different than code, such as requirements and design artifacts.

Another direction for future research is the replication of our experiments with larger samples, in order to provide a more reliable generalization of the findings. A replication of the experiment in Paper IV with a larger sample may provide significant data and could even enable regression analyses to verify how the intensity of affective states may impact on the problem-solving performance of software developers.

---

[1] Also the study in Kuzmickaja *et al.* (2015), not related to this dissertation, was published in a social-oriented venue and reviewed by researchers in psychology.

Besides replicating the experiments, future research is invited to test our process theory. As the study in Paper VII was qualitative, we suggest future research to test the proposed theory and to quantify the relationships using quantitative data. Although quantifying the impact of attractors was beyond the scope of this study, we feel that negative attractors triggered by non work-related events and positive attractors triggered by work-related events have the strongest impact on the performance of software developers. Furthermore, this study focused on the dimensions of positive and negative affects. It is expected that some types of affects and attractors matter more than others, and have different impact on the focus and performance. We leave future studies the option to study discrete affects, e.g., joy, anger, fear, and frustration. More generally speaking, we invite future studies to seek for a richer exploration of attractors, a better understanding of them, or even a refutation of them.

Our last specific suggestion for future studies is to focus on dynamic, episodic process models of affects and performance where time is taken into consideration. The underlying affects of developers change rapidly during a workday. The constituents and the effects of such changes should be explored. Additionally, exploring the dynamics of affects turning into attractors (and possibly vice-versa) and what causes such changes will provide further understanding of the effectiveness of interventions and of how to make developers feel happier, thus more productive.

It is of the present author's opinion that more studies are needed to understand the complex relationship of affects and software developers' creativity, motivation, commitment, job satisfaction, and well-being.

# Appendix A

# Reference sheet for the participants of the experiment in Chapter 4.

Daniel Graziotin, Xiaofeng Wang, Pekka Abrahamsson

Faculty of Computer Science, Free University of Bozen-Bolzano, Italy

{first.last}@unibz.it

Hi and thank you for participating to this experiment. This sheet contains the instructions for completing it. First, please do not logout/shutdown/reboot the PC. We will lose your data otherwise.

Your Reference Code is: **<Reference Number>**

Please, provide it when requested. The experiment is completely anonymous. We only need the Reference Code to connect your surveys with the data that you will provide us during the experiment.

If you have a question, feel free to call one of the supervisors whenever you want.

The following are the experiment phases.

## A.1 Survey

Please open the browser and go to <URL> to reach the survey. Answer to all the provided questions. Remember that the period is the past 4 weeks, including right now. Provide the Reference Code **<Reference Number>**.

Remember to submit the Survey once you have finished it. It should take you less than 5 minutes to complete it, but take your time.

## A.2 Photographs game

Go to the supervisors and provide your Reference Code **<Reference Number>**. You will receive two photographs, one at a time. Imagine that you are participating in the Best Caption of the Year contest. This contest is organized by a famous magazine and the winning captions will be published along with the photographs. Your job is to try to win this contest by writing the best captions possible for each of these two photographs. The captions can be absolutely anything you would like. You can write as many captions as you would like. Please, remember to write your Reference Code **<Reference Number>** in the photographs, too. This task should take you less than 20 minutes, but take your time.

## A.3 Survey

Please open the browser and go to <URL> to reach the survey. Answer to all the provided questions. Remember that the period is the past 4 weeks, including right now. Provide the Reference Code **<Reference Number>**.

Remember to submit the Survey once you have finished it. It should take you less than 5 minutes to complete it, but take your time.

## A.4   Tower of London game

Please open the PEBL software. As Participant Code (located in the top-center section of the user interface), enter your Reference Code **<Reference Number>**. Do not press the "+" button.

On the left side panel, follow this path: *battery/ -> tol/ -> TOL.pbl*. Select TOL.pbl with the mouse. Click the button labeled "Add to Chain". The TOL.pbl will appear in the Experiment Chain list. Click the button labeled "Launch Chain". A new window will appear. When requested, press key 3 on the keyboard to select Shallice Test ([1, 2, 3] pile heights, 3 disks, Shallice's 12 problems).

It should take you about 10 minutes to finish the game, but take your time.

When you finish the game, please call one of the supervisors of the experiment. Do not close the program. We remember you again; please do not logout/shutdown/reboot the PC.

Thank you for your collaboration.

# Appendix B

# Instructions for participants and interview skeleton for Chapter 5.

Daniel Graziotin, Xiaofeng Wang, Pekka Abrahamsson

Free University of Bozen-Bolzano, Bolzano, Italy

{daniel.graziotin, xiaofeng.wang, pekka.abrahamsson}@unibz.it

## B.1   Instructions for participants

The following section contains the guidelines, which were employed to administer the Self-Assessment Manikin (SAM) [1] and the productivity questionnaire to the participants. The guidelines have been written by following the technical manual by Lang et al. [2].

I appreciate your participation in this study. I am interested in observing human behaviors during the development of software. I will now describe you how this study works. I am going to interview you about your demographic data, the project you are working on and the task that you will face today. For about the next 90 minutes, you are going to work on your task. Each 10 minutes, you will be rating your task in terms of how do you feel while working on it. At the end of your task, I will interview you again about how the task went. There are no right or wrong answers, so simply respond as honestly as you can.

Now, let me explain your involvement in more detail. You are going to see the following on my tablet <*show a paper version of the survey*>. You can see 3 sets of 5 figures and a sentence. The three sets of figures are arranged along a scale. You will use these figures to rate how you feel while working on your code. They show three different kinds of feelings: Unhappy vs. Happy, Calm vs. Excited, and Controlled vs. In-control.

The first scale is the **unhappy-happy** scale, which ranges from a frown to a smile. If you feel completely unhappy, annoyed, unsatisfied, melancholic, despaired, bored, you can indicate this by choosing the figure at the left. The other end of the scale is when you feel completely happy, pleased, satisfied, contented, and hopeful. The figures also allow you to describe intermediate feelings of pleasure, by choosing any other pictures.

The **calm vs. excited** dimension is the second type of feeling displayed here. At one extreme of the scale you feel relaxed, calm, sluggish, dull, sleepy, and un-aroused. On the other hand, at the other end of the scale, you feel completely stimulated, excited, frenzied, jittery, wide-awake, aroused. You can represent intermediate levels by choosing any of the other figures.

The last scale of feeling that you will rate is the dimension of **controlled vs. in-control**. At one end of the scale you have feelings characterized as completely controlled, influenced, cared-for, awed, submissive, and guided. At the other extreme of this scale, you feel completely controlling, influential, in control, important, dominant, and autonomous. If you feel neither in control nor controlled you should chose middle picture. Your rating should reflect your immediate personal experience, and no more.

The fourth item of the survey is the sentence "My **productivity** is", followed by an ordered scale of endings for the sentence (very low, below average, average, above average, very high). You should complete the sentence by choosing the appropriate end that describes your current productivity. If you self-assess that your effort on the task is not what you expected, choose "very low". If you self-assess that your effort is a lot more than what you expected, choose "very high". If you think that your effort is what you expected to be, choose "average". You can also choose intermediate values such as "below average" and "above average".

I remember you that there are no right or wrong answers, so simply respond as honestly as you can. I show you how the survey works on my tablet. You simply touch the picture representing your answer and then press the "Submit" button. Are there any questions before we begin? Feel free to ask questions during the survey if you have any doubts.

## B.2 Interviews

Although the questions look to be closed-ended, they are employed to start an open conversation with the participant.

Pre-task Interview Starting Questions

- Can you tell me your name, age, current year of study or current degree?

- Can you describe the project you are currently working on?

- What is your actual role for this project?

- Have you worked on similar project in the past? How do you consider your experience with this domain on a scale between 1 to 3, where 3 is high?

- What is the programming language employed for the project? Any frameworks involved? How do you consider your experience with this programming language on a scale between 1 to 3, where 3 is high?

- What are you working on during the last days, for this project?

- Can you describe me the task that you are going to face today?

Post-task Interview Starting Questions

The reader is advised the penultimate question listed above has to be asked at last, in order to not influence the response to the previous one and to not reveal the participant what the experiment is about.

- Can you summarize the task you worked on for 90 minutes?

- Are you satisfied with the results of your performance?

- What difficulties have you faced?

- Have you been productive during these 90 minutes?

- What convinced you that your productivity at this point <shows a graph of productivity> was higher/lower than the previous measurement time?

- What had an impact on your productivity during the task?

- Did the survey annoy you in any possible way? Has it disturbed you to the point to reduce your productivity?

- Did your mood or your emotions have an impact on your productivity for this task?

- What do you think is productivity while developing software?

## B.3    Measurement instrument

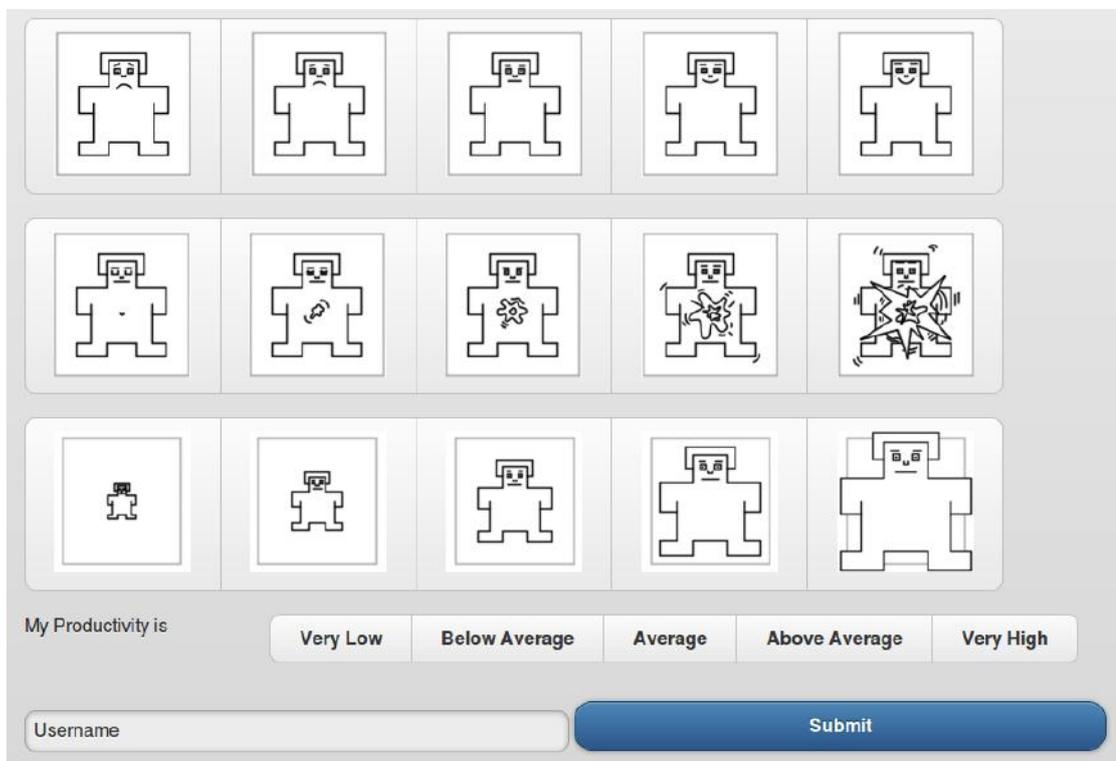

FIGURE B.1: *A screenshot for the measurement instrument*

Based on the work of Bradley & Lang [1] with permission of Elsevier; however obtained from http://irtel.uni-mannheim.de/pxlab/demos/index_SAM.html.

# Appendix C

# Guidelines for Psychoempirical Software Engineering

This appendix provides an excerpt of Paper II, which had the main contribution to provide the theoretical foundations of affect theories for software engineering but also the proposal of guidelines for conducting software engineering studies with psychology. We have called it "psychoempirical software engineering", which was first presented at the 2014 annual meeting of the International Software Engineering Research Network (ISERN) (Graziotin *et al.*, 2014d).

The present author decided to place the guidelines in the Appendix of this dissertation because they are an important output related to this PhD's research activities, yet they are not really part of the main dissertation.

## Proposal: theoretical background of affects and guidelines for Psychoempirical Software Engineering

While it would be preposterously arrogant on our side to claim the all-encompassing knowledge of the topic, we would like to share what we have learned so far in our journey to understanding software developers through their affect. This article builds upon our experience, the feedback collected at our talks and peer review processes, and the previously conducted research, to build some theoretical background for understanding the affect of software developers. We draw from research in psychology in the last decades, and offer a comprehensive review of the theory of affect (Chapter 2) and, as a follow-up of our ISERN 2014 workshop (Graziotin *et al.*, 2014d), we propose our guidelines for psychoempirical SE (Section C.1) for conducting studies in SE with psychological theory and measurement.

## C.1  Guidelines for Psychoempirical Software Engineering

A much requested feature in our previous discussions at recent academic venues such as ISERN 2014, CHASE 2015 (Begel *et al.*, 2015), and ICSE 2015 had been *How should one conduct research with psychological measurements?* By making sense of the hundreds of articles we reviewed on psychology and organizational behavior, we came up with a simple series of steps, listed below.

### C.1.1  Defining a research objective

As with any research activity, it is important to understand what we want to do in a study. Suppose two different, yet common scenarios with the affects of developers. They have been adapted from two of our previous studies (Graziotin *et al.*, 2014b, 2015a).

**Scenario A**  Assessing how happy developers are generally.

**Scenario B**  Assessing over a time frame the emotional reaction of a stimulus (e.g., employing a software tool) on developers.

Both of them require a deep understanding of the topic under study.

### C.1.2  Theoretically framing the research

*Scenario A*—From a comprehensive literature review, we would understand that we can call happy those developers who are in a strongly positive mood, or those who frequently have positive and meaningful experiences (see (Graziotin *et al.*, 2015c) for more), thus having a positive affect balance. We decide to focus on dimensions of affects, e.g. with the Positive and Negative Affect Schedule (PANAS) (Watson *et al.*, 1988a,b), which still lets us evaluate discrete affects before the aggregated scores.

*Scenario B*—Suppose that, instead of asking a developer what emotions she is feeling when using a tool, we are interested in knowing how she feels in terms of more aggregated dimensions like pleasure, energy, and dominance. We focus then on the dimensional theory of affects like the one in the PAD models (Russell and Mehrabian, 1977; Russell, 1980; Mehrabian, 1996).

### C.1.3 Selecting a validated measurement instrument

*Scenario A*—The PANAS dimensional model recommend employing the PANAS (Watson *et al.*, 1988a,b) measurement instrument which is one of the most notable measurement instruments for affective states. However, a deeper look at the literature shows that there are several shortcomings that have been criticized for this instrument. The PANAS reportedly omits core emotions such as *bad* and *joy* while including items that are not considered emotions, like *strong*, *alert*, and *determined* (Diener *et al.*, 2009a; Li *et al.*, 2013). Another limitation has been reported in its non-consideration of the differences in desirability of emotions and feelings in various cultures (Tsai *et al.*, 2006; Li *et al.*, 2013). Furthermore, a considerable redundancy has been found in PANAS items (Crawford and Henry, 2004; Thompson, 2007; Li *et al.*, 2013). PANAS has also been reported to capture only high-arousal feelings in general (Diener *et al.*, 2009a).

Recent, modern scales have been proposed to reduce the number of the PANAS scale items and to overcome some of its shortcomings. Diener *et al.* (2009a) developed the Scale of Positive and Negative Experience (SPANE). SPANE assesses a broad range of pleasant and unpleasant emotions by asking the participants to report them in terms of their frequency during the last four weeks. It is a 12-items scale, divided into two sub-scales. Six items assess positive affective states and form the SPANE-P scale. The other six assess negative affective states and form the SPANE-N scale. The answers to the items are given on a five-point scale ranging from 1 (*very rarely or never*) to 5 (*very often or always*). For example, a score of five for the *joyful* item means that the respondent experienced this affective state *very often* or *always* during the last four weeks. The SPANE-P and SPANE-N scores are the sum of the scores given to their respective six items. Therefore, they range from 6 to 30. The two scores can be further combined by subtracting SPANE-N from SPANE-P, resulting in the Affect Balance Score (SPANE-B). SPANE-B is an indicator of the pleasant and unpleasant affective states caused by how often positive and negative affective states have been felt by the participant. SPANE-B ranges from -24 (*completely negative*) to +24 (*completely positive*). The SPANE measurement instrument has been reported to be capable of measuring positive and negative affective states regardless of their sources, arousal level or cultural context, and it captures feelings from the emotion circumplex (Diener *et al.*, 2009a; Li *et al.*, 2013). The timespan of four weeks was chosen in SPANE in order to provide a balance between the sampling adequacy of feelings and the accuracy of memory (Li *et al.*, 2013), and to decrease the ambiguity of people's understanding of the scale itself (Diener *et al.*, 2009a).

*Scenario B*—The PAD dimensional models have been implemented in several measurement instruments. One of the most notable instruments is the Affect Grid (Russell *et al.*, 1989), which is a grid generated by intersecting the axes of valence and arousal accompanied by four discrete affects, i.e. depression-relaxation and stress-excitement, to guide the participant in pointing where the emotional reaction is located. The affect grid has been employed in SE research, e.g. in (Colomo-Palacios *et al.*, 2011). Yet, the grid was shown to have only moderate validity (Killgore, 1998), thus other measurement instruments would be more desirable. Thus comes the Self-Assessment Manikin (SAM, (Bradley and Lang, 1994; Lang *et al.*, 1999)). SAM is a pictorial, i.e. non-verbal, assessment method. SAM measures valence, arousal, and dominance associated with a person's affective reaction to an object (or a stimulus) (Bradley and Lang, 1994). As a picture is worth a thousand words, we reproduce SAM in figure C.1. The figures of the first row range from a frown to a smile, representing the valence dimension. The second row depicts a figure showing a serene, peaceful, or passionless face to an explosive, anxious, or excited face. It represents the arousal dimension. The third row ranges from a very little, insignificant figure to a ubiquitous, pervasive figure. It represents the dominance affective dimension. As reported in (Kim *et al.*, 2002), SAM has the advantage of eliminating the cognitive processes associated with verbal measures but it is still very quick and simple to use.

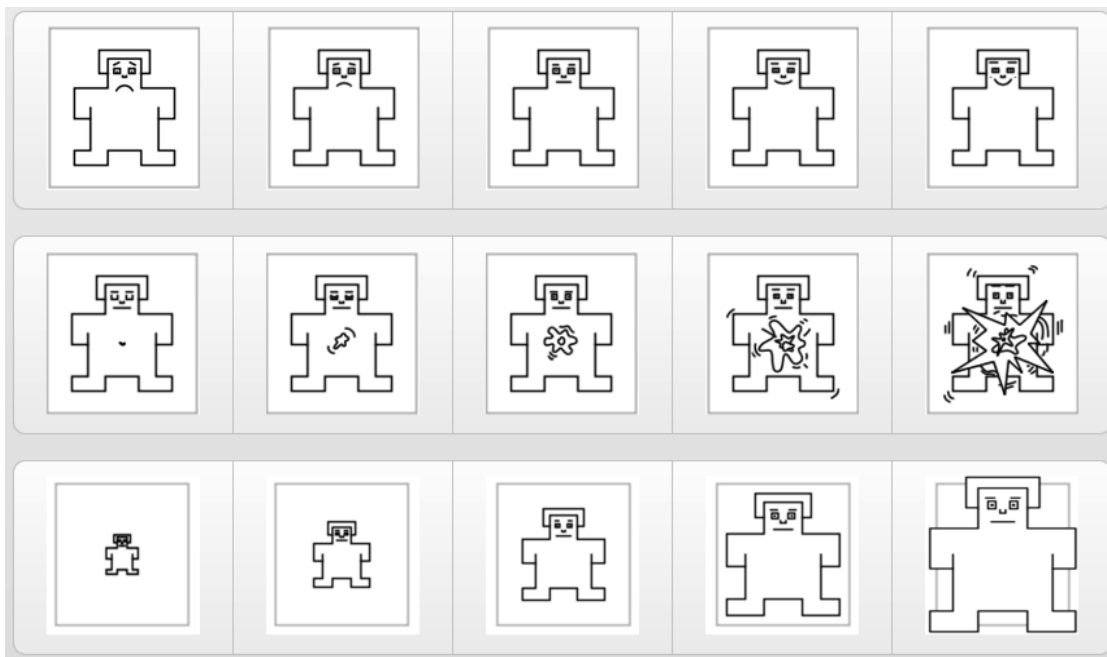

FIGURE C.1: *The Self-Assessment Manikin.*

### C.1.4 Considering psychometric properties

As we noted in a previous paper (Graziotin *et al.*, 2015c), a selected measurement instrument has to possess acceptable validity and reliability properties, which are provided in psychometric studies of the measurement instrument. Psychometrics is a term, which has been misused in SE including ourselves. It is a subfield of psychology that focuses on the theory and techniques of psychological measurements. Psychometric studies deal with the design, development and especially the validation of psychological measures.

A modification to an existing measurement instrument (e.g., adding, deleting, or rewording items) often requires a new psychometric study because the reliability of a measurement instrument can be compromised. Therefore, it is not advisable to modify validated psychological measurements or models as it happened in (Colomo-Palacios *et al.*, 2013).

*Scenario A*—The SPANE has been validated to converge with other affective states measurement instruments, including PANAS (Diener *et al.*, 2009a). The scale provided good psychometric properties in the introductory research (Diener *et al.*, 2009a) and in numerous follow-ups, with up to twenty-one thousand participants in a single study (Silva and Caetano, 2011; Dogan *et al.*, 2013; Li *et al.*, 2013). Additionally, the scale proved consistency across full-time workers and students (Silva and Caetano, 2011).

*Scenario B*—The SAM has been under scholarly scrutiny, as well. The original article describing SAM already reports good psychometric properties (Bradley and Lang, 1994). A very high correlation was found between the SAM items and those of other verbal-based measurement instruments (Morris and Waine, 1993; Morris, 1995), including high reliability across age (Backs *et al.*, 2007). Therefore, SAM is one of the most reliable measurement instruments for affective reactions (Kim *et al.*, 2002).

### C.1.5   Administering the measurement instrument correctly

The psychometric properties of a measurement instrument in psychology are also also calculated by administering the instrument in the same way in each study. This is because the instructions might influence the participants' responses. For this reason, any good measurement instrument is always accompanied with the instructions for the participants. We encourage administering a measurement instrument as it is reported in the accompanying instructions, and to further share the instructions with participants. Furthermore, the gained transparency ensures a higher reproducibility of the studies.

We strongly encourage the authors of SE studies to report the participants' instructions when publishing an article, preferably in an archived format. [1]

*Scenario A*—The SPANE instructions for participants are clearly stated in the original paper (Diener *et al.*, 2009a) and in the instrument itself, which is freely available. [2]

*Scenario B*—The SAM instructions for participants are exhaustively reported in the accompanying technical report (Lang *et al.*, 1999).

### C.1.6   Performing strong analyses

We encourage the authors in SE to spend some time to understand whether such complex and delicate constructs require accurate analyses.

---

[1] For the participants' instructions in (Graziotin *et al.*, 2014b), see `https://dx.doi.org/10.7717/peerj.289/supp-1`. For the participants' instructions in (Graziotin *et al.*, 2013, 2015a), see `http://dx.doi.org/10.6084/m9.figshare.796393`

[2] `http://internal.psychology.illinois.edu/~ediener/SPANE.html`

*Scenario A*—The SPANE scores can be considered as ordinal values or as discrete pinpoints of a continuous scale. Regression analyses on the aggregated SPANE-P, SPANE-N, and SPANE-B scores are possible given that the assumptions for linear regression are met. Otherwise, especially when groups have to be compared, the usual assumptions for employing the *t-test* or non-parametric tests should be taken into account. It is also important to report an effect size measure such as the Cohen's *d*.

*Scenario B*—Repeated measures within-subject that need a between subject comparison pose several issues. First, there is not a stable and shared metric for assessing the affects across persons. For example, a score of one in valence for a person may be equal to a score of three for another person. However, a participant scoring two for valence at time *t* and five at time *t+x* unquestionably indicates that the participant's valence increased. As stated by Hektner *et al.* (2007), "*it is sensible to assume that there is a reasonable stable metric within persons*" (p. 10). In order to have comparable measurements, the raw scores of each participant are typically transformed into *z-scores* (also known as standard scores). A z-score transformation is such that a participant's mean score for a variable is zero, and scores for the same variable that lie one standard deviation above or below the mean have the value equivalent to their deviation. One observation is translated to how many standard deviations the observation itself is above or below the mean of the individual's observations. Therefore, the participants' measurements become dimensionless and comparable with each other, because the z-scores indicate how much the values are spread (Larson and Csikszentmihalyi, 1983; Hektner *et al.*, 2007).

Second, the repeated measurements often present dependencies of data at the participants' level and the time level grouped by the participant. The analysis of variance (ANOVA) family provides rANOVA as a variant for repeated measurements. However, rANOVA and general ANOVA procedures are discouraged (Gueorguieva and Krystal, 2004) in favor of mixed-effects models, which are robust and specifically designed for repeated, within-participant longitudinal data (Laird and Ware, 1982; Gueorguieva and Krystal, 2004; Baayen *et al.*, 2008). A linear mixed-effects model is a linear model that contains both fixed effects and random effects (Robinson, 1991). The estimation of the significance of the effects for mixed models is an open debate (Bates, 2006; R Community, 2006). We encourage the reader to follow our reasoning in (Graziotin *et al.*, 2015a) for a deeper discussion.

# Appendix D

# Articles not included in this dissertation

Besides the articles included in this dissertation, the present author has co-authored the following articles that were not related to the PhD main research activities.

Kuzmickaja, I., Wang, X., Graziotin, D., Dodero, G., and Abrahamsson, P. In Need of Creative Mobile Service Ideas? Forget Adults and Ask Young Children. *SAGE Open*, 5(3):1–15 (2015). doi: 10.1177/2158244015601719

Graziotin, D., Wang, X., and Abrahamsson, P. A framework for systematic analysis of open access journals and its application in software engineering and information systems. *Scientometrics*, 101(3):1627–1656 (2014a). doi: 10.1007/s11192-014-1278-7

Graziotin, D. Recent trends in agile processes and software engineering research - XP 2014 conference report. *The Winnower*, 2:e141901.13372 (2014d). doi: 10.15200/winn. 141901.13372

Graziotin, D. Green open access in computer science - an exploratory study on author-based self-archiving awareness, practice, and inhibitors. *ScienceOpen Research*, 1(1):1–11 (2014b). doi: 10.14293/A2199-1006.01.SOR-COMPSCI.LZQ19.v1

Graziotin, D. An author-based review of the Journal of Open Research Software. *The Winnower*, 2:e140326.62772 (2014a). doi: 10.15200/winn.140326.62772